# Light-Emitting Diodes based on Metal Halide Perovskite and Perovskite Related Nanocrystals

*Ying Liu*[+]*, Zhuangzhuang Ma*[+]*, Jibin Zhang, Yanni He, Jinfei Dai*, Xinjian Li, Zhifeng Shi*, and Liberato Manna**

Y. Liu[+], Z. Ma[+], J. Zhang, X. Li, Z. Shi
Key Laboratory of Materials Physics of Ministry of Education
School of Physics
Zhengzhou University,
Zhengzhou 450052, China
E-mail: shizf@zzu.edu.cn
[+] These authors contributed equally to this work

Y. He, J. Dai
Key Laboratory for Physical Electronics and Devices of the Ministry of Education & Shaanxi Key Lab of Information Photonic Technique School of Electronic Science and Engineering
Xi'an Jiaotong University
Xi'an 710049, China
E-mail: daijinfei123@xjtu.edu.cn

J. Dai
Collaborative Innovation Center of Extreme Optics
Shanxi University
Taiyuan 030006, China

L. Manna
Nanochemistry
Istituto Italiano di Tecnologia
Via Morego 30, Genova 16163, Italy
E-mail: liberato.manna@iit.it

Keywords: light emitting diode, metal halide nanocrystals, luminescence improvement, emission mechanism, device optimization

**Abstract.**

Light-emitting diodes (LEDs) based on halide perovskite nanocrystals have attracted extensive attention due to their considerable luminescence efficiency, wide color gamut, high color purity, and facile material synthesis. Since the first demonstration of LEDs based on $MAPbBr_3$ nanocrystals were reported in 2014, the community has witnessed a rapid development in their performances. In this review, we provide a historical perspective of the development of LEDs based on halide perovskite nanocrystals and then present a comprehensive survey of current strategies to high-efficiency lead-based perovskite nanocrystals LEDs, including synthesis optimization, ion doping/alloying and shell coating. We then review the basic characteristics and emission mechanisms of lead-free perovskite and perovskite-related nanocrystals emitters in environmentally friendly LEDs, from the standpoint of different emission colors. Finally, we cover the progress in LED applications and provide an outlook of the opportunities and challenges for future developments in this field.

## 1. Introduction

Halide perovskites are materials whose structure is made of corner sharing metal halide octahedra that are charge-stabilized by large monovalent cations. Especially the lead halide perovskites have been under the spotlight for their appealing properties, such as strong optical absorption, low exciton binding energies, and large exciton diffusion lengths, which have triggered their exploitation in applications spanning across various fields, and primarily in photovoltaics and light emission.[1] The mechanism of emission from lead halide perovskites is generally based on recombination from free excitons, whereby loosely bound carriers (either photogenerated or electrically injected) are relatively free to move across the substantially rigid lattice of three-dimensionally interconnected octahedra.[2] This gives rise to emission bands that are narrow and with a small Stokes shift from the absorption onset, and which can be easily tuned spectrally by adjusting the halide composition.[3] Light-emitting diodes (LEDs) based on lead halide perovskites stand out as promising candidates due to their tunable spectral properties, considerable luminescence efficiency, low-cost manufacturing and rich libraries of materials. Perovskite LEDs have initially targeted polycrystalline thin films and single crystals as the emitting layers.[4] Typical issues with these approaches are the high density of grain boundaries in the films, acting as charge trapping centers, and the large thickness of single crystals, leading to ineffective carrier collection.[5] Subsequent research on LEDs based on halide perovskites has also targeted the corresponding nanocrystals (NCs), motivated by their high photoluminescence (PL) quantum yield (QY) often reaching near-unit values, tunable emissions, and high color purity. These NCs offer a wide color gamut (>130% National Television System(s) Committee, NTSC) and narrow PL linewidths (typically <20 nm), so they can easily meet the requirements of the Rec. 2020 standard. Perovskite NCs are akin to organic molecules/polymers and to the more traditional colloidal semiconductor quantum dots in terms of technology development, especially for LEDs applications,[6] and can be used straighforwardly for device manufacturing using techniques such as inkjet printing, transfer printing, photolithography and other patterning processes.[7] Compared to films of perovskite prepared by vacuum thermal evaporation, films based on colloidal NCs might be advantageous in terms of better control of phase purity, compositions and bandgaps. These merits place perovskite NCs LEDs among the most promising technologies for next-generation ultrahigh-definition display applications.

Within just a few years, perovskite NCs have emerged as contenders to the already well-developed organic materials and to II−VI semiconductor NCs for LEDs. To give one example, for electrically driven LEDs, the external quantum efficiency (EQE) has surpassed 25% for both green and red emissions.[7d, 8] Yet, many issues remain unresolved. One is related to material instability under operational conditions, and this is reflected in the remarkable drop in the EQEs at elevated current densities or long operational times. Moreover, the use of lead is increasingly restricted due to its toxicity. For example, the European Commission, through the Directive on 'Restriction of Hazardous Substances in Electrical and Electronic Equipment (RoHS)' has set stringent limits on the use of this hazardous heavy metal,[9] calling for more extensive research on halide perovskites and related materials and devices not containing lead in their formulation, an aspect that will be covered extensively in this review.

A systematic summary of LEDs exploiting metal halide NCs would be highly desirable both as a guide for beginners and to stimulate further developments. In this review we will first provide a short historical perspective of the field. We will then focus on strategies that have been developed to date to prepare high-efficiency lead-based perovskite NCs LEDs. After that, we will cover the current state of the art on lead-free metal halide NCs emitters in environmentally friendly LEDs from the standpoint of different emission colors, with ample discussion on their fundamental characteristics and emission mechanisms. We will then review the progress in LED applications including phosphor-converted WLEDs, electrically driven LEDs, micro-LED lighting and displays, active-matrix LED displays and the corresponding device optimizations. Finally, we will provide an outlook on the opportunities and challenges for future developments.

## 2. A short history of perovskite NCs LEDs

The first perovskite NCs LEDs were reported by Schmidt et al. in 2014 using $MAPbBr_3$ NCs as the emitters.[10] Although the luminance of the devices was extremely low (< 1 cd $m^{-2}$), the work opened a window of opportunities for perovskite NCs electroluminescent LEDs. In 2015, Zeng et al. first fabricated all-inorganic $CsPbX_3$ NCs-based LEDs, with an EQE in the 0.1%−0.4% range for orange, green, and blue emission.[11] Various additional works then followed in which NCs with various compositions and/or dimensionality were tested in LEDs.[12]

**Figure 1**a highlights the milestones in the development of perovskite NCs LEDs, while **Table 1** lists and compares representative LED materials and their optoelectronic performance. For high-definition displays based on the International Telecommunication Union Ultra High Definition Television standard (ITU Rec. 2020),[13] obtaining pure blue (460–470 nm), pure

green (525–535 nm) and pure-red (620–650 nm) emission from lead halide perovskite NCs is desirable. Therefore, strategies have been developed to prepare NCs with pure blue, green and pure red emissions, including i) synthesis of NCs with mixed halide compositions, based on the fact that the strong contributions of halide n*p* orbitals to the band edge states leads to an easy tuning of the band gap through mixed halide compositions,[14] ii) preparing perovskite NCs with one or mode dimensions smaller than the exciton Bohr diameter, which leads to strong quantum confinement effects and provides an additional handle to tune the band gap;[7d, 15] and iii) doping NCs with various ions,[16] by which it is possible to tailor the color of emission by modification of the band structure through the influence of the dopants' orbitals.[16a]

Advances in approaches to increase luminescence efficiency, NC stability, and charge transport properties of perovskite NCs (all aspects that will be covered in detail in later sections) have led over the years to improved device performances. Luminescence and charge transport properties of NCs determine the radiative recombination of electron-hole pairs and charge injection efficiency in the device, respectively, which are crucial to improve the internal quantum efficiency of LEDs. Just to give some examples, Wang et al. have recently developed a synergistic dual-ligand approach that greatly improved the conductivity of perovskite NCs, generating a luminance of 1,000 cd $m^{-2}$ at a record-low voltage of 2.8 V for pure-red LEDs.[17] Lee et al. instead developed a simple bar-coating method decoupling the crystallization of perovskites from film formation to achieve a highly efficient and large-area green NCs LEDs, with an EQE of 22.5% for a pixel size of 102 $mm^2$.[18] Apart from the hot-injection method for synthesizing perovskite NCs, direct synthesis-on-substrate have been identified as an effective approach that can avoid long-chain ligands and improve the electrical conductivity of the films, leading to high-performance LEDs.[12a, 19] The best blue, green and red LEDs using lead halide perovskite NCs to date have reached EQEs values of 26.4%, 30.2% and 28.5%,[7d, 20] basically meeting or close to the requirements of commercial products (~30%), see Figure 1b. NIR LEDs on the other hand have achieved EQE that only slightly higher than 20%.

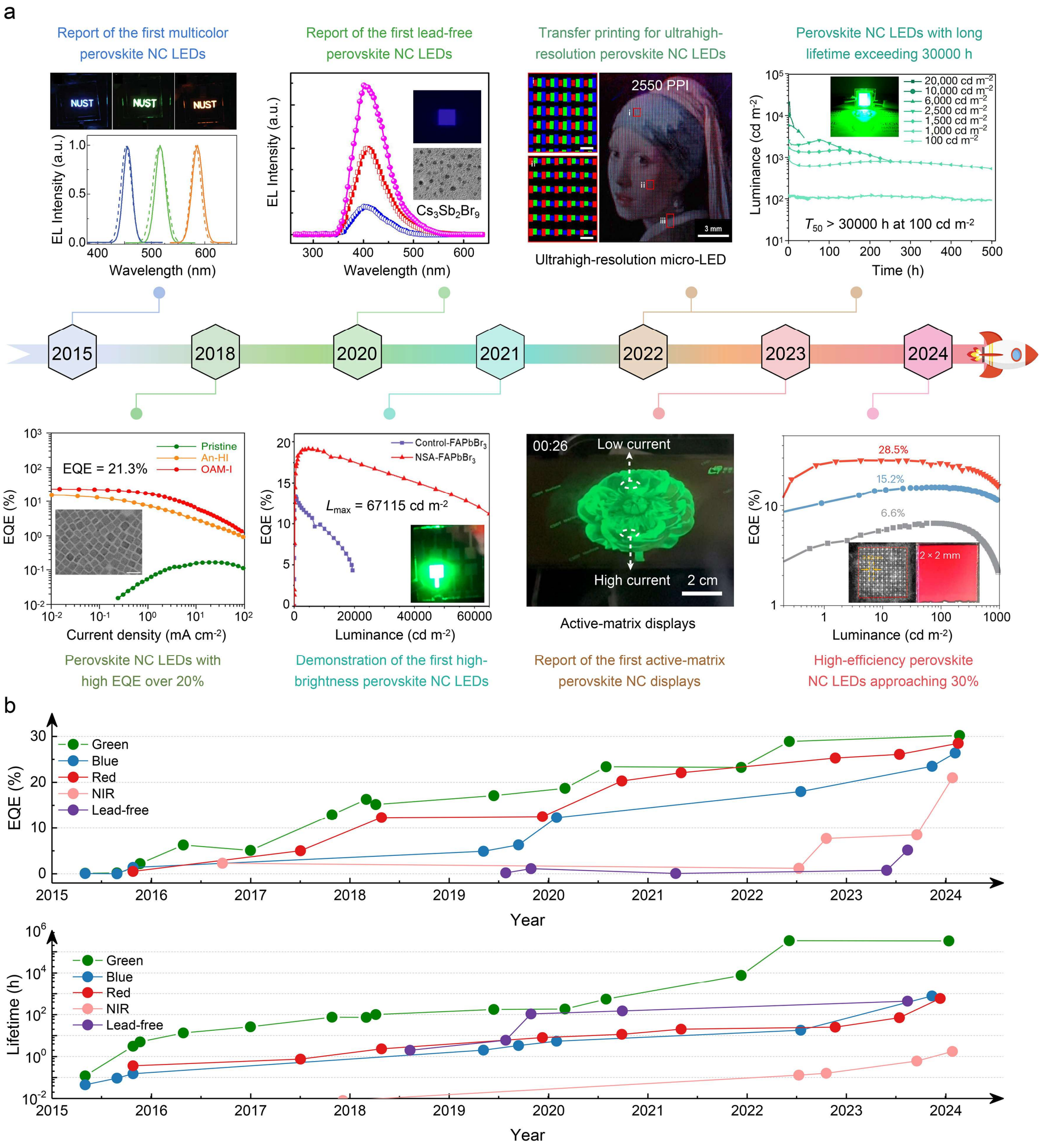

**Figure 1**. Milestones in the development of perovskite NCs LED systems. (a) Timeline of major breakthroughs in perovskite NCs LEDs. Upper part: reproduced and adapted from ref.[11-12, 21]. Left to right: Copyright 2015, Wiley; Copyright 2019, American Chemical Society; Copyright 2022, AAAS; Copyright 2022, Springer Nature. Bottom part: reproduced and adapted from ref.[7d, 22]. Left to right: Copyright 2018, Springer Nature; Copyright 2021, American Chemical Society; Copyright 2023, Springer Nature; Copyright 2024, Springer Nature. (b) Evolution over time of EQE and operational stability in perovskite NCs LEDs. The data are taken from the literature reports discussed in this review.

**Table 1.** Representative characteristics of different types of LED technologies.

| Parameter | Inorganic LEDs | Organic LEDs | II−VI quantum dots LEDs | Perovskite NCs LEDs |
|---|---|---|---|---|
| Representative materials | R: AlInGaP<br>G/B: InGaN | R/G: lr, Pt dopant/host<br>B: anthracene derivative | R/G: InP, CdSe<br>B: ZnSe, CdSe | R: $CsPb(BrI)_3$<br>G: $CsPb(Br)_3$<br>B: $CsPb(ClBr)_3$ |
| Fabrication | Material synthesis, sequential fabrication of each layer (epitaxial growth) | Material synthesis, sequential fabrication of each layer (solution process or vacuum deposition) | Material synthesis, sequential fabrication of each layer (solution process) | Material synthesis, sequential fabrication of each layer (solution process) |
| Emitter thickness | >1 μm | <100 nm | <100 nm | <100 nm |
| Color gamut | – | ~90% | ~120% | ~130–140% |
| State-of-the-art EQE | R/G/B: >50% | R/G: >40%<br>B: >35% | R/G: >30%<br>B: >25% | R/G: >30%<br>B: >25% |
| Max. luminance | ~$10^6$–$10^8$ (cd $m^{-2}$) | ~$10^2$–$10^4$ (cd $m^{-2}$) | ~$10^2$–$10^6$ (cd $m^{-2}$) | ~$10^2$–$10^6$ (cd $m^{-2}$) |
| FWHM | ~40–100 (nm) | ~40–70 (nm) | ~25–50 (nm) | ~15–40 (nm) |
| Operational stability | >$10^7$ h at 1000 cd $m^{-2}$ | >$10^6$ h at 1000 cd $m^{-2}$ | >$10^6$ h at 1000 cd $m^{-2}$ | 10–$10^6$ h at 100 cd $m^{-2}$ |
| Emitter cost | – | ~US$650–1300 per g | ~US$10 per g | ~US$2 per g |
| Development history | ~117 years | ~45 years | ~30 years | ~10 years |

Moving on from lab-scale, proof of concept LEDs, significant efforts have been devoted to develop luminescent applications of perovskite NCs from an industrial perspective.[23] In the display field, perovskite NCs have followed three main technological routes: i) Backlit displays, with NCs replacing traditional red/green inorganic phosphors, and combined with liquid crystal display (LCD) technology, in order to push the color gamut from 70-80% NTSC up to 130% NTSC (commercial stage);[23] ii) Micro-LEDs, using red/green perovskite NCs as color conversion layers combined with blue Micro-GaN-LED arrays to form red, green, and blue pixel display light sources (laboratory stage);[7b] iii) Active-matrix displays, due to the fact that perovskite NCs based LEDs have electroluminescence rise times on the order of milliseconds,[7a] utilizing the independent EL of red/green/blue perovskite NCs, combined with active-matrix thin-film transistor (TFT) driver circuits, to build a display array (laboratory stage).[7d] Yet, the industrialization of perovskite NCs displays have faced many challenges, which can be summarized as follows: i) Stability issues of perovskite NCs, along with a poor understanding of mechanisms such as charge injection/transport/leakage, exciton quenching, lifetime degradation, which fundamentally delay improvements in EL efficiency and stability. To give some numbers, the longest $T_{50}$ for green LEDs is 30,000 h (at 100 cd $m^{-2}$), while the corresponding values for the red and blue devices are only 780 h ($T_{90}$, at 100 cd $m^{-2}$) and 59 h (at 100 cd $m^{-2}$).[12a, 17, 24] ii) Low maturity of the manufacturing processes including inkjet printing, transfer printing, photolithography and other color pattern processes, which restricts the industrialization process of high-resolution full-color perovskite NCs LED displays.

Another area that was soon scrutinized is the development of lead-free perovskite NCs LEDs, driven by concerns about environmental pollution caused by the toxicity of lead.[25] $Sn^{2+}$ ions, for example, have been considered as one of the promising alternatives to $Pb^{2+}$ ions due to their outer electronic configurations (n$s^2$) matching that of $Pb^{2+}$ ions. $CsSnX_3$ NCs, with emission also originating from free excitons, have been tested as the emitters in both phosphor-converted and electrically driven LEDs.[26] The issues with these materials however are their high instability, as $Sn^{2+}$ ions are quickly oxidized to $Sn^{2+}$. Many other lead-free NCs have been investigated. Halide double perovskites, for example, are perovskite structures in which the octahedral sites are alternatively occupied by monovalent and trivalent cations (for example $Cs_2AgInCl_6$). As will be amply discussed in this review, some of these lead-free NCs, upon careful synthesis and often with proper alloying/doping strategies, have demonstrated luminescence efficiency comparable to that of lead halide perovskites, however their emission is generally characterized by a broad full width at half maximum (FWHM > 100 nm) and a large Stokes shift (of hundreds of meV) from the absorbing states.[27] While these features are clearly not desirable for color displays, they are better suited for phosphor converted white LEDs (WLEDs).[28]

Double perovskites share such broadband emission behavior with a plethora of other metal halides having a variety of different crystal structures. This type of emission is generally recognized as stemming from self-trapped excitons (STEs) recombination. STEs are localized excitons generated in a 'soft' lattice environment, typical of metal halides, in the presence of strong electron–phonon interactions, which lead to local lattice distortions in which excitons can be easily trapped.[29] Due to the spin-flip of carriers caused by strong orbital motion, the total angular momentum of STEs can be 0 or 1, corresponding to either singlet or triplet excitons, with a 1:3 ratio. If there is a moderate energy gap between the singlet and triplet states, the excitons can return from the triplet state to the singlet state through the reverse intersystem crossing (RISC) processes with the assistance of thermal energy, theoretically improving the utilization efficiency of excitons. In a ‘soft’ lattice, there can be intrinsic STEs and extrinsic STEs. Both can give contribution to the white emission. Intrinsic STEs are formed in a lattice without permanent lattice defects, in which self-trapped states can be considered as excited states. Once the photoexcitation is removed, the lattice returns to its original state and the excited states (self-trapped states) quickly decay. On the other hand, permanent lattice defects are required to generate extrinsic STEs.[30] Generally, dopants can induce extrinsic STEs.[31]

The STE recombination is very common in lead-free metal halides. In addition to double perovskite NCs,[32] it has been observed for example in many vacancy-ordered perovskites and in $Cu^{+}$-/$In^{3+}$-based halide NCs.[33] $Cu^{+}$-based halide NCs, in particular, can have different emission colors that depend on composition and/or crystal structure (e.g. $Cs_3Cu_2I_5$: deep-blue; $Cs_3Cu_2Cl_5$: green; $CsCu_2I_3$: yellow), with near-unity PLQYs and good stability.[34] Among them, $Cs_3Cu_2I_5$ NCs have been extensively tested in the electrically driven blue LEDs with an EQE of 1.12% and device lifetime of ~108 h ($T_{50}$, at 6.7 V).[35] Other ions with n$s^2$ configurations as in $Pb^{2+}$ or $Sn^{2+}$ ions, such as $Bi^{3+}$ and $Sb^{3+}$, lead to materials with $A_3B_2X_9$ compositions, crystal structures that are quite different from that of a perovskite, and with emissions that are generally in the violet and blue region.[21a, 36] In 2020, Shi et al. employed $Cs_3Sb_2Br_9$ NCs as light emitting layer for LEDs, realizing a violet emission at 408 nm with an EQE of 0.206%.[21a] This low value may be due to the inherent low conductivity of such materials with large band gap and the relative difficulty in injecting holes and electrons due to the large carrier effective masses.

Metal halide NCs based on rare-earth ions ($Eu^{2+}$, $Eu^{3+}$, $Ce^{3+}$ and $Tb^{3+}$) have been synthesized in recent years and have been explored as emitters in electrically driven LEDs.[37] They are characterized with narrowband emissions, stemming from atom-like *f*–*d* or *f*–*f* transitions[38] and in some cases due to Eu-5*d*→Eu-4*f*/Br-4*p* transitions, as for $CsEuBr_3$.[37, 39] A recent paper by Song et al. for example reported an efficient lead-free perovskite LED based on $Cs_3EuCl_6$ NCs with an EQE of 5.17%, through optimization of energy transfer and passivation of surface defects.[38b]

Overall, we can state that, despite the significant progress made on the emission properties of these lead-free metal halide NCs (PLQY >90% and tunable emissions in the visible region), the performance of lead-free metal halides LEDs (EQE <10%) is not yet comparable to that of lead-based devices. Moreover, patterned lead-free NCs and LEDs for display applications have not yet been reported, which may be attributed to the lack of ideal materials and technologies. However, this does not mean that they cannot be realized in the future. The fields remain wide open, and there is ample room for further developments in display technologies. For phosphor converted WLEDs, lead-free metal halide NCs are promising, since they usually have good stability, large Stokes shifts, and efficient broadband emission that are the prerequisites for commercialization. These characteristics are not present in lead-based perovskites.

## 3. A survey of current strategies to high-efficiency lead-based perovskite nanocrystals LEDs

### 3.1 General strategies adopted to tune the color of emission of Pb-based perovskite NCs

We begin this section by providing a more in-depth coverage of the current strategies aimed at tuning the emission color of Pb-based perovskite NCs which have then been used in LEDs, starting from blue emission (**Figure 2**). To this scope, four types of blue emitting lead halide perovskite NCs have been successfully developed to date: mixed halide NCs ($CsPb(Cl,Br)_3$), A/B-site alloyed NCs, ultra-small $CsPbBr_3$ NCs, and $CsPbBr_3$ nanoplatelets (NPLs) and nanorods (NRs). Mixed halide perovskite NCs are the most common blue emitters in LED devices. Their band gap can be easily adjusted by tuning halide ions ($X^-$) composition, due to the dominant contribution of the halide n*p* orbitals to the valence band maxima (VBM) (**Figure 2**a, b).[40] Examples are shown in Figure 2c for mixed Cl/Br perovskites, with emission colors ranging from deep blue to pure blue. Mixed halide perovskite NCs however suffer from the presence of halide vacancies and from halide migration, resulting in low PLQY and emission color instability.[41] Apart from halide ions, the *s* and *p* orbitals of B-site ions ($Pb^{2+}$) in the lead-based perovskites also participate in the electronic structure at the band edge.[40] Thus, choosing suitable ions ($Al^{3+}$, $Sr^{2+}$ and $Cd^{2+}$) to partially replace $Pb^{2+}$ ions can expand the band gap and lead to blue emission in the lead bromide NCs.[16] For example, the incorporation of $Al^{3+}$ ions in $CsPbBr_3$ NCs modifies their electronic structure, widening the band gap and shifting the emission peak from 515 to 456 nm, that is, in the deep-blue region;[16a] Introducing $Cd^{2+}$ in $CsPbBr_3$ NCs has also the effect of increasing the band gap from 2.34 to 2.49 eV, leading to a blue emission at 485 nm.[16c] A-site cations can also affect the band gap, not through direct contribution of the orbitals to band edge states, but through structural distortions. For example, $Rb^+$ alloying at the $Cs^+$ site in $CsPbBr_3$ NCs leads to a blue shift of the emission peak[42] to ~490 nm (that is, in the cyan light region) and this is rationalized as a tilting distortion of the $[PbX_6]^{4-}$ octahedra caused by the smaller $Rb^+$ ions (compared to the $Cs^+$ ones). The octahedral distortion leads to a decrease of the Pb-X-Pb angle and consequently a decreased anti-bonding orbital overlap between $Pb^{2+}$ and $X^{-}$, thus enlarging the band gap.[43] When reducing the size of perovskite NCs along one, two or three dimensions below the exciton Bohr diameter, to form nanorods (NRs), nanoplatelets (NPLs) or ultra-small NCs, blue emission can be achieved due to quantum confinement effects (Figure 2d).[44] For example, the exciton Bohr diameter for $CsPbBr_3$ and $MAPbBr_3$ is around 7 nm.[3a, 45] Therefore, blue emission in pure bromide-based perovskites can be obtained when NCs of these materials are significantly smaller than 7 nm. On the other hand, small-sized perovskite NCs suffer from a high density of surface trap states and easy degradation/aggregation, requiring stabilization procedures, as discussed later.

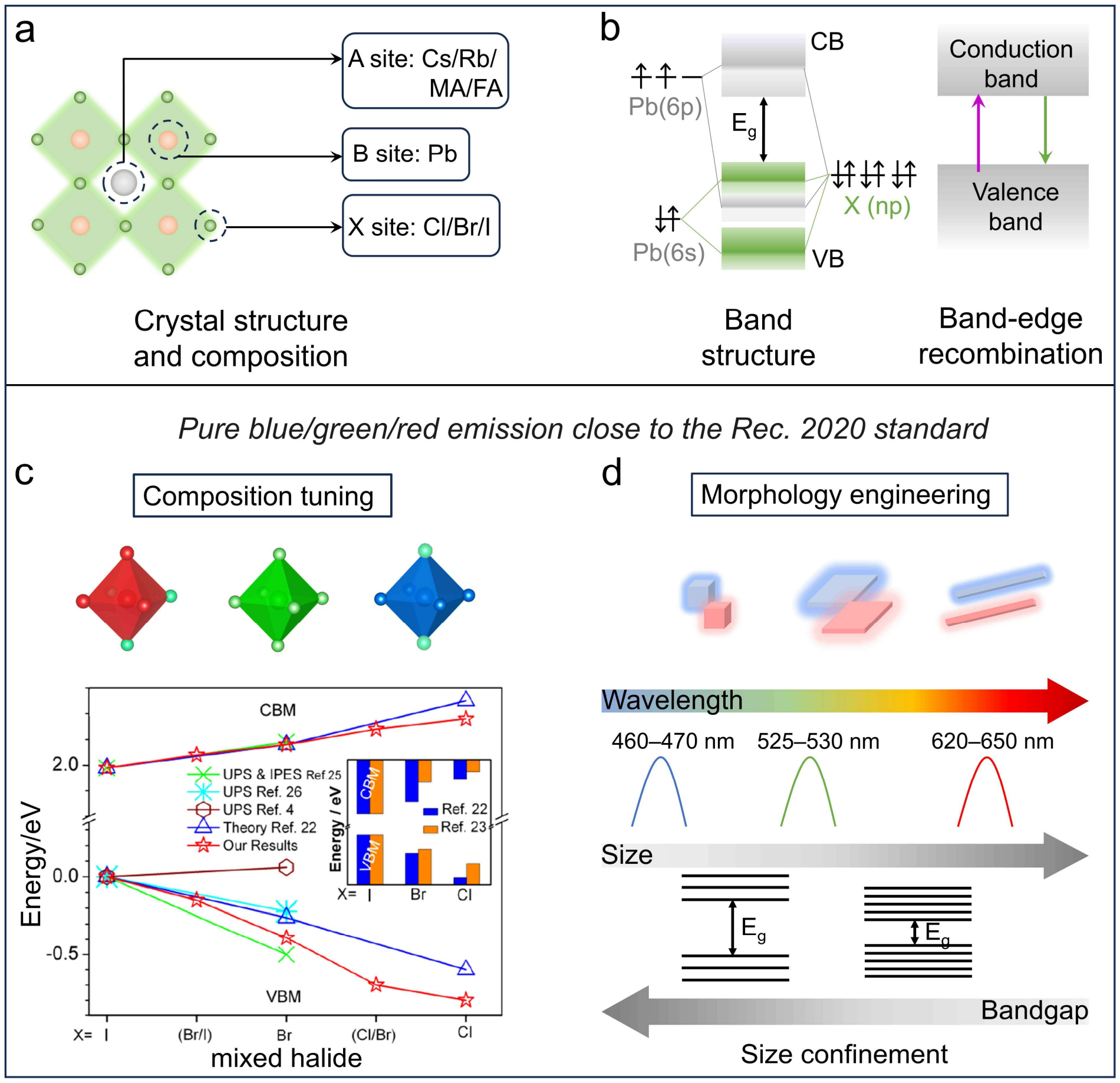

**Figure 2.** Lead-based perovskite NCs as LED emissive layers: (a) Scheme of crystal structure and composition of $APbX_3$ perovskites; (b) Band structure and PL mechanism; (c, d) Composition and morphology engineering aimed at achieving pure blue/green/red emission. Reproduced and adapted from ref.[46]. Copyright 2016, American Chemical Society.

$APbBr_3$ (A = Cs, Rb, MA, FA) NCs[12a, 47] with emissions centered at ~512, ~525, and ~530 nm, FWHM of ~20 nm and PLQYs that can be easily over 80% have been used for green emissive LEDs. When considering the Rec.2020 color standards, $FAPbBr_3$ NCs, with their emission peak closer to pure-green emission, stand out.[48] Yet it is the $CsPbBr_3$ NCs, with their comparatively higher stability, that have been mainly studied in LED applications, even though their emission chromaticity (515–520 nm) is not sufficiently close to the Rec. 2020 standard. Attempts to tune the green emission, for example by alloying with iodide, have led to instability issues, since mixed halide perovskite NCs (similarly to the case of polycrystalline films) tend to undergo halide segregation under an electrical bias during LED operation.[41] $CsPb_2Br_5$ NCs and NPLs have also been studied as the emitting layers in phosphor converted LED and electrically driven LED, respectively.[49] Differently from the three-dimensional (3D) orthorhombic structure of $CsPbBr_3$, $CsPb_2Br_5$ has a two-dimensional (2D) crystal structure in which a layer of $Cs^+$ ions is sandwiched between two layers of $[Pb_2Br_5]^-$.[50] Such crystal structure facilitates the synthesis of $CsPb_2Br_5$ NPLs. There are reports claiming that $CsPb_2Br_5$ NCs and NPLs have bright green emissions peaking at 510–520 nm with narrow bandwidths (12–20 nm).[49, 51] Some works have hypothesized that the emission of $CsPb_2Br_5$ originates from „amorphous lead bromide ammonium complexes" present on the NCs surface or alternatively from intrinsic defects in $CsPb_2Br_5$.[51-52] However, all these claims need to be carefully verified.

Lead-based perovskite NCs as the red emissive layers in LEDs include $CsPbI_3$ and $CsPb(Br/I)_3$ NCs in the deep-red (650–680 nm) and pure-red (620–650 nm) emission regions, respectively. $CsPbI_3$ has four different crystal structures at different temperatures. These $CsPbI_3$ phases (α-, β-, γ-, δ-$CsPbI_3$) are slightly different in symmetry and Pb−I−Pb bond angles, resulting in different band gaps. The desired phase is the emissive α-$CsPbI_3$ one, with a band gap of 1.73 eV. The structural stability in the perovskite structure can be evaluated by the Goldschmidt tolerance factor ($t$), which is defined as $(r_A + r_X)/\sqrt{2}\,(r_B + r_X)$ ($r_A$, $r_B$, $r_X$: the ionic radius of A-site, B-site and halide ions).[53] A stable perovskite structure has a tolerance factor in the range of 0.8–1.0. The cubic α-$CsPbI_3$ phase has a relatively low tolerance factor (0.81) and tends to convert over time to the non-emitting δ-$CsPbI_3$ phase characterized by an orthorhombic structure at room temperature,[54] and various strategies have been developed to improve its stability, as will be discussed later. Pure red LEDs with the emission wavelength of 620–650 nm are highly desired to match the ITU-R recommendation BT.2020 (Rec. 2020)[13] it order to be used for high-definition displays. Similarly to pure blue perovskite NCs, preparing mixed halide $CsPb(Br/I)_3$ NCs, $CsPbI_3$ NPLs or small-sized (~5 nm) $CsPbI_3$ NCs are all strategies that have been followed to achieve such pure red emission.

### 3.2 Synthesis optimization towards stable and efficient perovskite NCs

Generally, lead halide perovskite NCs are synthesized by hot injection or anti-solvent recrystallization methods.[55] In a typical hot injection synthesis, Cs-oleate is injected into a Pb-halide solution at a certain reaction temperature (140–200 ºC) under an inert gas (**Figure 3**a). Ligands include molecules such as oleylamine (OLA) and oleic acid (OA) that are added in the solution in order to solubilize $PbX_2$ and stabilize the NCs. A subsequent rapid quenching in an ice-cold water bath is needed to terminate the reaction and obtain NCs with narrow size distribution. Through optimization of the ligands, reaction temperature and precursor ratio/type, it is possible to control NCs size, morphology and improve their stability and luminescence efficiency (Figure 3b).[56] In the hot injection method, low reaction temperatures (80–130 ºC) are needed to slow the growth process and obtain strongly quantum confined systems (small sized NCs, NPLs or NRs).[44c, 57] A gradient purification method by multi-step centrifugation can be used to recover more uniform, small-sized NCs with a narrow size distribution.[58] To promote the NCs film formation in device applications, Wang et al. used lauryl methacrylate (LMA) instead of ODE in the synthesis of uniform small sized $CsPbI_3$ NCs (~4 nm), and their reason was that the methacrylate group of LMA enables radical polymerization and cross-linking between two adjacent NCs, which is beneficial for NCs film formation without further purification to remove ODE.[59]

For all these strongly-confined perovskite NCs, characterized by a large surface-area-to-volume ratio, preserving spectral and phase stability is key for LED applications, especially for $CsPbI_3$ NCs that suffer from structural instability. Replacing the weakly bound OA/OLA ligands on the NC surface with ligands that bind more strongly can help to stabilize them. Taking $CsPbI_3$ in LED applications as an example, these ligands include 1-hydroxy-3-phenylpropan-2-aminium iodide (HPAI)/tributylsulfonium iodide (TBSI), amino acids and 3-phenyl-1-propylamine (PPA)/tetrabutylammonium iodide (TBAI).[8a, 60] These are additionally relatively short-chain ligands which can also promote electronic conductivity, hence improved electroluminescence. Other stabilization routes in the hot injection method have been explored to improve the stability of $CsPbI_3$ NCs. For example, Xie et al. reported that using sodium dodecyl benzene sulfonate (SDBS) as an additive in the synthesis process could improve the stability of $CsPbI_3$ NPLs.[61] The benzene ring of SDBS was hypothesized to suppress $[PbI_6]^{4-}$ octahedral distortions at the surface of the NCs through a steric hindrance effect, thus helping to preserve phase stability, similar to the effect observed when using PPA.[61] Ebe et al. developed a post-synthetic guanidinium iodide (GAI) treatment to stabilize small sized $CsPbI_3$ NCs with a pure-red emission at 629 nm.[62] The calculated binding energy between $GA^+$ and $I^-$ ions is higher than that between $OLA^+$ and $I^-$ ions, thus in principle explaining a more stable NCs surface.[62] Li et al. reported a nanosurface-reconstruction strategy to stabilize $CsPbI_3$ NCs and obtain 5.4 nm $CsPbI_3$ NCs with pure-red emission peaking at 642 nm:[7d] in essence, they used diisooctylphosphinic acid (DSPA) instead of OA in the hot injection method and injected hydriodic acid together with a solvent (e.g., xylene or toluene) into the reaction flask as this was cooled to 100–120 °C. Surface etching operated by hydriodic acid was claimed to remove 'imperfect' octahedra from the surface of $CsPbI_3$ NCs, thus reducing the NCs size and improving the stability through the coordinating ligand DSPA.[7d]

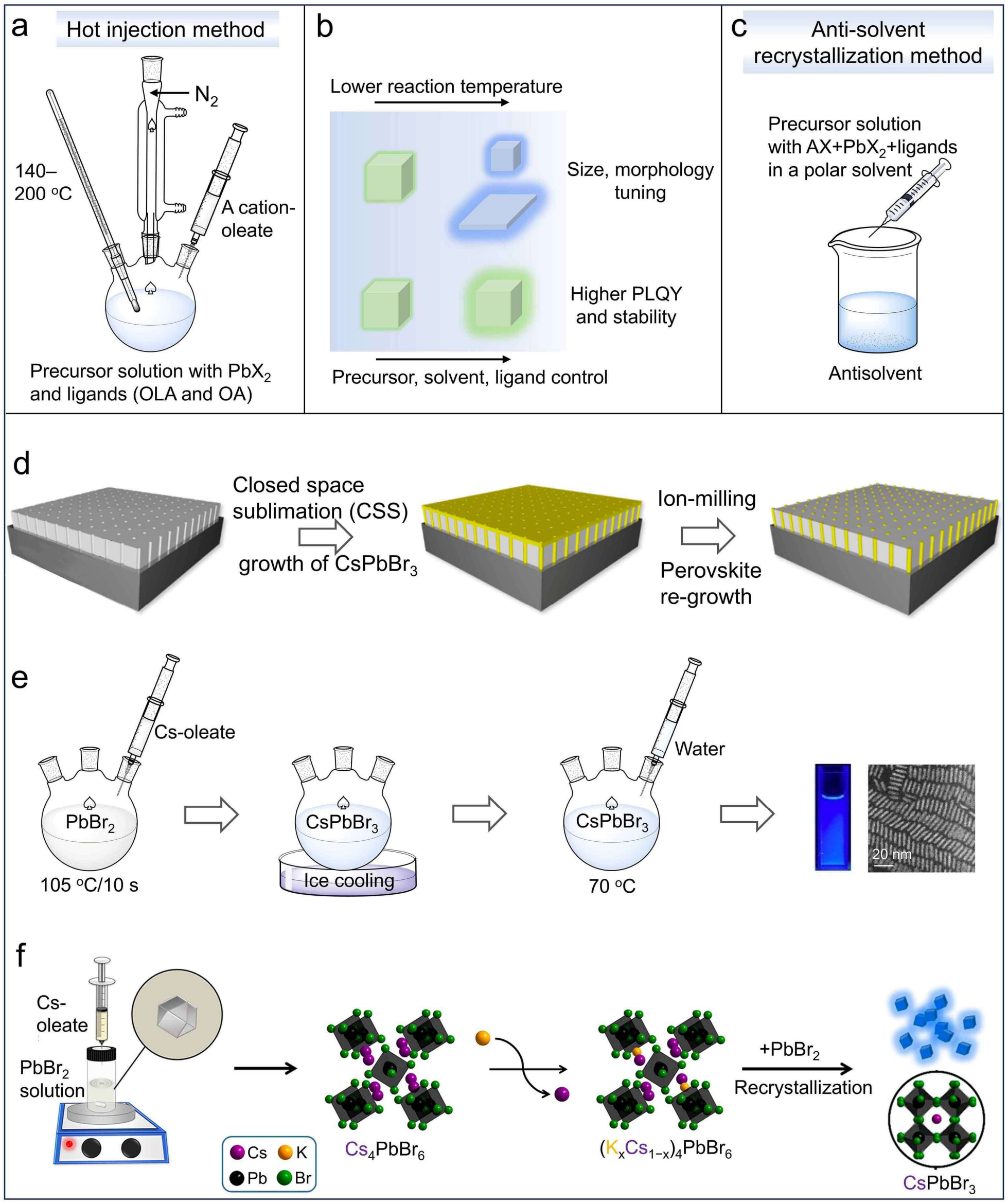

**Figure 3.** (a) Scheme of the hot injection method. (b) Synthesis condition control for perovskite NCs with different sizes/morphologies and improved PLQY. (c) Scheme of the antisolvent recrystallization synthesis. (d) Low-pressure closed space sublimation process to grow $CsPbBr_3$ nanowire arrays in nanoporous anodic aluminum oxide templates. Reproduced and adapted from ref.[63]. Copyright 2022, American Chemical Society. (e) Sketch of the water-driven synthesis process for $CsPbBr_3$ nanowires. Reproduced and adapted from ref.[64]. Copyright 2023, Wiley. (f) Scheme of the synthesis for $CsPbBr_3$ NCs by the conversion of $(K_xCs_{1-x})_4PbBr_6$. Reproduced and adapted from ref.[15]. Copyright 2024, American Chemical Society.

The synthesis of NCs by the anti-solvent recrystallization method is a good approach for scale-up, as this is fast and is usually carried out in ambient conditions at room temperature.[55] In the synthesis, all precursors and ligands are dissolved in a polar solvent, and then injected into an antisolvent to promote precipitation of the NCs (Figure 3c). The type of precursor, nonpolar-phase solvent, ligand and reaction temperature are all parameters that can be used to precisely control the size of NCs, or the NPL thickness in the case of NPL synthesis.[65] For example, Yang's group used liquid nitrogen in the reprecipitation method to synthesize ultrasmall $CsPbBr_3$ and $MAPbBr_3$ NCs.[65c, 65d] The cryogenic temperature reaction lowered the nucleation and growth rates.[65c, 65d] The prepared small sized $CsPbBr_3$ (3 nm) and $MAPbBr_3$ (2.12 nm) NCs both presented sharp deep blue emission at ~460 nm with a PLQY of ~98%.[65c, 65d] Bi et al. synthesized $CsPbI_3$ nanowire arrays (with a width of 5–10 nm) by a modified anti-solvent recrystallization method in which the precursor-ligand complexes were mixed in nonpolar organic media with a large amount of amines.[66] The nanowires self-assembled into nanowire arrays at room temperature and under ambient conditions, suggesting that the synthesis is suitable for large-scale production. The assembled nanowire arrays showed emission peaking at 600 nm with a PLQY of 91%.[66]

Additional approaches have been pursued to synthesize halide perovskite NCs for LED applications. For example, Fu et al. grew $CsPbBr_3$ nanowire arrays in nanoporous anodic aluminum oxide templates by a low-pressure closed space sublimation process.[63] $CsPbBr_3$ nanowires with variable diameters (2.8–6.6 nm) and tunable emissions from 467 to 512 nm were obtained by tuning the pore sizes (Figure 3d).[63] Xu et al. cut $CsPbBr_3$ NPLs into small sized blue emitting NCs by ultrasonication and hydrobromic acid processing.[67] The ultrasonication treatment was reported to break the NPLs into small sized segments and the addition of hydrobromic acid contributed to solubilize the NPLs and passivate the surface;[67] Zhang et al. developed a water-driven synthesis variant of the hot injection method to synthesize uniform $CsPbBr_3$ nanowires:[64] the water was injected into the synthesized $CsPbBr_3$ nanowires reheated at 70 °C, under stirring, for 30 s. The injected water was claimed to control the orientation of the surface ligands, inducing oriented nanowire growth in the Ostwald ripening regime by phagocytizing the unstable ultra-small NCs. The synthesized $CsPbBr_3$ nanowires showed a blue emission at 456 nm with FWHM of 20 nm and PLQY of 94% (Figure 3e).[64] More recently, Otero-Martínez et al. were able to grow 3.5 nm sized $CsPbBr_3$ NCs from the slow decomposition of $(K_xCs_{1-x})_4PbBr_6$ NCs (Figure 3f).[15] Contrary to the case of $Cs_4PbBr_6$ NCs, which react quickly with $PbBr_2$ and deliver large (> 6–7 nm) NCs, the presence of $K^+$ slowed down their reaction, and essentially the $(K_xCs_{1-x})_4PbBr_6$ NCs acted as reservoirs for the slow nucleation and growth (occurring over hours) of the $CsPbBr_3$ NCs.[15]

### 3.3 Surface passivation

Surface halide vacancies are the predominant defect species in lead halide perovskite NCs. They act as defect sites which trap charge carriers and therefore compete with radiative recombination. Hence, passivation of surface halide vacancies is essential to saturate the uncoordinated $Pb^{2+}$ dangling bound and increase the PLQY. In the case of mixed halide NCs, passivation of surface halide vacancies is necessary to mitigate halide ion migration.[68]

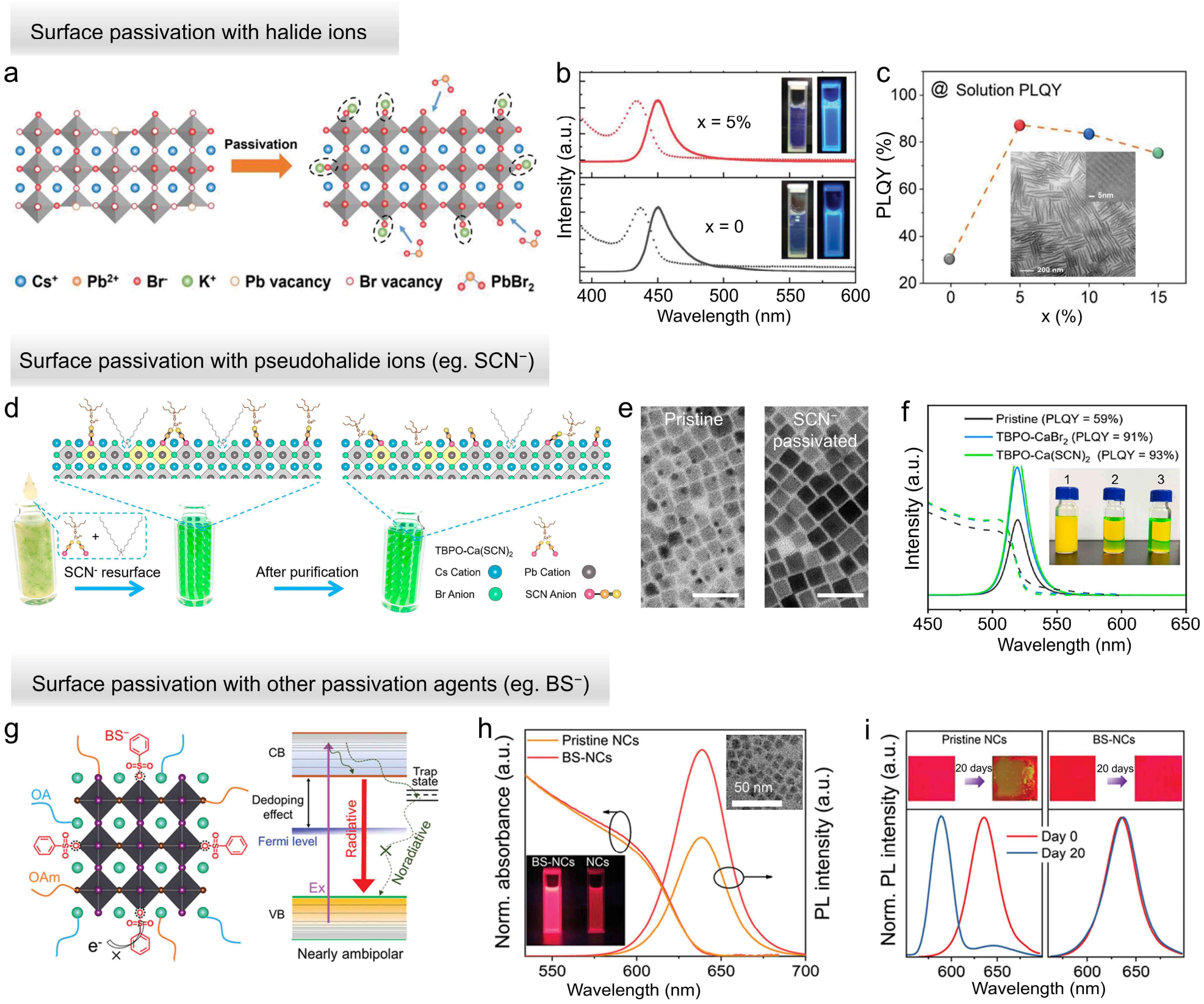

**Figure 4.** (a) Scheme of KBr passivation on the surface of $CsPbBr_3$ NPLs; (b) Absorption and photoluminescence (PL) spectra of pristine ($x = 0$) and $K^+$ passivated ($x = 5\%$) $CsPbBr_3$ NPLs ($x$: K/Pb ratio); (c) PLQYs of $CsPbBr_3$ NPLs by addition of different amounts of $K^+$ ions. The inset shows a TEM image of KBr-passivated $CsPbBr_3$ NPLs. Reproduced and adapted from ref.[69]. Copyright 2021, Wiley. (d) Scheme of the introduction of $SCN^-$ ions to passivate the surface of $CsPbBr_3$ NCs; (e) TEM images of pristine (left) and $SCN^-$ passivated (right) NCs, respectively. Scale bars: 50 nm; (f) Absorption and PL spectra (excitation at 365 nm) of pristine, TBPO-$CaBr_2$ and TBPO-$Ca(SCN)_2$ treated $CsPbBr_3$ NCs solutions. The insets are the corresponding photographs under daylight. Reproduced and adapted from ref.[70]. Copyright 2023, American Chemical Society. (g) Sketch of the $BS^-$ passivation $CsPb(Br/I)_3$ NCs and its effect on the optical properties; (h) Absorption and PL spectra of pristine and $BS^-$ passivated $CsPb(Br/I)_3$ NCs solution; (i) PL stability of pristine and $BS^-$ passivated $CsPb(Br/I)_3$ NCs films. Reproduced and adapted from ref.[71]. Copyright 2022, Wiley.

Halide vacancies are generally filled by addition of extra halide ions, often in combination with other cations, and their effect can be seen through an increase in PLQY. This enhancement is usually more pronounced and widely reported in NPLs, most likely since the high surface-to-volume ratio of NPLs leads to a high density of surface halide vacancies. Examples of metal halide salts added to 'repair' the surface of $CsPbBr_3$ NPLs used in LED applications include Ni-Br, $ZnBr_2$, KBr and OLA-Br.[69, 72] Among them, $K^+$ ions were reported to stabilize $Br^-$ on the surface of NPL by complexing with $Br^-$ ions. By adding potassium oleate and excess $PbBr_2$ to the precursor solutions, KBr-passivated $CsPbBr_3$ NPLs with blue emission at ~450 nm and a PLQY of 87% could be synthesized (**Figure 4**a-c);[69] $Ni^{2+}$ and $Zn^{2+}$ were reported to form strong bonds with $Br^-$ ions.[72a, 72b] In mixed halide NCs, surface passivation using halide ions can mitigate halide ion migration and improve spectral stability under electric field and/or UV light irradiation. For $CsPb(Br,I)_3$ NCs, iodide and bromide salts (such as benzyl iodide, guanidinium iodide (GAI) and potassium bromide, K-Br) were used to passivate surface defects and mitigate halide ion migration in $CsPb(Br,I)_3$ NCs based LEDs under operational conditions.[73] Shao et al. added phenethylammonium chloride (PEACl) in the hot injection synthesis of $CsPbCl_3$ NCs and were able to enhance the PLQY from 5.4% to 62.3% due to suppressed nonradiative recombination, most likely by decreasing the number of $Cl^-$ vacancies.[74] Then they exchanged $Cl^-$ with $Br^-$ at room temperature, obtaining a blue emission at 470 nm with a PLQY of 80.2%.[74] To prepare small-sized, stable $CsPbBr_3$ NCs, Bi et al. introduced hydrobromic acid to promote the formation of a halide-rich surface and added didodecylamine and phenethylamine ligands, which in the reaction environment became protonated and were able to passivate the surface uncoordinated sites.[75] The 48% PLQY of the pristine small sized NCs increased to 97% upon such treatment. These NCs were then used as the emitting layer in a blue LED.[75] Feng et al. grew a $PbCl_x$ shell on the surface of 6.6 nm $CsPbI_3$ NCs (PL peak: 633 nm) by injecting acyl chloride, leading to a near-unity PLQY.[8b] Compared to other additives used to prepare Cl-passivated $CsPbI_3$ NCs (metal chlorides or ammonium chloride), acyl chlorides could react with excess ligands and release $Cl^-$ ions, resulting in a complete surface passivation by $Cl^-$ ions without dismantling the whole NCs.[8b]

The case of fluoride ions is also noteworthy. These ions have a higher electronegativity than the other halide ions ($Cl^-$, $Br^-$, $I^-$), hence they bind stronger to surface $Pb^{2+}$ ions and effectively passivate the surface halide vacancies.[76] In addition, they have also been used successfully to reduce thermal quenching in LEDs. This effect, as already mentioned earlier, is a typical problem of LEDs based on lead halide perovskite NCs when operated at high temperatures, and is attributed to halide vacancies: the shallow nature of the energy states associated to them (and located above the conduction band minimum (CBM)), causes a low trapping activation energy and non-radiative recombination.[76b] Liu et al. proposed a fluoride-based post-synthesis treatment to suppress thermal quenching and increase the emission efficiency in $CsPbBr_3$ NCs.[76b] The introduced $F^-$ could increase the trapping activation energy. A similar fluoride post-treatment has also been applied to red-emitting $CsPbBr_{x-3}I_3$ and $CsPbI_3$ NCs.[76b]

Surface passivation with pseudohalide ions, including thiocyanate ($SCN^-$) and tetrafluoroborate ($BF_4^-$) anions can also be exploited to saturate halide-related defect sites in perovskite NCs used in LEDs.[70, 77] Compared to halide ions, $SCN^-$ ions form a stronger bond with $Pb^{2+}$, hence they tend to replace/fill halide ions/vacancies at the surface of the NCs, leading to increased stability and improved optical properties.[77a] For example, Yang et al. and Bhatia et al. both developed a 'resurfacing' strategy that consists in adding $Ca(SCN)_2$ (Yang) or KSCN (Bhatia) to green emitting $CsPbBr_3$ (Yang) or $FAPbBr_3$ (Bhatia) NCs, leading to enhanced PLQY (>90%) and stability.[70, 77a] To obtain a high concentration of $SCN^-$ ions on the $CsPbBr_3$ NCs surface, Yang et al. also added tributylphosphine oxide (TBPO) to promote the dissolution of $Ca(SCN)_2$ (Figure 4d-f).[70] To effectively passivate the surface of $CsPb(Cl,Br)_3$ NCs, Zheng et al. chose n-dodecylammonium thiocyanate (DAT; $C_{12}H_{25}NH_3SCN$) with a high solubility (>100 mg/mL) in a nonpolar solvent such as toluene.[78] The DAT post-treatment was reported to fill the $Cl^-$ vacancies and remove electron traps in the band gap, leading to blue emitting $CsPb(Cl,Br)_3$ NCs with an emission peak at 468.8 nm and a near-unity PLQY.[78] $NH_4SCN$, NaSCN and Guanidine thiocyanate (GASCN) were used to achieve effective NCs surface passivation for high-performance $CsPbI_3$ NCs based LEDs.[79] Lu et al. reported NaSCN/GASCN passivated $CsPbI_3$ NCs with a PLQY of 95.1% and then fabricated LEDs with an external quantum efficiency (EQE) of 24.5%.[79b] The introduced $Na^+$ and $SCN^-$ ions passivated the uncoordinated sites on the NCs surface. Also, the $GA^+$ ions, with their three amino groups, most likely enhanced the stability of the perovskite crystal structure by forming hydrogen bonds with $I^-$ ions.[79b]

The use of surface passivation agents to reduce the density of surface trap states in perovskite NCs and improve the corresponding LED performance has been extensively reported also for red and deep-red emitting perovskite NCs which were then used in LEDs. For $CsPbI_3$ NCs, passivating agents include $NH^{4+}$, $Zr^{4+}$, $Mg^{2+}$, $AcO^-$, $SO_4{}^{2-}$, or $OAc^-$.[80] For $CsPb(Br/I)_3$ NCs, they include benzenesulfonate ($BS^-$)[71] and trichloroacetic acid (TCA).[81] For $BS^-$, it was found that it can both passivate the uncoordinated $Pb^{2+}$-related defects on the NC surface and increase the formation energy of halide vacancies (Figure 4g).[71] The $BS^-$ passivation enhanced the PLQY from 63.5% to 93% of $CsPb(Br/I)_3$ NCs and improved the stability in air of $CsPb(Br/I)_3$ NCs films (Figure 4h, i).[71] Overall, the passivated $CsPb(Br/I)_3$ NCs could reach a PLQY over 90% and exhibited increased stability. For $CsPb(Cl,Br)_3$ NCs that were used to fabricate blue LEDs, Ma et al. utilized hydrazinium cations ($Hz^{2+}$) to remove chlorine vacancies and increase the PLQY of the blue emission from 58% to 94%.[82] The introduced $Hz^{2+}$ ions were reported to effectively suppress the formation of chlorine vacancies by establishing a Hz-Cl-Cs bridge on the NCs surface.[82] For $CsPbBr_3$ NCs based green LEDs, Jeong et al. reported that the surface defects on $CsPbBr_3$ NCs can be passivated by the addition of $Cd^{2+}$, $Zn^{2+}$ or $Hg^{2+}$ acetate salts.[83] With didodecyldimethylammonium ($DDA^+$)-organic ligands and cadmium acetate, the PLQY of green emitting $CsPbBr_3$ NCs was improved from 41.8% to 96%.[83]

### 3.4 Ion doping/alloying

Ion doping/alloying, commonly employed in traditional semiconductor NCs, has been also extensively applied in perovskite NCs to regulate the optical and electronic properties, as also discussed earlier.[84] Additionally, when this is done at the A and/or B sites of perovskite NCs, it can improve the phase and environmental stability of the perovskite NCs, which is of key relevance also for LED applications. Compared to $MAPbX_3$ and $FAPbX_3$ that have a cubic structure at room temperature, $CsPbX_3$ usually adopt the lower symmetry orthorhombic structure due to the octahedral tilting induced by the smaller $Cs^+$ ions. Thus, $CsPbX_3$ (X = Cl, Br, I) have relatively small tolerance factors (Cl: 0.87; Br: 0.86; I: 0.85).[85] A-site ions doping/alloying with larger cations in $CsPbX_3$ can increase the tolerance factor to values closer to 1 and thus improve the structural stability (**Figure 5**a).The incorporation of guanidinium ($GA^+$) and formamidinium ($FA^+$) cations having larger ionic radii (2.78 Å and 2.70 Å, respectively) than $Cs^+$ cations (1.88 Å) can improve both the stability and optical properties of perovskite NCs for better LED performance.[86] For example, Serafini et al. alloyed $GA^+$ into $CsPbI_3$ NCs to tailor the tolerance factor for a stable cubic α-$CsPbI_3$ phase.[86a] Gao et al. synthesized $FA^+$ alloyed $CsPb(Cl_{0.5}Br_{0.5})_3$ NCs and increased PLQY from 10% of the pure $Cs^+$ phase to 65% of the $FA^+$ alloyed phase (Figure 5b-e).[86b] The PL peak of $(Cs,FA)Pb(Cl_{0.5}Br_{0.5})_3$ NCs red-shifted with the increase in the ratio of $FA^+/Cs^+$ due to the influence of $FA^+$ on the band structure (Figure 5c). The increased PLQYs values were instead attributed to the suppression of non-radiation decay channels from reduced electron-electron and electron-phonon scattering in the carrier relaxation process (Figure 5e).[86b] In another work, $Rb^+$ ions were alloyed into $CsPb(Cl,Br)_3$ NCs, leading to an increased PLQY, from 2.2% to 20.4%, possibly due to the suppressed nonradiative recombination by the likely formation of a Rb-related passivation layer on the NCs surface.[42b]

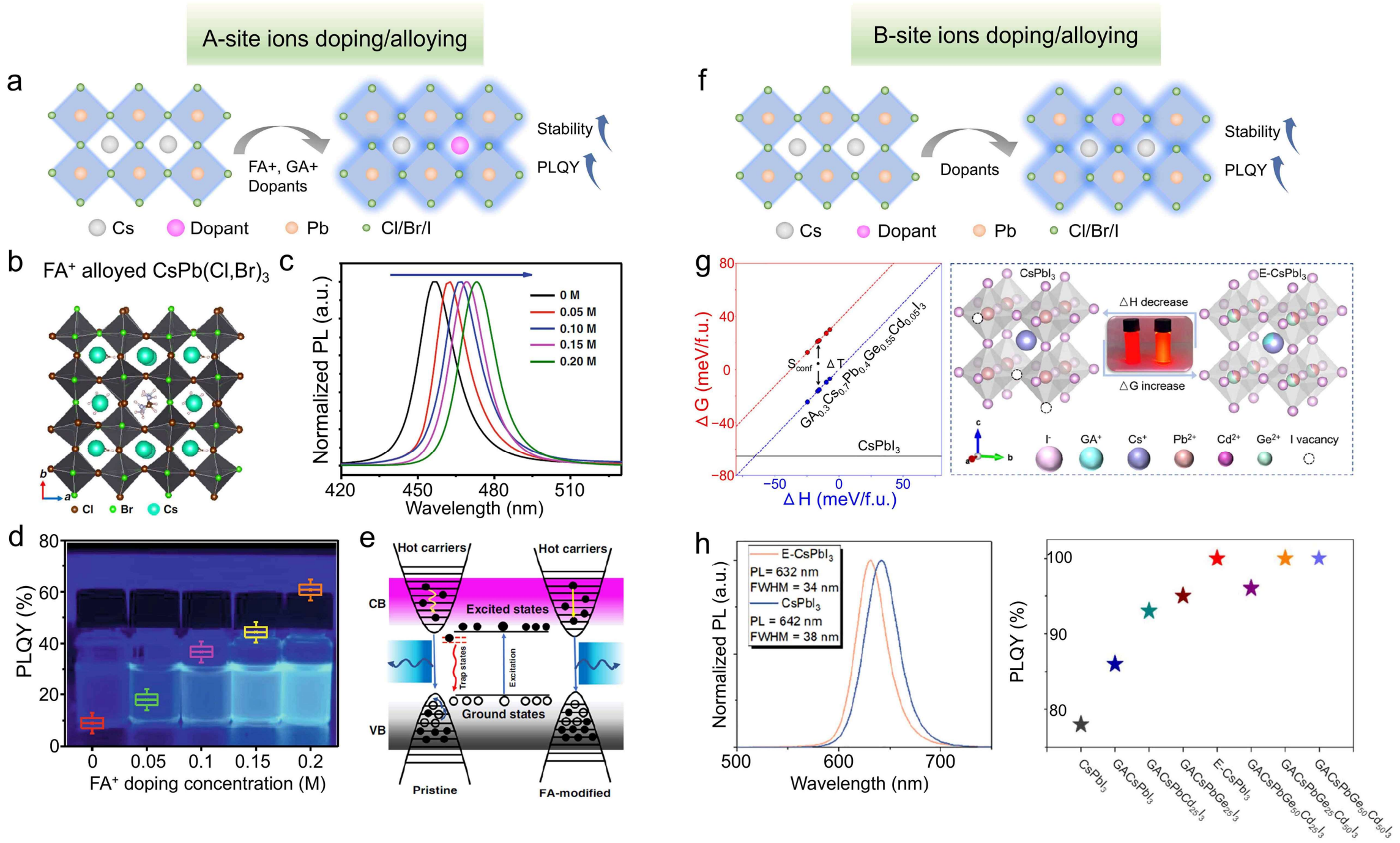

**Figure 5.** (a) Scheme of A-site ions doping/alloying in perovskite NCs; (b) Structure of $FA^+$ alloyed $CsPb(Cl,Br)_3$; (c) PL spectra and (d) PLQYs of $FA^+$ alloyed $CsPb(Cl,Br)_3$ NCs with $FA^+$ feeding ratio of 0, 0.05, 0.1, 0.15, and 0.2 M; (e) Mechanism sketch of $FA^+$ on the carrier relaxation of $CsPb(Cl,Br)_3$ NCs. Reproduced and adapted from ref.[86b]. Copyright 2022, Springer Nature. (f) Scheme of B-site ions doping/alloying in perovskite NCs; (g) Left: Calculated decomposition enthalpy ($\Delta H$) and Gibbs free energy ($\Delta G$) for alloyed $CsPbI_3$ NCs ($\Delta H$, blue line; $\Delta G$, red line) and $CsPbI_3$ NCs ($\Delta H$ and $\Delta G$: black line); Right: crystal structure

of unalloyed and alloyed $CsPbI_3$ NCs (The inset: a photograph of NCs under 365 nm excitation); (h) PL spectra and PLQYs of unalloyed and alloyed $CsPbI_3$ NCs. Reproduced and adapted from ref.[87]. Copyright 2023, American Chemical Society.

B-site ion doping/alloying has also been studied extensively to improve the optical properties of perovskite NCs and achieve a better LED performance (Figure 5f). The cations doped/alloyed into perovskite NCs for LED applications include monovalent (such as $Cu^{+}$),[88] divalent ($Mg^{2+}$, $Ca^{2+}$, $Sr^{2+}$, $Mn^{2+}$, $Co^{2+}$, $Ni^{2+}$, $Cu^{2+}$, $Zn^{2+}$, $Ge^{2+}$),[42b, 77a, 87, 89] and trivalent ones ($La^{3+}$, $Ce^{3+}$, $Sb^{3+}$).[90] They can improve the PLQY and stabilize the phase for a series of potential reasons (by reducing the density of halide vacancies, or by modulating the lattice formation energy/tolerance factor). For example, $Cu^{+}$ doping in $CsPb(Cl,Br)_3$ NCs led to a halide-rich surface due to a stronger Cu-Cl bond compared to the Pb-Cl one.[88] $Cu^{2+}$, $Mn^{2+}$, $Ni^{2+}$ or $La^{3+}$ doped/alloyed in blue emitting $CsPb(Cl,Br)_3$ NCs can inhibit the formation of defect states by increasing the defect formation energy.[42b, 89a-c, 90a] The incorporation of $Cu^{2+}$ ions in $CsPb(Br,I)_3$ NCs was also reported to enhance the lattice formation energy and hence stabilize the cubic phase.[89h] As for $CsPbI_3$ NCs, the cubic α-phase tends to convert to the orthorhombic δ-phase due to its relatively low tolerance factor, as discussed earlier. Hence, the incorporation of B-site ions smaller than $Pb^{2+}$ in $CsPbI_3$ can increase the tolerance factor and improve the phase stability. Examples in this case include $Sr^{2+}$, $Mn^{2+}$, $Ni^{2+}$, $Zn^{2+}$ or $Sb^{3+}$ alloying in $CsPbI_3$ NCs.[89d-g, 89i-m, 90c] Guo et al. synthesized small sized $CsPbI_3$ NCs alloyed with multiple ions ($Cd^{2+}$, $Ge^{2+}$ and $GA^{+}$) and characterized by a narrow FWHM of 34 nm and a PLQY of basically 100% (Figure 5g, h).[87] Calculations have shown an increased decomposition enthalpy ($\Delta H$) and Gibbs free energy ($\Delta G$) in the alloyed NCs, thus hinting at an improved stability compared to the pure $CsPbI_3$ NCs (Figure 5g).[87] As another example, Yao et al. doped $Ce^{3+}$ ions into green emitting $CsPbBr_3$ NCs, increasing the PLQY up to 89%.[90b] The incorporation of $Ce^{3+}$ ions is more likely to create new near band-edge energy states, resulting in increased electron density in the conduction band. Such an increase in density of states was reported to facilitate exciton relaxation and recombination.[90b]

### 3.5 Shell coating

Developing a core/shell structure by encapsulating a protective shell around the NCs core is a popular approach to improve the ambient stability and passivate the surface defects, and this strategy has been exploited also for perovskite NCs which were then used in LED applications. Shell materials tested so far include perovskites, inorganic salts (oxide, sulfide, zeolite or glass), or organic molecules (polymer, metal–organic frameworks). In the perovskite/perovskite core-shell NCs cases, the shells can be a 0D, 2D or a 3D perovskite phase, or even amorphous $CsPbBr_x$ (**Figure 6**a). Among them, 0D $Cs_4PbX_6$(X = Br, I) has been widely reported as an effective matrix to encapsulate $CsPbX_3$(X = Br, I) NCs, forming $CsPbBr_3/Cs_4PbBr_6$ nanocomposites used in both electrically driven LEDs and WLEDs.[91] Some works have reported that tuning the Cs/Pb feed ratio or the capping ligand can lead to the desired $CsPbX_3/Cs_4PbX_6$ architecture.[91-92] For example, the Cs/Pb molar ratio in the precursors mixture was found to be paramount to regulate the formation of the $CsPbBr_3/Cs_4PbBr_6$ core/shell structure: for in situ NCs film synthesis, a molar ratio of Cs/Pb > 1 was used;[91a, 91d] for the solution-based synthesis (hot injection method or one-pot synthesis in the DMF solvent with ultrasonication and evaporation), a molar ratio of Cs/Pb ≥ 4 was used instead.[91b, 91c] The non-emissive 0D $Cs_4PbBr_6$ shell has a larger band gap (~3.9 eV) than that of $CsPbBr_3$ (~2.4 eV), resulting in the formation of type-I heterojunctions. This band alignment can promote carrier confinement in the $CsPbBr_3$ core region and facilitate radiative recombination of the carries inside that region.[91a, 93] Additionally, the $Cs_4PbBr_6$ shell can passivate surface defects and thus enhance the PLQY and stability.[91b, 94] As an example, Kim et al. synthesized mixed ultra-small $CsPbBr_3$-$Cs_4PbBr_6$ NCs with an efficient and stable blue emission and a PLQY over 90%.[94] Similarly, $CsPbBr_3$ NCs were encapsulated in a 2D $CsPb_2Br_5$ shell, forming a type-I heterojunction, leading to an enhanced PLQY and stability.[95] As a note of caution, the careful structural characterization of these materials by advanced electron microscopy tools is extremely challenging, due to degradation of the materials under the beam as well as their inherent instability and transformative behavior even under ambient conditions. Hence, all claims of successful formation of heterostructures (for example core-shells) will need to be carefully validated.

As other examples of shell-coating, Liu et al. coupled quantum-confined $Cs_xFA_{1-x}PbBr_3$ (~5.8 nm) NCs and quasi-2D perovskites with higher band gaps to promote charge injection and suppressed nonradiative recombination for higher performance blue perovskite LEDs.[12c] The efficient blue emission, with a PLQY over 60%, originated from the almost complete transfer of photogenerated excited states from the quasi-2D perovskites to the quantum-confined NCs.[12c] Ye et al. developed $MAPbBr_3$/chiral layered perovskites core−shell NCs with a type-II band alignment.[96] The core−shell NCs exhibited spin-polarized luminescence from the shell-to-core electron injection and electron-hole recombination in the $MAPbBr_3$ core.[96] A case of all-perovskite core-shell structure, used then in the fabrication of LEDs, was reported by Zhang et al., who grew a $CsPbBr_3$ shell epitaxially on the surface of $FAPbBr_3$ NCs cores.[97] To address the lattice mismatch issue between core and shell, an alloyed layer, $FA_xCs_{1-x}PbBr_3$, was claimed to be formed at the core–shell interface. The obtained core/alloyed-shell/shell $FAPbBr_3$/$FA_xCs_{1-x}PbBr_3$/$CsPbBr_3$ NCs exhibited enhanced stability.[97] Wang et al. was also able to synthesize blue emitting core-shell $CsPbBr_3$@amorphous $CsPbBr_x$ NCs by a hot injection method:[98] during the synthesis, the nucleation of small sized $CsPbBr_3$ NCs was reported to be followed by the overgrowth of an amorphous $CsPbBr_x$ shell due to the specific synthesis conditions. The amorphous $CsPbBr_x$ shell protected the NCs core (2 nm size) against light, oxygen and moisture induced degradation and improved the blue light PLQY from 54% to 84%.[98]

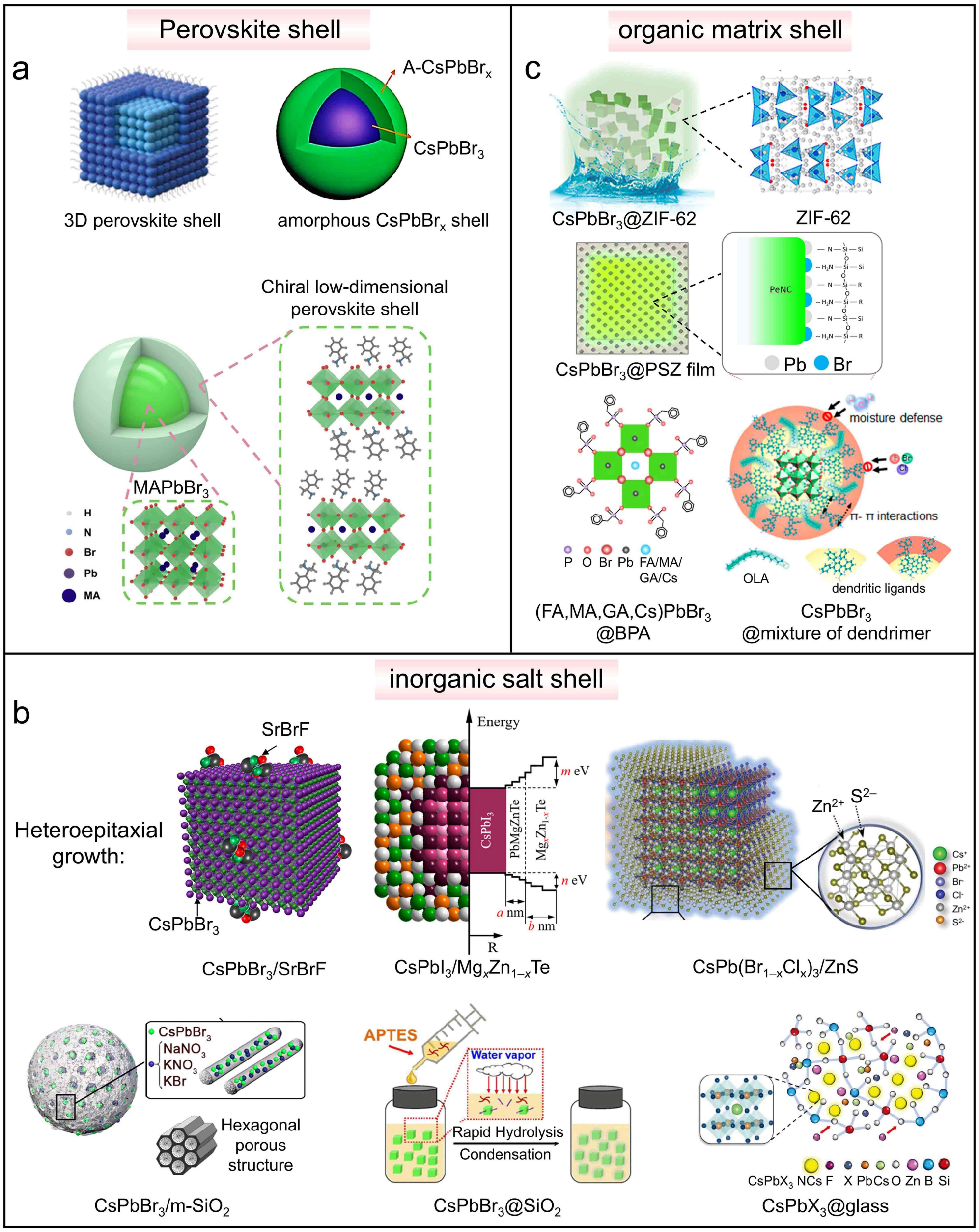

**Figure 6.** (a) Perovskite shell,[96-98] (b) Inorganic salt shell,[99] and (c) Organic matrix shell[12a, 100] for perovskite NCs used in LED applications. (a): Reproduced and adapted from ref.[97] Copyright 2020, Wiley. Reproduced and adapted from ref.[98]. Copyright 2017, American Chemical Society. Reproduced and adapted from ref.[96]. Copyright 2022, American Chemical Society. (b): Reproduced and adapted from ref.[99a]. Copyright 2023, American Chemical Society. Reproduced and adapted from ref.[99b]. Copyright 2020, American Chemical Society. Reproduced and adapted from ref.[99c]. Copyright 2023, Wiley. Reproduced and adapted from ref.[99d]. Copyright 2021, American Chemical Society. Reproduced and adapted from ref.[99e]. Copyright 2023, Royal Society of Chemistry. Reproduced and adapted from ref.[99f] Copyright

2023, Wiley. (c): Reproduced and adapted from ref.[100a]. Copyright 2023, American Chemical Society. Reproduced and adapted from ref.[100b]. Copyright 2019, American Chemical Society. Reproduced and adapted from ref.[12a]. Copyright 2022, Springer Nature. Reproduced and adapted from ref.[100c]. Copyright 2023, American Chemical Society.

Aside from a metal halide shell, various other inorganic matrices have been proposed as a means to encapsulate green emitting perovskite NCs which were then used in LEDs: these include sulfide (CdS, ZnS, PbS),[99c, 101] $Mg_xZn_{1-x}Te$,[99b] α-$BaF_2$,[102] $CaI_2$,[103] $SiO_2$,[99e, 104] ZnO,[104a] $ZrO_2$,[104b] $NiO_x$,[105] $AlO_x$,[106] SrBrF,[99a] $PbWO_4$,[107] $PbSO_4$,[108] mesoporous silica (or all-silicon molecular sieves, abbreviated as m-$SiO_2$),[99d, 109] porous alumina membranes (PAMs),[14] zeolites,[110] and glass (Figure 6b).[99f, 111] The growth of an epitaxial shell generally requires minimal lattice mismatch between the shell material (<5%) and the underlying perovskites.[99b] For example, the similarity in lattice constants for $CsPbBr_3$ (5.85 Å) and CdS (5.83 Å) result in a $CsPbBr_3$/CdS interface with low strain, although the mechanism by which atomic alignment at the interface is achieved remains unclear.[112] As stated by the authors of the work, the suppression of surface defects in the core $CsPbBr_3$ NCs upon CdS shell growth led to an enhanced performance of the LED that were fabricated using these $CsPbBr_3$/CdS NCs.[101a, 101b] Similarly, the epitaxial growth of semiconductor materials such as $Mg_xZn_{1-x}Te$, α-$BaF_2$ and $CaI_2$ on $CsPbI_3$ NCs was exploited to passivate surface defects and optimize the energy level structure to reduce the carrier injection barrier, leading to a better charge injection balance within the devices. It must be emphasized that in some works the successful growth of an epitaxial shell was not sufficiently backed by convincing structural analysis.

Metal oxides have also been reported as effective shelling materials to stabilize perovskite NCs due to their excellent oxygen/moisture resistance and good thermal stability. Encapsulation of $CsPbBr_3$ NCs in $SiO_2$ is usually achieved by the hydrolysis of (3-aminopropyl)triethoxysilane (APTES),[99e, 104b, 104d, 104f, 104g] tetramethoxysilane (TMOS),[104a, 104e] or tetraethoxysilane (TEOS)[104c] either as a post-synthesis treatment or carried out in a one-pot synthesis approach. M-$SiO_2$ or all-silicon molecular sieves have been widely used to improve the stability of $CsPbBr_3$ NCs for WLED/LCD applications.[99d, 99e, 104a, 104c-e, 109a-d, 109f-i, 109k, 109l] M-$SiO_2$ has a large surface area, tunable pore size and a long-range ordered pore structure that can[113] act as a space-confined growth template for $CsPbBr_3$ NCs. $CsPbBr_3$ NCs embedded in m-$SiO_2$ can be prepared using a post-synthesis treatment, by in situ hot-injection, or by in situ solid-state sintering synthesis.[99d, 109c, 109f] The most frequently reported all-silicon molecular sieve is MCM-41.[99d, 109a, 109e-g, 109j, 109n] The small pore size of MCM-41 (2.5 nm) can also limit the growth of nanoparticles and thus promote the formation of strongly confined blue emitting $CsPbBr_3$ NCs.[109n] Notably, the pore collapse in MCM-41 at high temperature can lead to sealing of the pores, which then completely encapsulate the NCs by forming a compact shell that protected the $CsPbBr_3$ NCs.[109e] On the other hand, the pore collapse could also result in a subsequent uncontrollable NCs growth, with a resulting broad size distribution and relatively low PLQYs.[109f] An et al. and Liu et al. developed molten-salts-based ($KNO_3$−$NaNO_3$−KBr) methods for the synthesis of $CsPbBr_3/SiO_2$ nanocomposites at a relatively milder temperature (~350 ºC) compared to the previously discussed approach that was carried out at a higher sintering temperature (600-900 ºC).[99d, 109a] The resulting $CsPbBr_3/SiO_2$ composites, in Liu's case consisting of beads with diameters of the order of one hundred nanometers, had also their surface pores sealed and exhibited PLQY close to 80%, as well as good stability against moisture, aqua regia, high temperature and high flux irradiation.[99d, 109a] Zhang et al. used a m-$SiO_2$ template, SBA-15, with uniform pore size (6.5 nm), which could be stable even at 1200 ºC:[109f] a pre-hydrolysis sealing-pores strategy by TMOS hydrolysis was used to prepare $CsPbBr_3$/SBA-15/$SiO_2$ with sealed pores and PLQY up to 93%.[109f] Similar to m-$SiO_2$, porous alumina membranes (PAMs) with ultra-small pore size (~6.4 nm) were used as templates to grow perovskite quantum wires. Due to the quantum confinement effect induced by the thin diameter of the nanowires, enhanced light out-coupling efficiency and surface passivation operated by the PAMs, perovskite quantum wires-based LEDs with stable electroluminescence could be fabricated.[14]

Embedding $CsPbBr_3$ NCs in an inorganic glass is another promising strategy to improve their thermal and ambient stability for WLED applications.[113] In this case, perovskite NCs are formed by in situ crystallization in the glass matrix. Because of the suppressed growth of NCs in the dense glass network structure, the $CsPbBr_3$ NCs embedded in the glass have small size and a high density of surface defects, thus they tend to feature lower PLQYs than their colloidal counterparts.[111a, 114] On the other hand, increasing the Br/Pb precursor molar ratio or adjusting the perovskite concentrations of $CsPbBr_3$ NCs can enhance the PLQYs up to 90%.[99f, 111a-c] Also, a tunable green emission (517-528 nm) could be achieved by varying the NCs sizes when regulating the temperatures of the thermal treatment.[111b] To improve the performance of electrically driven LEDs, Ni$O_x$, Al$O_x$, SrBrF, MABr or $PbWO_4$ capping layers were exploited to resurface the $CsPbBr_3$ NCs and passivate their surface traps.[99a, 105-107, 115] Ni$O_x$ can also help to improve charge balance by facilitating the hole injection/transport.[105] The SrBrF capping layer was reported to form a heterostructure with $CsPbBr_3$ NCs and helped to impede $Br^-$ migration and reduce the number of halide vacancies due to the presence of $F^-$.[99a] To passivate the exposed $Pb^{2+}$ ions on the surface of $CsPbBr_3$ NCs, Ru et al. introduced $WO_4^{2-}$ ions by a microemulsion-assisted post-treatment.[107] The $WO_4^{2-}$ ions and the exposed $Pb^{2+}$ ions induced the formation of a $PbWO_4$ shell on the NCs surface.[107] Ru et al. were also able to grow a $PbSO_4$ layer on the surface of $CsPb(Br/I)_3$ NCs.[108] The $SO_4^{2-}$ ions, with strong binding on the surface can protect the perovskite thin film from water and inhibit phase separation, thus enhancing operational stability.[108] An $AlO_x$ shell was deposited on the $CsPbBr_3$ NCs to enhance the PLQY and stability of the core, but its insulating character was the most likely cause of a relatively low current density for the LED device.[106]

Another type of core-shell structure, built by encapsulating $CsPbBr_3$ or $MAPbBr_3$ NCs within an organic materials matrix, has attracted much attention due to the effective suppressed diffusion of moisture and oxygen into the NCs. Examples of organic shell include poly(methyl methacrylate) (PMMA),[116] polyethylene (PE),[117] poly(maleic anhydride-alt-1-octadecene) (PMAO),[118] Si–N/Si–O-based polysilazane (PSZ),[100b] self-dried acrylic resin (SDAR),[119] benzylphosphonic acid (BPA),[12a] polyacrylonitrile (PAN),[120] poly (ethyl α-cyanoacrylate) (PECA),[121] poly(2-ethyl-2-oxazoline) (PEOXA),[122] octavinyloctasilasesquioxane (OVS),[121] poly (maleic anhydride-alt-1-octadecene) (PMA),[123] a dendrimer (dendritic-structured ligands with a stiff structure and strong hydrophobic tail),[100c] metal stearate,[124] and metal–organic frameworks (MOFs) (Figure 6c).[47, 100a, 125] Also, an organic matrix containing an hydrophobic group (PAN, PMAO, PE, PECA, OVS, dendrimers (a family of dendritic ammonium ligands synthesized by Wang et al.)) can effectively block the $H_2O$ penetration in the NCs core.[100c, 117-118, 120-121] Other molecules with a functional group such as C=O have been exploited to passivate the uncoordinated surface $Pb^{2+}$ ions. PEOXA was reported for example to promote the growth of $CsPb(Br,I)_3$ NCs and lower the crystallization temperature of $CsPb(Br,I)_3$ NCs films as its functional C=O group can coordinate the $PbI_2/PbBr_2$ precursors.[122] The encapsulating processes can be conducted *in situ* during the modified one-pot hot-injection synthesis procedure,[100c, 123] in the room temperature recrystallization,[12a, 116c, 122] or by a post-synthetic treatment.[116d, 117-119, 121] For example, Lee et al. reported an innovative in-situ core/shell perovskite NCs synthesis strategy using a benzylphosphonic acid (BPA) treatment.[12a] During the reaction, large 3D crystals were split into smaller NCs, and BPA effectively coated the NCs, acting as a passivation ligand by forming bonds with undercoordinated lead atoms. This significantly reduced the trap density in the core/shell NCs film while preserving good charge-transport properties. As a result, an ultra-bright, efficient, and stable core/shell NCs-based LEDs could be fabricated.[12a]

Additional coating strategies involve organic polymers. For example, Yu et al. used a drop-casting method to obtain a uniform and stable $CsPbBr_3$ NCs/PMMA film.[116b] No aggregates were formed that could act as PL quenching centers and decrease the device performance.[116b] Yoon et al. prepared $CsPbBr_3$ NCs/polysilazane (PSZ) films by electrospray deposition of a silazane oligomer-treated $CsPbBr_3$ NCs solution.[100b] The atmospheric moisture during the electrospray deposition promoted the cross-linking of the silazane oligomers on the NCs surface, forming a PSZ polymer matrix.[100b] Li et al. developed a fiber spinning chemistry strategy to prepare *in situ* $MAPbBr_3$ NCs/polyacrylonitrile (PAN) nanofiber films at room temperature.[120] In their synthesis, spinning fibers acted as the reactor: with DMF solvent evaporation, the PAN polymer solidified into nanofibers and the formed NCs were encapsulated in such PAN matrix. The method could be easily upscaled and is additionally appealing as it does not require organic ligands, plus it avoids the production of intermediate waste.[120]

## 4. Lead-free metal halide nanocrystals for LED applications

### 4.1 Violet-blue emitting lead-free nanocrystals

Driven by the motivation to replace lead with less toxic elements, various lead-free metal halide NCs have been studied widely in recent years as potential alternatives. In these sections, we aim to summarize the advances in this direction, focusing on NCs that have then been tested as emitters in LEDs applications. Their crystal structure, electronic structure, PL properties and mechanisms are discussed based on the emission color, starting here with violet-blue emitting NCs. Those reported to date include vacancy ordered double perovskite $Cs_2ZrCl_6$ NCs, double perovskite NCs, $Eu^{2+}$ based/doped NCs ($CsEuBr_3$, $CsBr:Eu^{2+}$ and $CsCaCl_3:Eu^{2+}$), as well as NCs of other non-perovskite phases (such as $Cs_3(Sb/Bi)_2Br_9$, $Cs_3InX_6$, or $Cs_3Cu_2X_5$). The PL performances of lead-free NCs with different emission colors are listed in Table 2. What follows in a non-exhaustive overview of the most representative cases. The much-investigated metal halide $Cs_2ZrCl_6$ NCs have a direct band gap and exhibit excellent optical properties. Liu et al. pioneered the synthesis of blue-emitting colloidal $Cs_2ZrCl_6$ NCs with a high PLQY of 60.37% (Figure 7a, b).[126] They found that the efficient PL stems from thermally activated delayed fluorescence (TADF)-related STEs emission, which involves reverse intersystem crossing (RISC) from long-lived triplet back to short-lived singlet states. Due to the small singlet-triplet splitting energy ($\Delta E$ = 0.064 eV), the RISC process occurs efficiently at room temperature, and this promotes the emission of triplet excitons.[126] Recently, Wei et al. increased the PLQY (from 57.3% to 65.0%) of the blue emission at 430 nm in $Cs_2ZrCl_6$ NCs by adding $InCl_3$ in the hot injection synthesis,[127] an effect that was attributed to the passivation of surface defects. The NCs were then used to fabricate an efficient phosphor-converted blue LED.[127] Blue emitting double perovskite $Cs_2NaYCl_6$:$Sb^{3+}$ NCs have been synthesized by Wang et al. and Bai et al. by a hot-injection method and then used in WLEDs.[128] $Cs_2NaYCl_6$ NCs have a broadband blue emission centered at 440 nm under 320 nm excitation, originating from STEs.[128] Like most double perovskites, the emission from $Cs_2NaYCl_6$ NCs is weak due to the parity-forbidden character of the transition, even though $Cs_2NaYCl_6$ has a direct band gap.[128] $Sb^{3+}$ has been a popular dopant since its outermost $5s^2$ electrons can contribute to the CBM and VBM in the band structure and thus modify considerably the electronic structure.[129] $Sb^{3+}$ doping has been exploited for example in $Cs_2NaYCl_6$ NCs, leading to an increased PLQY from 5.6% to 22.5%.[128a] The $Sb^{3+}$ ions induce a new transition ($^1S_0 \rightarrow {}^1P_1$), resulting in a higher probability of population of emissive STEs and thus enhanced emission (in the blue region in this case).[128]

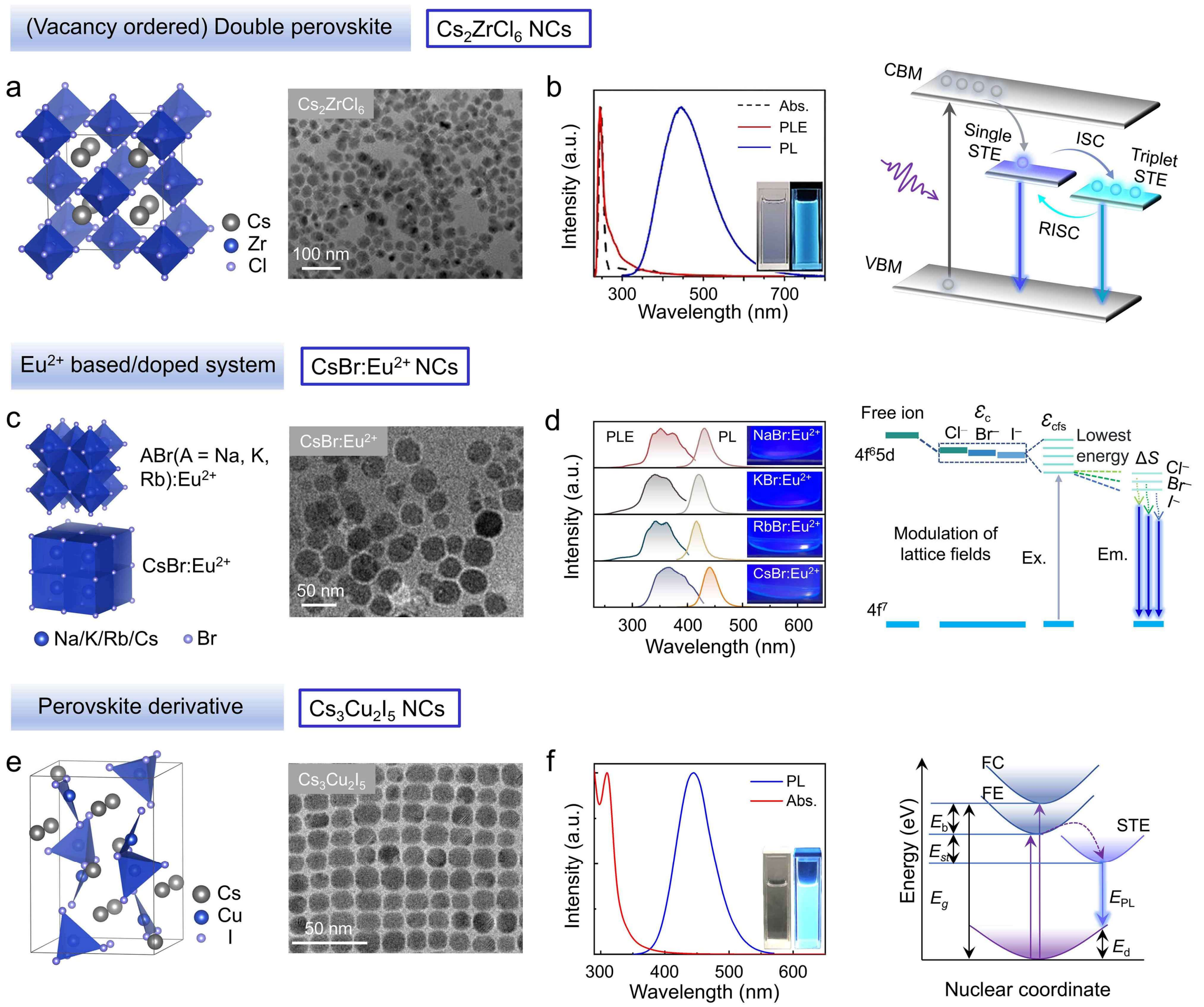

**Figure 7.** (a) Crystal structure and TEM images of $Cs_2ZrCl_6$ NCs; (b) Absorption, PLE, and PL spectra of $Cs_2ZrCl_6$ NCs. Reproduced and adapted from ref.[130]. Copyright 2020, Wiley. (c) Crystal structures of ABr:$Eu^{2+}$ (A = $Na^+$, $K^+$, $Rb^+$, $Cs^+$) NCs and TEM image of CsBr:$Eu^{2+}$ NCs; (d) PLE/PL spectra of ABr:$Eu^{2+}$ NCs and photophysical processes in AX:$Eu^{2+}$ NCs. Reproduced and adapted from ref.[38a]. Copyright 2024, American Chemical Society. (e) Crystal structure and TEM images of $Cs_3Cu_2I_5$ NCs; (f) PL /absorption spectra of $Cs_3Cu_2I_5$ NCs and the configuration diagram of the STEs dynamic mechanism. Reproduced and adapted from ref.[35]. Copyright 2020, American Chemical Society.

$Eu^{2+}$ based metal halide NCs can be promising candidates for lead based NCs in LED applications due to their high PLQYs (> 90%) and narrow FWHM (< 30 nm).[38a, 39, 131] Various groups have synthesized $CsEuX_3$ (X = Cl, Br) NCs, however only $CsEuBr_3$ NCs have been explored in LEDs to date, most likely due to their higher PLQY (> 90%) compared to that of $CsEuCl_3$ NCs (< 20%). $CsEuBr_3$ has an orthorhombic perovskite-type structure with $[EuBr_6]^-$ octahedra surrounded by $Cs^+$ cations, similar to that of lead halide perovskites, possibly due to the similarity in ionic radii for $Eu^{2+}$ (1.17 Å) and $Pb^{2+}$ (1.19 Å).[37, 39, 131a, 132] $CsEuBr_3$ has a direct band gap of ~3.75 eV[39] and its VBM is composed mainly of Eu 4*f* and Br 4*p* orbitals, while the CBM is dominated by Eu 5*d* orbital.[37, 39] In the high-temperature (280 ºC) hot injection synthesis of $CsEuBr_3$ NCs, $Eu^{2+}$ ions are produced in-situ from the reduction of $Eu^{3+}$ by oleylamine (OLA).[39] The reaction time and temperature need to be controlled carefully to obtain pure phases.[39] $CsEuBr_3$ NCs exhibited a narrowband deep-blue PL emission centered at 443 nm, deriving from Eu-5*d*→Eu-4*f*/Br-4*p* transitions, with a FWHM of 28.5 nm and a PLQY of 93.5%.[39]

$Eu^{2+}$ doped metal halide NCs, including AX (A = Na, K, Rb, Cs; X = Cl, Br, I) and $CsCaX_3$ (X = Cl, Br, I) NCs can also have bright narrow-band blue emissions and have been explored in LED applications.[38a, 39, 131] Differently from $Eu^{2+}$ based metal halide NCs, the PL emissions from $Eu^{2+}$ doped NCs originates from the $4f^6 5d^1$–$4f^7$ transitions of $Eu^{2+}$ ions instead of free carrier recombination.[131a] For example, CsBr:$Eu^{2+}$ NCs have blue emission at 441 nm with a PLQY of 91.1% and a FWHM of 30 nm (Figure 7c, d);[38a] A synthesis of $CsCaX_3$ (X = Cl, Br, I) NCs reported particles with blue emissions peaking from 435 to 458 nm, with a maximum PLQY of 81.9% and FWHM of 22–30 nm.[131b] When varying the cations and halide anions in $Eu^{2+}$ doped hosts, the blue emission shifts due to the influence of the crystal fields on the 5*d* orbital of the $Eu^{2+}$ ions.[38a, 131b] The peak shift is actually determined by the crystal field splitting effect and by the centroid shift.[38a, 131b, 133] As a reminder, the crystal field splitting energy represents the energy difference between the lowest and highest 5*d* levels of $Eu^{2+}$; the centroid shift is instead the downward shift of the average energy of the 5*d* energy levels compared to that of the free ion $Eu^{2+}$, which is also affected by the coordinating anions.[134] For example, when changing the A site cations from $Na^+$ to $K^+$ to $Rb^+$, the emission peak of $Eu^{2+}$-doped ABr (A = $Na^+$, $K^+$, $Rb^+$) NCs with the same octahedral coordination showed a blue shift (Figure 7c, d).[38a] This is because the (calculated) crystal field splitting energy of $Eu^{2+}$ gradually decreases from $Na^+$ to $K^+$ to $Rb^+$, leading to an increase in the energy gaps between the $Eu^{2+}$ 5*d* and 4*f* energy levels, hence a blue shift of the emission;[38a] In $Eu^{2+}$ doped $CsCaX_3$ (X = Cl, Br, I) NCs, when varying the halide anions from $Cl^-$ to $Br^-$ to $I^-$, the emission peak red shifts due to the centroid shift of the 5*d* orbital affected by the coordinating halide anions.[131b] The $Eu^{2+}$ 5*d* energy level exhibits smaller centroid shift when coordinated by halogens with higher electronegativity,[135] resulting in reduced energy gaps between $Eu^{2+}$ 5*d* and 4*f* energy levels, hence a red shift of the emission peak.[38a, 131b]

Apart from blue emitting $Eu^{2+}$ based metal halide NCs, $La^{3+}$ based metal halide $Cs_3LaCl_6$ NCs can also achieve blue emission and have been investigated in electrically driven LED applications.[38b] $La^{3+}$ ions are non-luminous rare earth ions because $La^{3+}$ ions have a $3d^{10}4f^{0}$ electronic configuration with a completely filled *d* shell and empty f shell.[136] Hence, the broadband emission at 442 nm from $Cs_3LaCl_6$ NCs with a PLQY of 57% was reported to originate solely from the host STE.[38b] Violet emitting NCs of another rare earth-based metal halide, $Cs_3CeCl_6·3H_2O$, characterized by a 1D crystal structure, were recently synthesized via a hot injection method by Sun et al.[137] $Cs_3CeCl_6·3H_2O$ NCs exhibited a doublet emission peak at 373 and 406 nm with a PLQY of ~100% arising from the Ce-5*d*→Ce-4*f* transition.[137] Recently, rare earth-based $Cs_3TmCl_6$ NCs with blue emission at 440 nm and a PLQY of 23% were synthesized following a hot injection method and were then used as color converters in a WLED.[138] The broadband blue emission of $Cs_3TmCl_6$ NCs was reported to originate from STEs due to the excited-state structural distortion of the $[TmCl_6]^{3-}$.[138]

**Table 2**. PL performance of lead-free NCs with different emission colors in LED applications.

| Composition | $\lambda_{abs}$ (nm) | $\lambda_{em}$ (nm) | FWHM (nm) | PLQY (%) | PL mechanism | Application | Refs. |
|---|---|---|---|---|---|---|---|
| Violet emitting lead-free nanocrystals | | | | | | | |
| $Cs_3Sb_2Br_9$ | 368 | 408 | / | 51.2 | / | Electrically driven LED | [21a] |
| $Cs_3CeCl_6·3H_2O$ | 334 | 373, 406 | / | 100 | Ce-5*d*→Ce-4*f* transition | Electrically driven LED | [137] |
| Blue emitting lead-free nanocrystals | | | | | | | |
| $InCl_3$ treated $Cs_2ZrCl_6$ | 250 | 430 | 121 | 65 | STEs | Phosphor-converted LED | [127] |
| $Cs_3Bi_2Br_9$ | 396 | 410 | 48 | 19.4 | exciton recombination | WLED | [139] |
| $FA_3Bi_2Br_9$ | 404 | 437 | 65 | 52 | exciton recombination | Phosphor-converted LED | [140] |
| | 300 | 445 | 63 | 87 | | Electrically driven LED | [35] |
| $Cs_3Cu_2I_5$ | 295 | 440 | / | 96.6 | STE | Phosphor-converted LED | [32a] |
| | ~300 | 441 | 94 | ~100 | | | [141] |
| $CsBr:Eu^{2+}$ | ~360 | 441 | 30 | 91.1 | $4f^65d^1$→$4f^7$ of $Eu^{2+}$ | Electrically driven LED | [38a] |
| | 272, 345 | 440 | 31 | 32.8 | | WLED | [131a] |
| | 340 | 444 | 30 | 53.4 | | | [131c] |
| $CsEuBr_3$ | 360 | 443 | 28.5 | 93.5 | $4f^65d^1\rightarrow4f^7$ of $Eu^{2+}$ | WLED | [39] |
| $CsCaCl_3:Eu^{2+}$ | 365 | 435 | 22 | 81.9 | $4f^65d^1\rightarrow4f^7$ of $Eu^{2+}$ | Phosphor-converted LED | [131b] |
| $Cs_3InBr_6$ | 370 | 440 | 80 | 46 | STE | WLED | [32b] |
| $Cs_2NaYCl_6:Sb^{3+}$ | 320 | 440 | / | 22.5 | STE | Phosphor-converted LED | [128a] |
| $Cs_3LaCl_6$ | 362 | 450 | / | 57.2 | STEs | Electrically driven LED | [38b] |
| $Cs_3Tb_{0.5}Ce_{0.5}Cl_6$ | 340 | 412, 546 | / | 92% | *d–f* transition of $Ce^{3+}$ and *f–f* transition of $Tb^{3+}$ | WLED | [142] |
| Green emitting lead-free nanocrystals | | | | | | | |
| $Cs_3TbCl_6$ | 365 | 545 | / | 76.3% | $^5D_4\rightarrow{}^7F_J$ (6,5,4,3) of $Tb^{3+}$ | Electrically driven LED | [38b] |
| $Cs_2ZnBr_4:Mn^{2+}$ | 365 | 528 | 54 | 26.4% | $Mn^{2+}$ in tetrahedral coordination | WLED | [143] |
| Yellow emitting lead-free nanocrystals | | | | | | | |
| $CsCu_2I_3$ | 330 | 580 | / | 47.3% | STE | Phosphor-converted LED | [144] |
| | 312 | 556 | 100 | 17 | | WLED | [141] |
| $Cs_2SnBr_6$ | 340 | 600 | 121 | 31 | STE | Phosphor-converted LED | [33] |
| Red emitting lead-free nanocrystals | | | | | | | |
| $Cs_2NaYCl_6$:3%$Sb^{3+}$, 7%$Mn^{2+}$ | 320 | 440, 630 | / | 38.7 | $Mn^{2+}$ ions | Phosphor-converted LED | [128a] |
| $Cs_3EuCl_6$ | 365 | 617 | 18 | 92.4 | $^5D_0\rightarrow{}^7F_J$ (1,2) of $Eu^{3+}$ | Electrically driven LED | [38b] |
| $Cs_2ZnBr_4:Mn^{2+}$ | 365 | 660 | 115 | 11.7 | $Mn^{2+}$ in octahedral coordination | WLED | [143] |
| $CsMnCl_3:Sn^{2+}$ | 420 | 669 | 98 | ~12 | $^4T_1\rightarrow{}^6A_1$ transition of $Mn^{2+}$ | Phosphor-converted LED and micro-scale pixel array | [145] |
| $CsMnBr_3$ | 365, 412 | 649 | 130 | 65.1% | $^4T_1 \rightarrow {}^6A_1$ transition of $Mn^{2+}$ | WLED | [146] |
| $Cs_2AgBiCl_6:Al^{3+}$ | 365 | 630 | 196 | 17.2% | STE | WLED | [147] |
| NIR emitting lead-free nanocrystals | | | | | | | |
| $CsSnBr_3$ | / | 683 | 50 | / | / | Electrically driven and Phosphor-converted LED | [26] |
| $CsSnI_3$ | / | 983 | 83 | / | / | | |
| White emitting lead-free nanocrystals | | | | | | | |
| $Cs_2NaYCl_6:3\%Sb^{3+}, 4\%Mn^{2+}$ | 320 | 440, 630 | / | ~35 | STE and $Mn^{2+}$ ions | WLED | [128a] |
| $Cs_3LaCl_6:Sb^{3+}$ | 320 | 575 | / | 30 | STE ($^3P_1 \rightarrow {}^1S_0$ of $Sb^{3+}$) | | [148] |
| $Cs_3ScCl_6:Sb^{3+}$ | 365 | 420&450, 480&555 | / | 48 | STE ($^3P_1 \rightarrow {}^1S_0$ of $Sb^{3+}$) and carbon dots | | [149] |
| $Cs_2NaInCl_6:Sb^{3+}, Sm^{3+}$ | 320 | 460, 602 | / | 80.1 | STE ($^3P_1 \rightarrow {}^1S_0$ of $Sb^{3+}$) and $Sm^{3+}$ | | [150] |
| $K_3SbCl_6:Mn^{2+}$ | 365 | 440, 600 | 102, / | 37.2 | STE and ($^4T_1 \rightarrow {}^6A_1$) of $Mn^{2+}$ | | [151] |
| $Cs_2AgInCl_6:30\% Bi^{3+}$ | 380 | 580 | / | 4 | STE | | [152] |
| $Cs_2AgIn_{0.9}Bi_{0.1}Cl_6$ | 375 | 650 | / | 31.4 | STE | Electrically driven WLED | [153] |
| $CsAgCl_2$ | 254 | 628 | ~240 | 39 | STE | | [154] |

Among the blue emitting metal halide NCs not based on perovskite structures, $A_3B_2X_9$ (A = Cs, FA; B = Sb, Bi; X = Cl, Br) NCs,[21a, 36, 139-140] $Cs_3Cu_2X_5$ (X = Br, I) NCs,[32a, 34-35, 141, 144, 155] and $Cs_3InX_6$ (X = Cl, Br, I) NCs[32b] with bright blue emission and relatively good stability have been extensively studied and have also tested as LED emitters. $Bi^{3+}$ and $Sb^{3+}$ ions have outermost electronic configuration ($Bi^{3+}$: $6s^2$ and $Sb^{3+}$: $5s^2$) similar to that of $Pb^{2+}$ ions ($6s^2$).[129a] Differently from lead-based perovskites with the $ABX_3$ chemical formula, to preserve the charge balance, $Bi^{3+}$ or $Sb^{3+}$ based metal halides adopt a $A_3B_2X_9$ chemical formula and their structure consists either of stacks of two-dimensional layers or isolated octahedral units.[156] Blue emitting $A_3B_2X_9$ (A = Cs, FA; B = Sb, Bi; X = Cl, Br) NCs can be synthesized by a facile ligand-assisted recrystallization method.[21a, 36, 139-140] Their PLQYs can be improved by synthesis optimization, including adjusting the precursor concentration, amount of ligands and reaction temperature.[21a, 36, 139-140] Additional ways to boost the PLQY have been also reported. For example, Ma et al. developed a treatment that improves the PLQY (from 20.2% to 46.4%) consisting in the addition of deionized water to a $Cs_3Bi_2Br_9$ NCs toluene solution.[36] The treatment promotes the formation of a BiOBr matrix that passivates the $Cs_3Bi_2Br_9$ NC surface defects.[36] By controlling the halide composition from Cl to I, the PL emissions can be finely tuned in the visible region (393–587 nm for $Bi^{3+}$ based NCs and 385–640 nm for $Sb^{3+}$ based NCs).[21a, 36, 140] Notably, the violet-blue emission from $A_3Bi_2X_9$ (A = Cs, FA; X = Cl, Br) NCs originates from free exciton recombination, featuring narrow bandwidths and small Stokes shift. Stable blue emitting $Cs_3Bi_2Br_9$/silica NCs (with a PLQY of 46.4%) and $FA_3Bi_2Br_9$/polymer polystyrene NCs (with a PLQY of 52%) were reported as emitters in phosphor-converted LEDs.[139-140] The violet broadband emission of $Cs_3Sb_2Br_9$ NCs originates from STEs. NCs of this material, with PLQY of 51.2%, were implemented in an electrically driven LED.[21a]

$Cs_3Cu_2X_5$ (X= Br, I) NCs have a 0D structure formed by isolated $[Cu_2X_5]^{3-}$ units (Figure 7e).[155a] In the band structure of $Cs_3Cu_2X_5$ (X = Br, I), the VBM has contributions from Cu 3*d* and halogen *p* orbitals, while the CBM is mainly composed of Cu 4*s* and halogen *p* orbitals.[32a, 141, 155a] The direct band gap decreases in going from Cl to Br to I, and thus the PL emission gradually blue-shifts (from green to blue) when varying the halide ions from Cl to Br to I, contrary to the case of $APbX_3$ discussed earlier.[141] Many groups have synthesized $Cs_3Cu_2X_5$ NCs, either by a hot injection method or by ligand-assisted recrystallization.[32a, 34, 141, 144, 155] $Cs_3Cu_2I_5$ NCs have bright deep-blue emission at ~440 nm with a very efficient PL (close to 100%) originating from STE recombination (Figure 7f).[32a, 141, 144] Benefiting from the near-unity PLQY and good stability, $Cs_3Cu_2I_5$ NCs were tested as emitters in both phosphor converted and electrically driven LEDs.[32a, 34-35, 141, 144, 155b] Similar to $Cs_3Cu_2X_5$, $Cs_3InX_6$ compounds with a 0D structure have direct band gaps.[32b, 157] Zhang et al. used a room-temperature recrystallization method to synthesize blue emissive $Cs_3InBr_6$ NCs with an emission peak at 440 nm and a PLQY of 46%.[32b] The obtained $Cs_3InBr_6$ NCs exhibited good stability and were used as blue emitters in WLEDs.[32b]

### 4.2 Green-yellow emitting lead-free nanocrystals

Among the different green emitting lead-free metal halide NCs, all-inorganic copper(I) halide ($Cu_3Cu_2Cl_5$) NCs and rare earth-based metal halide ($Cs_3TbCl_6$) NCs with relatively high PLQYs have been exploited as green emitters in phosphor converted and electrically driven LEDs. $Cu_3Cu_2Cl_5$ NCs reported by Halpert et al. have a green PL emission at ~520 nm with near-unity PLQY.[34] The PL peaks of $Cs_3Cu_2X_5$ (X = Cl, Br, I) NCs can cover a spectral region of 440–530 nm through halogen substitution (Figure 8a, b). These NCs with spherical shape had uniform sizes (below 10 nm in diameter) and excellent crystallinity (**Figure 8**a, b). In the 0D crystal structure of $Cu_3Cu_2Cl_5$, $Cu^+$ ions are tetrahedrally coordinated to $Cl^-$ ions, forming isolated $[Cu_2Cl_5]^{3-}$ dimers units separated by $Cs^+$ ions (Figure 8a).[158] Similar to other copper halides, $Cu_3Cu_2Cl_5$ displays structural distortion and strong electron-phonon coupling upon photoexcitation, resulting in the formation of STEs.[159] Thus, the green PL emission from these materials generally features large Stokes shifts (>200 nm) and broad FWHM values (~108 nm).[159a, 160] $Cu_3Cu_2Cl_5$ has a direct band gap of ~4.4 eV[34, 160] and the VBM of its band structure is mainly contributed by Cu 3*d* orbitals, whereas the CBM is mainly comprised of Cu 4*s* orbitals.[34, 160] $Cu_3Cu_2Cl_5$ NCs can be synthesized by both antisolvent recrystallization and hot injection methods.[34] Unfortunately, the easy oxidization of $Cu^+$ in the bright emitting $Cu_3Cu_2Cl_5$ NCs limits their application in LEDs, hence additional strategies for stabilizing $Cu^+$ species should be investigated in the future.[159a]

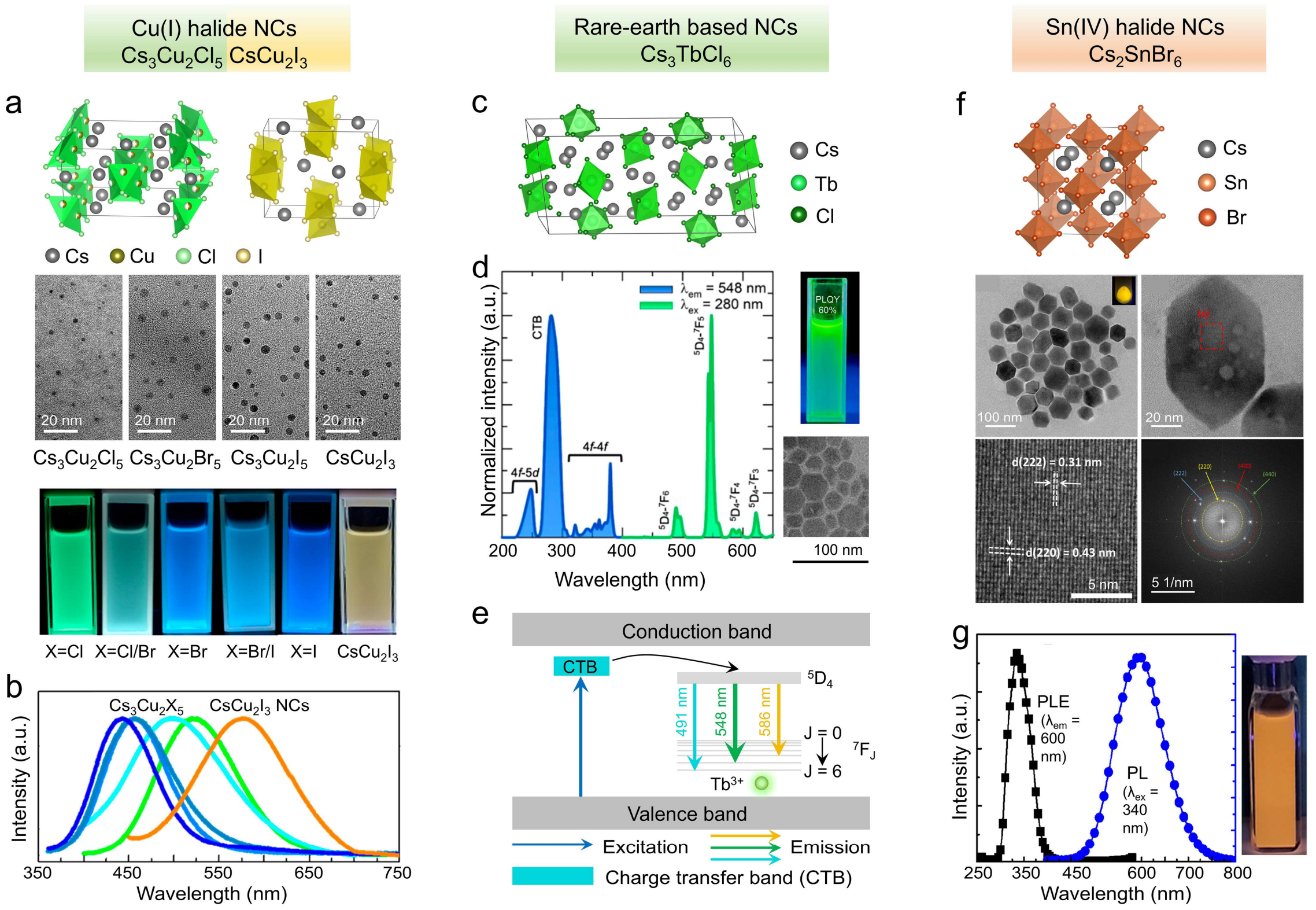

**Figure 8.** (a) Crystal structure of $Cs_3Cu_2X_5$ (X = Cl, Br, I) and $CsCu_2I_3$; (b) TEM images of $Cs_3Cu_2X_5$ (X = Cl, Br, I) and $CsCu_2I_3$ NCs, along with photographs (under UV light excitation)

and PL spectra of their colloidal suspension. Reproduced and adapted from ref.[34]. Copyright 2020, American Chemical Society. (c) Crystal structure of $Cs_3TbCl_6$; (d) PL and PLE spectra (left part) of $Cs_3TbCl_6$ NCs; the right part reports the corresponding photographs of colloidal suspensions of NCs under UV light excitation and a representative TEM image; (e) Schematic diagram of the PL mechanism. Reproduced and adapted from ref.[161]. Copyright 2022, American Chemical Society. (f) Crystal structure and TEM images of $Cs_2SnBr_6$ NCs; (g) PLE/PL spectra and photographs of a colloidal suspension of $Cs_2SnBr_6$ NCs. Reproduced and adapted from ref.[33]. Copyright 2021, American Chemical Society.

Recently developed green emitting $Cs_3TbCl_6$ NCs with high PLQY and good thermal stability have been tested in both down-converted phosphor-converted LEDs and electrically driven LEDs.[38b, 142] $Cs_3TbCl_6$ possesses a 0D structure formed by isolated $[TbCl_6]^{3-}$ octahedra that are charge-balanced by $Cs^+$ ions (Figure 8c). This compound has a direct band gap of 5.1 eV.[161-162] In the band structure of $Cs_3TbCl_6$, the Cl 3*p* orbital mainly contributes to the valence band states and the Tb 5*d* orbitals mainly contribute to the conduction band states.[161-162] $Cs_3TbCl_6$ NCs exhibit a weak broadband blue emission with emission peaking at 430 nm stemming from the STEs, and several sharp emission peaks at 489, 546, 588, and 621 nm assigned to $^5D_4 \rightarrow {}^7F_6$, $^5D_4 \rightarrow {}^7F_5$, $^5D_4 \rightarrow {}^7F_4$, and $^5D_4 \rightarrow {}^7F_3$ transitions of $Tb^{3+}$ ions, respectively (Figure 8d, e).[38b, 163] Sun et al. introduced a Lewis base molecule, ethyl 7-hydroxycoumarin-3-carboxylate (EHC) in the hot-injection synthesis of $Cs_3TbCl_6$ NCs of this material, and was able to increase their PLQY to 76.3%.[38b] It was hypothesized that the enhanced PLQY originated from efficient energy transfer process from EHC molecules to the $^5D_4$ energy level of $Tb^{3+}$ ions.[38b] The EHC modified $Cs_3TbCl_6$ NCs were used as the green emitter in an electrically driven LED.[38b] Samanta et al. incorporated $Ce^{3+}$ ions into $Cs_3TbCl_6$ NCs, increasing their PLQY to 86% in the case of NCs with $Cs_3Tb_{0.1}Ce_{0.9}Cl_6$ composition.[142] The $Cs_3Tb_{0.1}Ce_{0.9}Cl_6$ NCs displayed broadband luminescence peaking at 380 and 412 nm from $Ce^{3+}$ ions and sharp emission lines from $Tb^{3+}$ ions, making it the blue and green emitter in the WLED. The increased PLQY was due to the efficient energy transfer from the $5d^1$ excited state of $Ce^{3+}$ ions to the $^5G_1$ energy state of $Tb^{3+}$ ions.[142]

Moving to a different material, $Cs_2ZnBr_4$:$Mn^{2+}$ NCs in glass were also studied as the green emitters in WLEDs.[143] As-prepared $Cs_2ZnBr_4$:$Mn^{2+}$ NCs in glass had a single broadband emission at 660 nm stemming from emission from the octahedrally coordinated $Mn^{2+}$ ions in the glass matrix. Following a thermal treatment at 540 °C, the $Mn^{2+}$ ions were incorporated into the $Cs_2ZnBr_4$ NCs, resulting in an additional green emission at 528 nm with a FWHM of 54 nm from the tetrahedrally coordinated $Mn^{2+}$ ions.[143]

Lead-free yellow emitters with emission peaks in the 550–600 nm region that have been explored for LEDs include $CsCu_2I_3$ NCs and $Cs_2SnBr_6$ NCs, both with broadband emissions stemming from STEs. Yellow emitting $CsCu_2I_3$ NCs/NRs have been primarily studied. $CsCu_2I_3$ NCs/NRs can be easily synthesized by methods such as hot injection,[34, 144, 155b, 164] room-temperature antisolvent recrystallization,[34] or by a water post-treatment of blue emitting $Cs_3Cu_2I_5$ NCs.[141] In their 1D crystal structure, edge-sharing and face-sharing copper iodide tetrahedra form double 1D chains.[155b, 165] $CsCu_2I_3$ has a direct band gap and 1D electronic structure, which endows relatively low carrier effective masses along the chain direction.[166] The VBM is contributed by Cu 3*d* orbitals and I 5*p* orbitals, and the CBM is composed of Cu 4*s* and I 5*p* orbitals.[164b, 166] The reported $CsCu_2I_3$ NCs/NRs showed broadband PL emission at ~560 nm with a FWHM of ~100 nm.[34, 155b, 164b] The emission peak from the NRs (550–560 nm) was slightly blue shifted with respected to that of NCs (560–570 nm).[34, 155b, 164b] Compared to other copper halide NCs, $CsCu_2I_3$ NRs/NCs have relatively low PLQY values (the highest reported PLQY is ~11%) possibly due to a high density of surface defects.[141, 155b]

Vacancy-ordered double perovskite $Cs_2SnBr_6$ NCs with an orange emission peak at ~600 nm and good stability were also reported as emitters in a phosphor-converted candlelight LED.[33] $Cs_2SnBr_6$ crystalizes in a 0D structure formed by isolated $[SnBr_6]^{2-}$ octahedra units charge-balanced by $Cs^+$ ions (Figure 8f). In its band structure, characterized by a direct band gap, the VBM is composed mostly of Br 4*p* orbitals, and the CBM is contributed by the Br 4*p* and, to a minor extent, Sn 5*s* orbitals.[167] The $Cs_2SnBr_6$ NCs (prepared by a hot injection method) exhibited broadband emission at 600 nm from STEs recombination with a FWHM of ~121 nm, Stokes shift of 255 nm and a PLQY of 31% (Figure 8g).[33]

**4.3 Red and NIR emitting lead-free NCs**

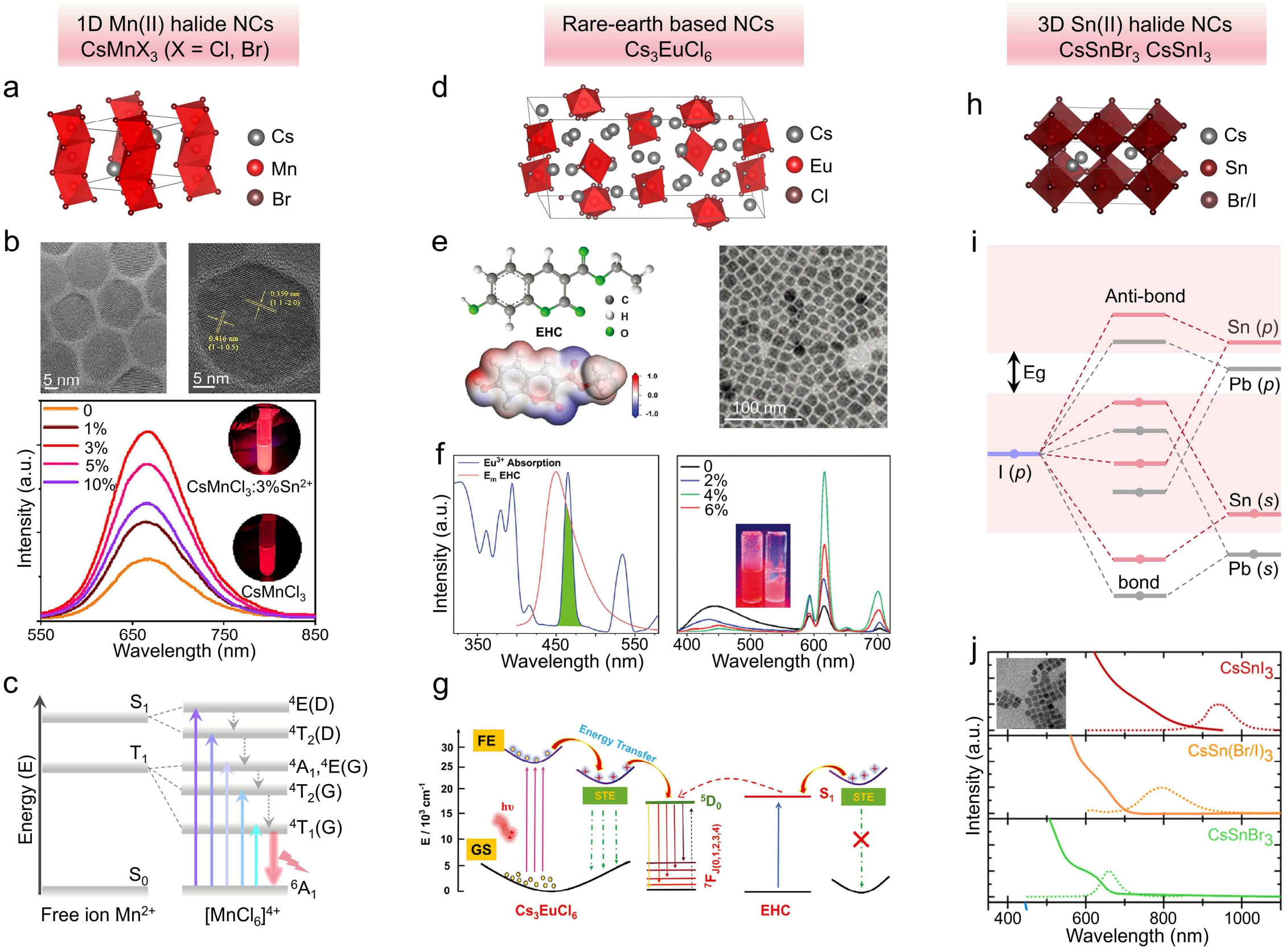

**Figure 9.** (a) Crystal structure of $CsMnX_3$; (b) TEM images and PL spectra of $CsMnCl_3$:$x$$Sn^{2+}$ NCs; (c) PL emission mechanism of $CsMnCl_3$ NCs. Reproduced and adapted from ref.[145]. Copyright 2024, Elsevier. (d) Crystal structure of $Cs_3EuCl_6$; (e) molecular structure and electrostatic potential map of the EHC molecule, and TEM image of $Cs_3EuCl_6$ NCs; (f) left: EHC emission (red line) and $Eu^{3+}$ absorption spectra (blue line); right: PL spectra of $Cs_3EuCl_6$ NCs treated with different amounts of EHC. (g) PL mechanism of EHC modified $Cs_3EuCl_6$ NCs. Reproduced and adapted from ref.[38b]. Copyright 2024, Wiley. (h) Crystal structure of $CsSnX_3$ (X = Br, I); (i) electronic band structure of $CsSnX_3$. Reproduced and adapted from ref.[168]. Copyright 2021, American Chemical Society. (j) PL spectra of $CsSnX_3$ (X = Br, I) NCs and TEM image of $CsSnI_3$ NCs. Reproduced and adapted from ref.[169]. Copyright 2016, American Chemical Society.

Different types of red-emitting lead-free metal halide NCs have been studied as the red emitters in LED, including 3D double perovskite $Cs_2AgBiCl_6$:$Al^{3+}$ NCs, 3D $CsSnBr_3$ NCs, 2D Ruddlesden−Popper-type $(C_{18}H_{35}NH_3)_2SnBr_4$, 1D $Mn^{2+}$ based metal halide $CsMnX_3$ (X = Cl, Br) NCs, and 0D $Eu^{3+}$ based metal halide NCs. The red emissions, characterized by either broad

or a narrow emission band can stem from STE recombination or from transitions between $Mn^{2+}$ or $Eu^{3+}$ states, depending on the specific cases.

STEs tend to form in metal halides having either a low electronic dimensionality (low connectivity of the atomic orbitals) or a low structural dimensionality (low connectivity of the polyhedra).[27] $Cs_2AgBiCl_6$ crystalizes in a 3D perovskite structure with $[AgCl_6]^{5-}$ and $[BiCl_6]^{3-}$ octahedra alternately connecting to each other, thereby $Cs_2AgBiCl_6$ has a low dimensional electronic structure.[170] Li et al. doped $Al^{3+}$ ions into $Cs_2AgBiCl_6$ NCs, obtaining an additional broadband red emission at 630 nm with a FWHM of 196 nm and a PLQY of 17.2%.[147] This is to be compared to the case of undoped $Cs_2AgBiCl_6$ NCs with indirect band gap, which showed only a blue-violet emission with a PLQY of 0.98%.[147] Upon $Al^{3+}$ doping, the Al *s* orbitals participate in the CBM, leading to a direct band gap in $Cs_2AgBiCl_6$:$Al^{3+}$, thus enhancing the radiative STEs emission in the red region.[147] $(C_{18}H_{35}NH_3)_2SnBr_4$ crystalizes in a 2D Ruddlesden−Popper-type structure with layers of $[SnBr_6]^{4-}$ octahedra alternating with layers of $C_{18}H_{35}NH_3^+$ ligands that are partially interdigitated and bound to the inorganic layers through interactions of the ammonium head groups with the $Br^-$ ions.[171] Zhang et al. used a hot injection method to synthesize $(C_{18}H_{35}NH_3)_2SnBr_4$ NCs with a bright broadband emission at 620 nm (88% PLQY).[171] The broad FWHM (140 nm) and large Stokes shift (307 nm) supported an emission mechanism stemming from STEs.[171]

$Mn^{2+}$ based/doped metal halide NCs have also been reported to have red emission arising from *d–d* transitions of $Mn^{2+}$ ions and have been used as the red emitting component in phosphor-converted WLEDs.[145-146] $Mn^{2+}$ ions in an octahedral coordination environment usually exhibit orange-red emission arising from the $^4T_1 \rightarrow {}^6A_1$ transition of $Mn^{2+}$ ions.[146, 172] For example, Bai et al. prepared red-emitting $Cs_2NaYCl_6$:$Sb^{3+}$,$Mn^{2+}$ NCs by varying the doping amount of $Mn^{2+}$ ions.[128a] The red emission of the $Cs_2NaYCl_6$:$Sb^{3+}$ was also accompanied by a broadband blue emission at 440 nm from the host STEs. $Mn^{2+}$ doping endowed an additional emission band peaking at ~630 nm. The red emission with PLQY of 38% was attributed to the energy transfer from the STEs of the $Cs_2NaInCl_6$ host to the $Mn^{2+}$ ions.[128a] $Mn^{2+}$ based metal halide $CsMnX_3$ (X = Cl, Br) crystallize in 1D chain structure with both corner- and face-sharing $[MnX_6]^{4-}$ octahedra (**Figure 9**a). A recent work reported $CsMnX_3$ (X = Cl, Br) NCs possessing good spectral stability under ambient conditions, however it was hypothesized that the small Mn-Mn distance in $CsMnX_3$ NCs results in strong Mn-Mn coupling interactions and PL quenching, similar to the concentration quenching effect.[173] Synthesis optimization and ions doping have been amply explored to improve the PLQY of these systems for LED applications.

Xu et al. optimized the amount of cesium carbonate in their solid-state method to synthesize red-emitting $CsMnBr_3$ NCs embedded in glasses with good stability and improved PLQYs as the red emitters in WLED.[146] They optimized the amount of cesium carbonate to improve the PLQYs. $CsMnBr_3$ NCs showed a broadband red emission peaked at 649 nm with a PLQY of 65.1%.[146] Ren et al. doped $Sn^{2+}$ into $CsMnCl_3$ NCs by the hot-injection method, resulting in an increased PLQY of the red emission (at 630 nm) from 12% to 49% (Figure 9b, c).[145] The incorporation of $Sn^{2+}$ with its $5s^2$ outer electronic configuration was reported to cause the appearance of a new absorption peak at 336 nm from the $^1S_0 \rightarrow {}^3P_1$ transition. The enhanced absorption and efficient energy transfer process from $Sn^{2+}$ to $Mn^{2+}$ led to an increased PLQY.[145] $Mn^{2+}$ doped in a $Cs_2ZnBr_4$ NC host with tetrahedral units embedded in glass without thermal treatment were reported to show a broadband red emission at 660 nm, an apparently counterintuitive finding, given that tetrahedrally coordinated $Mn^{2+}$ ions have green emission.[143] It was demonstrated that, without heat treatment, $Mn^{2+}$ ions could not be incorporated into $Cs_2ZnBr_4$ NCs. The red emission was therefore originating from the octahedrally coordinated $Mn^{2+}$ ions in the glass matrix.[143]

Rare earth $Eu^{3+}$ ions with $4f_6$ configuration exhibit sharp emission lines in the red region, originating from their $4f$–$4f$ transitions, hence $Eu^{3+}$ based metal halide NCs can show red emission arising from such transitions. However, such emission lines have insufficient PLQY for LED applications due to their parity-forbidden nature. Recent studies reported the synthesis of 0D metal halide $Cs_3EuCl_6$ NCs with $[EuCl_6]^{3-}$ octahedral units by a hot injection method which tested them as the emitting layer in electrically driven red LED (Figure 9d).[38b, 174] $Cs_3EuCl_6$ NCs exhibited a weak blue emission band from host STEs and a relatively intense red emission at 616 nm from $Eu^{3+}$ ions. Similar to green-emitting $Cs_3TbCl_6$ NCs mentioned above, Sun et al. introduced coumarin-based small molecules (ethyl 7-hydroxycoumarin-3-carboxylate, EHC) in the synthesis of $Cs_3EuCl_6$ NCs (Figure 9e).[38b] Apparently, the EHC molecules can serve as sensitizers and their singlet excited state ($S_1$) matched well with energy level of $Eu^{3+}$, thus facilitating an efficient energy transfer (ET) from host STEs → ($S_1$ state of EHC) →$Eu^{3+}$ (Figure 9f, g). Besides, the EHC were also hypothesized to passivate the NC surface defects, since their presence led to an enhanced PLQYs of 92.4% from the NCs.[38b]

$Sn^{2+}$ ion-based halide perovskite ($CsSnX_3$ (X = Br, I)) NCs as the emitters in both phosphor-converted and electrically driven LEDs have emissions in the red and near-infrared (NIR) region (depending on the halide composition) arising from band edge recombination. $CsSnX_3$ (X = Br, I) NCs have crystal structure and band structures similar to those of lead-based

perovskites (Figure 9h).[26] In the band structure of $CsSnX_3$, $Sn^{2+}$ *s* orbitals and halide X *p* orbital contribute to the VBM, and the $Sn^{2+}$ *p* orbital is dominant in the CBM, resulting in a direct band gap.[175] Notably, compared to $CsPbX_3$, the weaker spin-orbit coupling (SOC) in $CsSnX_3$ leads to an increased band dispersion, thus shifting the band gap into the red-NIR region (Figure 9i, j).[169, 175a, 175b] $CsSnX_3$ NCs are generally synthesized via a modified hot injection method in which tri-n-octyl phosphine is added as a reducing and a coordinating solvent to dissolve the $SnX_2$.[168, 176] Mahesh et al. synthesized $CsSnX_3$ (X = Br, I) NCs by a hot-injection method with NIR emissions at 683 and 938 nm, respectively, and reported their application in phosphor converted and electrically driven NIR LEDs.[26] Most studies for LED applications deal with $CsSnX_3$ (X = Br, I) bulk films. As a note, the instability from easily oxidized $Sn^{2+}$ in $CsSnX_3$ should be addressed for high-performance LED applications.

### 4.4 White emitting lead-free NCs

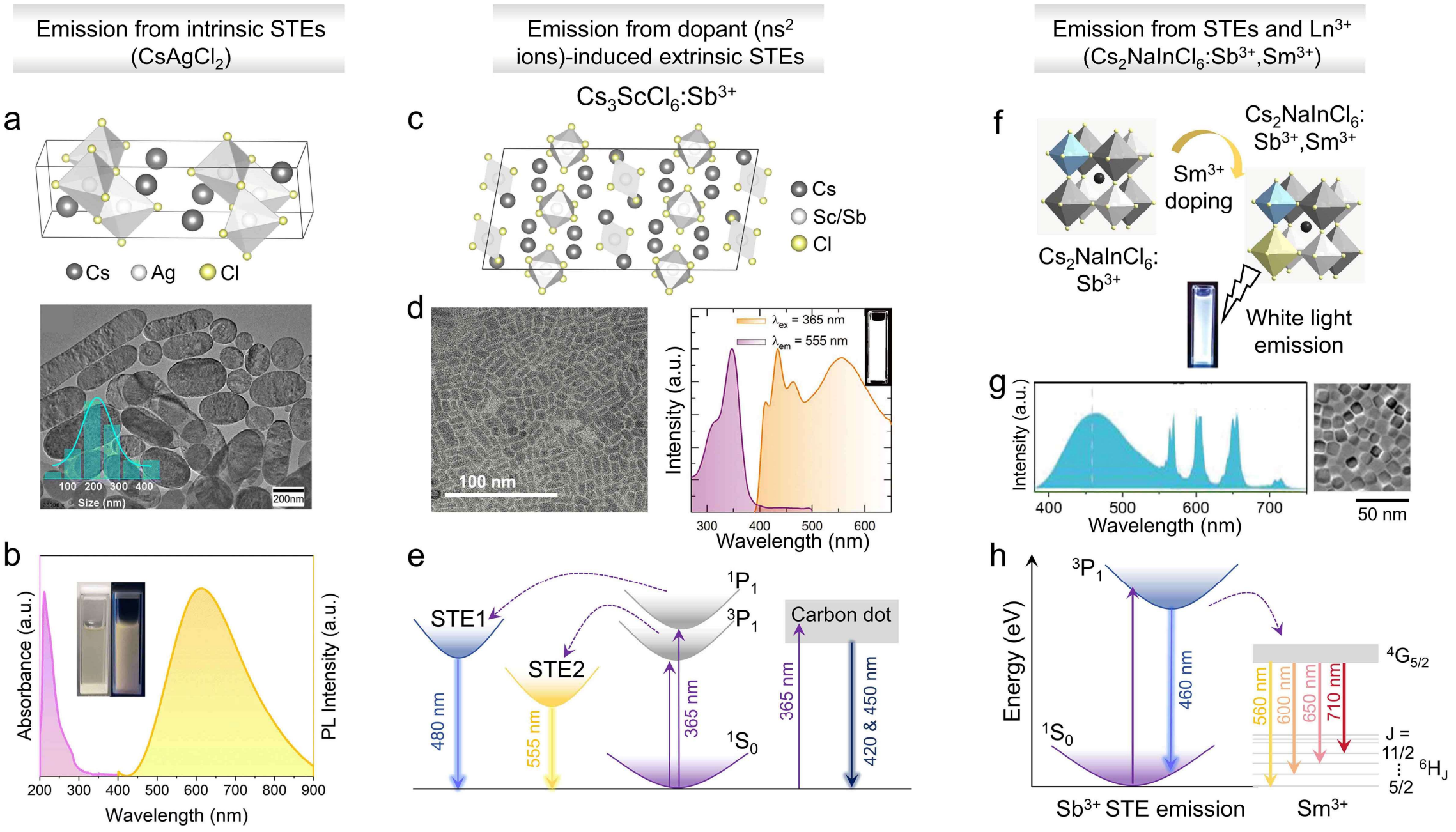


**Figure 10.** (a) Crystal structure and representative TEM image of $CsAgCl_2$ NCs; (b) absorption and PL spectra of $CsAgCl_2$ NCs. Reproduced and adapted from ref.[154]. Copyright 2022, Royal Society of Chemistry. (c) Crystal structure of $Cs_3ScCl_6$:$Sb^{3+}$; (d) representative TEM image and PLE/PL spectra of $Cs_3ScCl_6$:$Sb^{3+}$ NCs; (e) PL mechanism of $Cs_3ScCl_6$:$Sb^{3+}$ NCs. Reproduced and adapted from ref.[149]. Copyright 2023, Wiley. (f) Scheme of white light emission from $Cs_2NaInCl_6$:$Sb^{3+}$,$Sm^{3+}$ NCs; (g) PL spectra and representative TEM image of

$Cs_2NaInCl_6$:$Sb^{3+}$,$Sm^{3+}$ NCs; (h) white light emission mechanism from $Cs_2NaInCl_6$:$Sb^{3+}$,$Sm^{3+}$ NCs. Reproduced and adapted from ref.[150]. Copyright 2023, Wiley.

Lead-free single-component white-light NC emitters are ideal candidates for environmental-friendly WLEDs as they avoid undesirable energy transfer processes and elaborate LED fabrication processes. Various groups have developed different types of white emitting lead-free NCs and used them in WLED applications. The white emission in these cases can stem from intrinsic STEs, extrinsic STEs, or can be achieved by additional transitions between dopant states ($Ln^{3+}$, $Mn^{2+}$). Dopant ions can help to achieve white emission by providing blue, green or red emission components. For example, $Sm^{3+}$ or $Mn^{2+}$ doping brings red emission, in addition to any emission from the host, thus leading to white light emission in specific cases.[128a, 150-151, 177]

Emission from intrinsic STEs usually occurs in metal halides with low electronic dimensionality and soft lattice.[27b, 178] The broadband emission and a large Stokes shift from STEs recombination enable the achievement of single-component white-light emission.[30b] Silver and indium based double perovskite NCs have isolated $[AgCl_6]^{5-}$ and $[InCl_6]^{5-}$ octahedra. The $[AgCl_6]^{5-}$ octahedra are not connected to each other, and so do $[InCl_6]^{5-}$ octahedra, leading to a low electronic dimensionality. Upon light excitation, Ag-Cl bonds are elongated in the axial direction and compressed in the equatorial plane, resulting in a strong Jahn-Teller distortion of the $[AgCl_6]^{5-}$ octahedra and the formation of STEs.[179] Notably, the parity-forbidden transition in $Cs_2AgInCl_6$ results in a low PLQY, even though this material has a direct band gap. To improve the PLQY of $Cs_2AgInCl_6$ NCs, $Bi^{3+}$ and $K^+$ ions have been alloyed into $Cs_2AgInCl_6$ NCs. $Bi^{3+}$ ions were reported to passivate the $In^{3+}$ vacancies and suppressed non-radiative recombination loss.[179] The partial substitution of $Ag^+$ with $K^+$ can break the parity-forbidden transition by changing the symmetry of the STE wavefunction, thus promoting radiative recombination and improving the PLQY.[152-153, 179-180] The highest PLQY of $Cs_2AgInCl_6$ NCs is 31.4% for the emission from $Cs_2AgIn_{0.9}Bi_{0.1}Cl_6$ NCs covering almost the whole visible range.[153] $CsAgCl_2$ has 1D structure with isolated edge-sharing $[AgCl_5]^{4-}$ distorted trigonal bipyramidal units (**Figure 10**a).[181] Ji et al. synthesized $CsAgCl_2$ NCs with a broadband emission in the region of 450–900 nm and a Stokes shift of 313 nm from STEs recombination (Figure 10a, b).[154] $Sb^{3+}$ doping was used to improve the PLQY of $CsAgCl_2$ NCs from 16% to 39%, thus enhancing their potential for the electrically driven LED applications.[154] The PLQY improvement was possibly attributed to surface modification by $Sb^{3+}$ ions.[154] In metal halide

NCs with broadband emission in the blue light region from STEs, $Mn^{2+}$ ions can be alloyed in these hosts with octahedral coordination to offer red emitting component in the white light PL spectrum. For example, Shao et al. doped $Mn^{2+}$ ions into 0D $K_3SbCl_6$ NCs, obtaining double emission bands peaking at 440 nm and 600 nm, the former from the host STEs and the latter from the $^4T_1 \rightarrow {}^6A_1$ transition of $Mn^{2+}$ ions.[151] The red emitting component from $Mn^{2+}$ ions was essentially achieved through energy transfer from the host STEs. Also, a tunable white light emission color was realized by tuning the $Mn^{2+}$ ion doping concentration.[151]

Dopant-induced extrinsic STEs from n$s^2$ ions can also lead to white light emission. The n$s^2$ ions including $Te^{4+}$ and $Sb^{3+}$ have 5$s^2$ electron configuration. The ground state of $Te^{4+}$ and $Sb^{3+}$ ions is $^1S_0$. Upon light absorption, an electron transfer from 5*s* to 5*p* states takes place, resulting into an excited state with 5*s*5*p* electronic configuration.[129a] The excited state splits into four energy states: $^1P_1$, $^3P_0$, $^3P_1$, and $^3P_2$. The $^1S_0 \rightarrow {}^3P_0$ and $^1S_0 \rightarrow {}^3P_2$ transitions are forbidden, while the $^1S_0 \rightarrow {}^3P_1$ and $^1S_0 \rightarrow {}^1P_1$ transitions are parity-allowed.[129b] Under photoexcitation, the electrons of $Sb^{3+}$ or $Te^{4+}$ were excited from the 5*s* state to 5*p* state and trapped in $Te^{4+}$ or $Sb^{3+}$ halide polyhedrons, inducing broadband emissions from dopant STEs.[182] The broadband emissions from dopant-induced extrinsic STEs (sometimes together with host emissions or $Mn^{2+}$ ions emission) can lead to white light emission.[128a, 148-149, 183] For example, Yang et al. synthesized $Te^{4+}$ doped $Cs_2ZrCl_6$ NCs with broadband emission centered at 575 nm with a FWHM of ~115 nm from STEs formed by the $^3P_1 \rightarrow {}^1S_0$ transition of $Te^{4+}$.[183] The authors grew a polar alkyl-terminated silica-oligomer shell around the $Cs_2ZrCl_6$:$Te^{4+}$ NC core to passivate the surface defects, resulting in an increased stability and PLQY (>96%) for better-performance WLED.[183] Samanta et al. synthesized $Cs_3ScCl_6$:$Sb^{3+}$ NCs with bright white light emission (48% PLQY) (Figure 10c, d). The white light emission was composed of blue emitting components from carbon dots and yellow emitting components from STEs formed by $^3P_1 \rightarrow {}^1S_0$ and $^1P_1 \rightarrow {}^1S_0$ transition of $Sb^{3+}$ (Figure 10e).[149]

The incorporation of rare-earth ions is a common strategy to obtain white emission in phosphors due to their full-color luminescence in the visible region.[184] The emissions from the 4*f*–4*f* transitions of trivalent rare-earth ions feature sharp emission lines, long lifetime and good stability. Rare-earth ions can also be introduced into metal halide NCs with octahedral coordination, helping to realized single-component white light emission.[185] For example, Li et al. and Zhou et al. both doped rare earth $Sm^{3+}$ into $Cs_2NaInCl_6$:$Sb^{3+}$ NCs, achieving white light emissions (Figure 10f).[150, 177] The white emission consisted of one broadband emission in the

blue region from STEs emission of $Sb^{3+}$ and several sharp emission lines in the orange-red region from intrinsic *f–f* transitions of the $Sm^{3+}$ ions (Figure 10g, h).[150, 177] $Sm^{3+}$ ions doping could also passivate $In^{3+}$ vacancy defects and suppress nonradiative recombination, thus achieving a relatively high PLQY (65%) for white light emissions in $Cs_2NaInCl_6$:$Sb^{3+}$,$Sm^{3+}$ NCs.[150]

## 5. Perovskite nanocrystals lighting and display technology routes

In this section, we summarize the lighting and display technology routes that have been followed on perovskite NCs. We also discuss the key tactics to improve device performance. As optimizing NCs quality has been amply covered in previous sections, we will only marginally touch upon this topic and will instead focus our attention on engineering device architecture. The lighting and display related devices here mainly involve electroluminescent LEDs, phosphor-converted WLEDs, micro-LEDs, and active-matrix LEDs.

### 5.1 Electroluminescent LED

The device performance of perovskite NC-based electroluminescent LEDs has been substantially increased in just a few years, a much shorter time span compared to the conventional organic or II-VI quantum dots LEDs, which however had already paved the way over the previous decades through in-depth understanding of the basic device physics and design of device structure.[186] Here, we briefly introduce the device structure and working mechanism of perovskite NCs LEDs and discuss the different strategies to achieve high-performance EL devices. To meet the Rec.2020 color standards, LEDs with ideal emission wavelengths (blue: 460–470 nm; green: 525–535 nm; red: 620–650 nm) are more desired. Using mixed halide compositions, strongly confined perovskite NCs, or doping NCs with various ions can achieve ideal emissions, as anticipated in the previous sections. However, most of them suffer from instability and result in a poor operational lifetime of several minutes at an initial luminance of 100 cd $m^{-2}$. To improve the chemical and electrical stability, ligands engineering, core-shell structure, surface passivation, synthesis optimization have been used (all these aspects have been covered in the previous sections of this review). A summary of works on LEDs with relatively high operational stability (hours of operational lifetime) based on perovskite NCs emitters is presented in **Table 3**.

**Table 3.** Summary of reports on LEDs with relatively high operational stability based on ideal red, green, and blue emitting halide perovskite NCs.

| Nanocrystals (Strategies to improve the material/device performance) | EL peak [nm] | Max. luminance [cd $m^{-2}$] | EQE [%] | Operational lifetime | Year | Ref. |
|---|---|---|---|---|---|---|
| $CsPb(Cl/Br)_3$@MOF | 475 | 1,260 | 5.67 | $T_{50}$ = 2.23 h @4.2 V | 2022 | [125] |
| $CsPb(Cl/Br)_3$ (Rubidium compensation in the A-site of the perovskite) | 460 | 255 | 12 | $T_{50}$ = ~75 min @1 mA $cm^{-2}$ | 2024 | [20a] |
| | 465 | 430 | 16.7 | / | | |
| | 470 | 635 | 21.3 | $T_{50}$ = ~80 min @1 mA $cm^{-2}$ | | |
| | 475 | 1175 | 24.3 | / | | |
| | 480 | 1601 | 26.4 | $T_{50}$ = ~110min @1 mA $cm^{-2}$ | | |
| $CsPbBr_3$ (4 nm size) (HBr, didodecylamine and phenethylamine introduced) | 470 | 3,850 | 4.7 | $T_{50}$ = 12 h @102 cd $m^{-2}$ | 2021 | [75] |
| $CsPbBr_3$ (4.3 nm size) (With difunctional ZnO obtained by a ligand strategy of phenethylammonium bromide as the ETL) | 470 | 11,100 | 8.7 | $T_{50}$ = 35 h @100 cd $m^{-2}$ | 2022 | [187] |
| $CsPbBr_3$ (4.2 nm size) (Ligand $\alpha$-methyl-4-bromide-benzylammonium added) | 480 | 2910 | 17.9 | $T_{50}$ = 126 min @100 cd $m^{-2}$ | 2022 | [19a] |
| $CsPbBr_3$ (3.5 nm size) (Ligand $\alpha$-methyl-4-bromide-benzylammonium added) | 465 | 576 | 10.3 | $T_{50}$ = 18 min @100 cd $m^{-2}$ | 2022 | [19a] |
| $CsPbBr_3$ (4 nm size) (A ultrathin interlayer of $ZnCl_2$ was constructed between two NCs layers to form a sandwich panel) | 469 | 10,410 | 5 | $T_{50}$ = 59 h @100 cd $m^{-2}$ | 2022 | [24] |
| $CsPbBr_3$ (3.4 nm size) (an in-situ PL system to well-control the nucleation and growth of NCs) | 470 | 11,610 | 10.1 | $T_{50}$ = 21 h @102 cd $m^{-2}$ | 2023 | [57e] |
| $CsPbBr_3$ (3.5 nm size) ($ZnBr_2$ introduced) | 469 | 12,060 | 10.3 | $T_{50}$ = 25 h @115 cd $m^{-2}$ | 2023 | [188] |

| | | | | | | |
|---|---|---|---|---|---|---|
| $FA_{1-x}GA_xPbBr_3$ | 530 | / | 23.4 | $T_{50}$ = 132 min @100 cd $m^{-2}$ | 2021 | [86c] |
| $Cd^{2+}$ doped $FAPbBr_3$ | 534 | 9783 | 29.4 | $T_{50}$ = 111.8 min @100 cd $m^{-2}$ | 2024 | [189] |
| ($Cs_xFA_{1.3-x}PbBr_3$, $x = 0.7$) (Incorporation of MPC and $GA^+$ into the perovskite emissive layer to suppress nonradiative trap formation, ion migration and Auger recombination) | 528 | ~$4.6 \times 10^5$ | 21.2 | $T_{50}$ = 125 h @1000 cd $m^{-2}$<br>$T_{50}$ (extrapolated) = 4900 h @100 cd $m^{-2}$ | 2024 | [190] |
| $GA_{0.28}Cs_{0.72}Pb_{0.40}Ge_{0.55}Cd_{0.05}I_3$ | 633 | 4,932 | 20.1 | $T_{50}$ = 2.5 h @290 cd $m^{-2}$ | 2023 | [87] |
| $CsPb(Br/I)_3$ (In situ inorganic ligand strategy) | 642 | 290 | 24.4 | $T_{50}$ = 20 h @290 cd $m^{-2}$ | 2022 | [191] |
| $CsPb(Br/I)_3$ (Benzenesulfonate passivated) | 638 | 1,510 | 23.5 | $T_{50}$ = 97 min @100 cd $m^{-2}$ | 2023 | [71] |
| $CsPb(Br/I)_3@PbSO_4$ | 630 | 5,000 | 12.6 | $T_{50}$ = 5.2 h @100 cd $m^{-2}$ | 2023 | [108] |
| $CsPb(Br/I)_3$ (4-fluorobenzenesulfonamide anchored) | 630 | 21,590 | 21.8 | $T_{50}$ = 11.7 h @100 cd $m^{-2}$ | 2024 | [192] |
| $CsPb(Br/I)_3$ (Aniline hydroiodide and oleylammonium iodide utilized to achieve joint ligand exchange) | 636 | 1621 | 21.2 | $T_{50}$ = 4 h @130 cd $m^{-2}$ | 2024 | [193] |
| $CsPb(Br/I)_3$ (Diphenylphosphoryl azide-mediated regulation of the NCs surface) | 640 | 23,480 | 24.8 | $T_{50}$ = 13.1 h @100 cd $m^{-2}$ | 2024 | [194] |
| $CsPbI_3$ (5.4 nm size) (Synthesis mediated by diisooctylphosphinic acid and nanosurface reconstruction driven by etching with hydriodic acid) | 644 | 4,140 | 28.5 | $T_{50}$ = 30 h @100 cd $m^{-2}$ | 2024 | [7d] |

| | | | | | | |
|---|---|---|---|---|---|---|
| $CsPbI_3$ (4.4 nm size) (Replacing ODE with strong electrostatic potential benzene series solvents and using short-chain ligand post-treatment strategy) | 630 | 1,300 | 25.2 | $T_{50}$ = 2 h @100 cd $m^{-2}$ | 2024 | [8a] |
| $CsPbI_3$ ($PbCl_x$-modified) | 638 | 2,511 | 26.1 | $T_{50}$ = 7.5 h @100 cd $m^{-2}$ | 2024 | [8b] |

### 5.1.1 Device structure and working mechanism

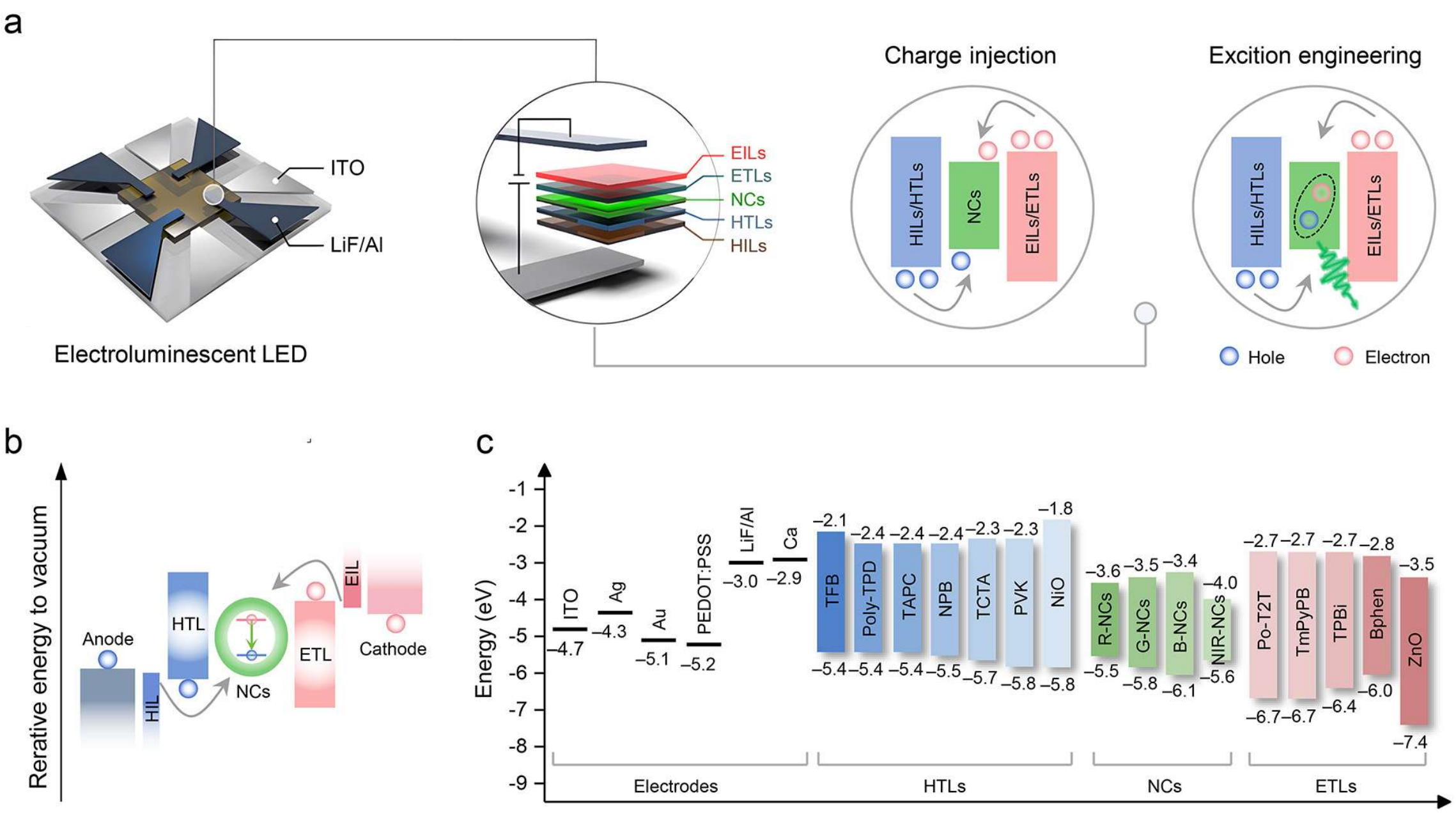


**Figure 11. Electroluminescent LEDs based on halide perovskite NCs.** (a) Schematic diagram of the general perovskite NC LEDs structure and working mechanism. Reproduced and adapted from ref.[195]. Copyright 2024, Tsinghua University Press. (b) Flat-band energy level diagram of perovskite NCs devices (c) Energy levels of some commonly used electrodes, HTLs, perovskite NCs, and ETLs.

A typical device architecture of a perovskite NCs LED consists of a transparent conducting cathode, typically indium tin oxide (ITO) glass, a *p*-type hole injecting/transport layer (HTL), a perovskite NCs emitting layer, an *n*-type electron injecting/transport layer (ETL), and an anode. According to the stacking mode of each functional layer and the direction of emitted

light, the structure can be divided into four main types: bottom emitting structure, top emitting structure, inverted bottom-emitting structure, and tandem structure.[196] **Figure 11**a, b shows the fundamental device structure and working mechanism of perovskite NCs LEDs with a standard *p-i-n* multilayer heterojunction structure. When an external voltage is applied to the device, charge carriers (electrons and holes) are injected in the device and ultimately recombine radiatively in the NCs emitting layer, resulting in EL. This process involves various dynamical processes, including carrier injection and transport, exciton recombination, and energy transfer.[186] Therefore, the selection of suitable materials for each functional layer, taking into account the electrical conductivity and band energy alignment with perovskite NCs, is the key to achieving efficient EL. Figure 11c presents some of the most common materials used to fabricate the multilayer structure of perovskite NCs LEDs. The band energy structure of perovskite NCs, that is, the positions of CBM and VBM, varies with their emission colors and compositions. Generally, the VBM of blue emitting NCs is deeper than that of green and red counterparts,[197] which means that different strategies are needed when designing perovskite NCs LEDs with different types of NCs. For lead-free perovskite NCs, due to the large band energy gap between the electrode and emitters, constructing stepped paths for charge carriers is crucial for realizing efficient charge injection into NCs. This requires the insertion of charge injection and/or transport layers with appropriate energy levels on the path from the electrodes to the NCs.[198] The performances of lead-free electroluminescent LEDs including EQE, operation stability and maximum radiance are summarized in **Table 4**.

**Table 4.** Summary of device performance for electroluminescent LEDs based on lead-free metal halide NCs.

| Material | EQE (%) | EL peak (nm) | Operation stability | Max. luminance (cd m$^{-2}$) | Device structure | Year | Refs. |
|---|---|---|---|---|---|---|---|
| $Cs_3Sb_2Br_9$ | ~0.206 | 408 | 6 h @ 7 V ($T_{90}$) | 29.6 | ITO/ZnO/PEI/NCs/TCTA/$MoO_3$/Al | 2019 | [21a] |
| $Cs_3Cu_2I_5$ | ~1.12 | 445 | ~108 h @ 6.7 V ($T_{50}$) | 263.2 | Al/LiF/TPBi/NCs/NiO/ITO | 2020 | [35] |
| $CsAgCl_2$ | 0.02 | 628 | 1 h @ 20 V ($T_{50}$) | 9.6 | ITO/PEDOT:PSS/NCs/TPBi/LiF/Al | 2022 | [154] |
| $Cs_2AgIn_{0.9}Bi_{0.1}Cl_6$ | 0.08 | 426, ~620 | 48.5 min @ 15 mA cm$^{-2}$. ($T_{50}$) | 158 | ITO/PVK/NCs/TPBi/LiF/Al | 2022 | [153] |
| $CsBr:Eu^{2+}$ | 3.15 | 443 | ~1 h @ 100 cd m$^{-2}$ ($T_{50}$) | 311.8 | ITO/PEDOT:PSS/TFB/PVK@NCs/TPBi/LiF/Al | 2024 | [38a] |

| | | | | | | | |
|---|---|---|---|---|---|---|---|
| $Cs_3CeCl_6\cdot 3H_2O$ | 0.13 | 380 | / | ~2.3 | ITO/PEDOT:PSS/ PVK+TCTA/NCs/ TPBi/LiF/Al | 2024 | [137] |
| $Cs_3LaCl_6$ | 0.53 | 442 | / | 127 | | 2024 | [38b] |
| $Cs_3TbCl_6$ | 1.58 | 515 | / | 370 | ITO/PEDOT/PVK/ NCs/TPBi/LiF/Al | 2024 | [38b] |
| $Cs_3EuCl_6$ | 5.17 | 616 | ~2 h @ 2300 cd m$^{-2}$ ($T_{50}$) | 2373 | | 2024 | [38b] |

To achieve efficient EL, it is crucial to control carrier dynamics and promote radiative recombination. From a device perspective, several conditions must be met to obtain an efficient NCs LED. First, efficient charge injection and charge balance are essential, meaning that carriers should be easily injected into the device with low barriers, and the net ratio of electrons to holes injected into the diode should be as close to 1 as possible.[199] Second, radiative recombination should be maximized relative to nonradiative recombination events in the NCs emitters, an aspect that is related to PLQY of NCs. Third, carrier leakage, where carriers flow away from the device without recombination, should be minimized to suppress leakage current. Finally, the photons generated within the NCs emissive layer should be efficiently extracted (outcoupling). Taking all the above effects into consideration, the external quantum efficiency of NCs LED can be described by the following Equation:[105, 200]

$$EQE=\gamma\cdot\eta_r\cdot\Phi_{\mathrm{PL}}\cdot\eta_{\mathrm{out}}$$

where $\gamma$ is the balance ratio of electrons and holes injected from the electrode, $\eta_r$ is the ratio of the radiative exciton to the total generated exciton in the device, $\Phi_{\mathrm{PL}}$ is the PLQY of NCs emitters, and $\eta_{\mathrm{out}}$ is the light outcoupling ratio from the glass. In addition, key parameters used to evaluate LEDs performance also include current efficiency (*CE*), power efficiency (*PE*), maximum luminance ($L_{\mathrm{max}}$), turn-on voltage ($V_{\mathrm{on}}$), and operating lifetime ($T_{50}$).

From the above introduction and discussion, there are three key factors that significantly affect the EL performance of perovskite NCs, namely, NC emitter synthesis, device architecture construction, and light outcoupling management.[86c] Previous studies have shown that the PLQY and conductivity balance of NCs layer not only have a great impact on the EQE of devices, but also affect the operational stability.[17] Regarding the device architecture, the interface quality and the charge injection balance of HTLs and ETLs are the primary considerations.[201] In terms of light outcoupling management, various nanostructures have been developed to improve the light outcoupling efficiency.[200] In the next section, based on the three factors discussed above, we will discuss some representative studies on performance optimization of perovskite NCs LEDs in the past several years. Our goal is to highlight current challenges and propose viable methods for future development.

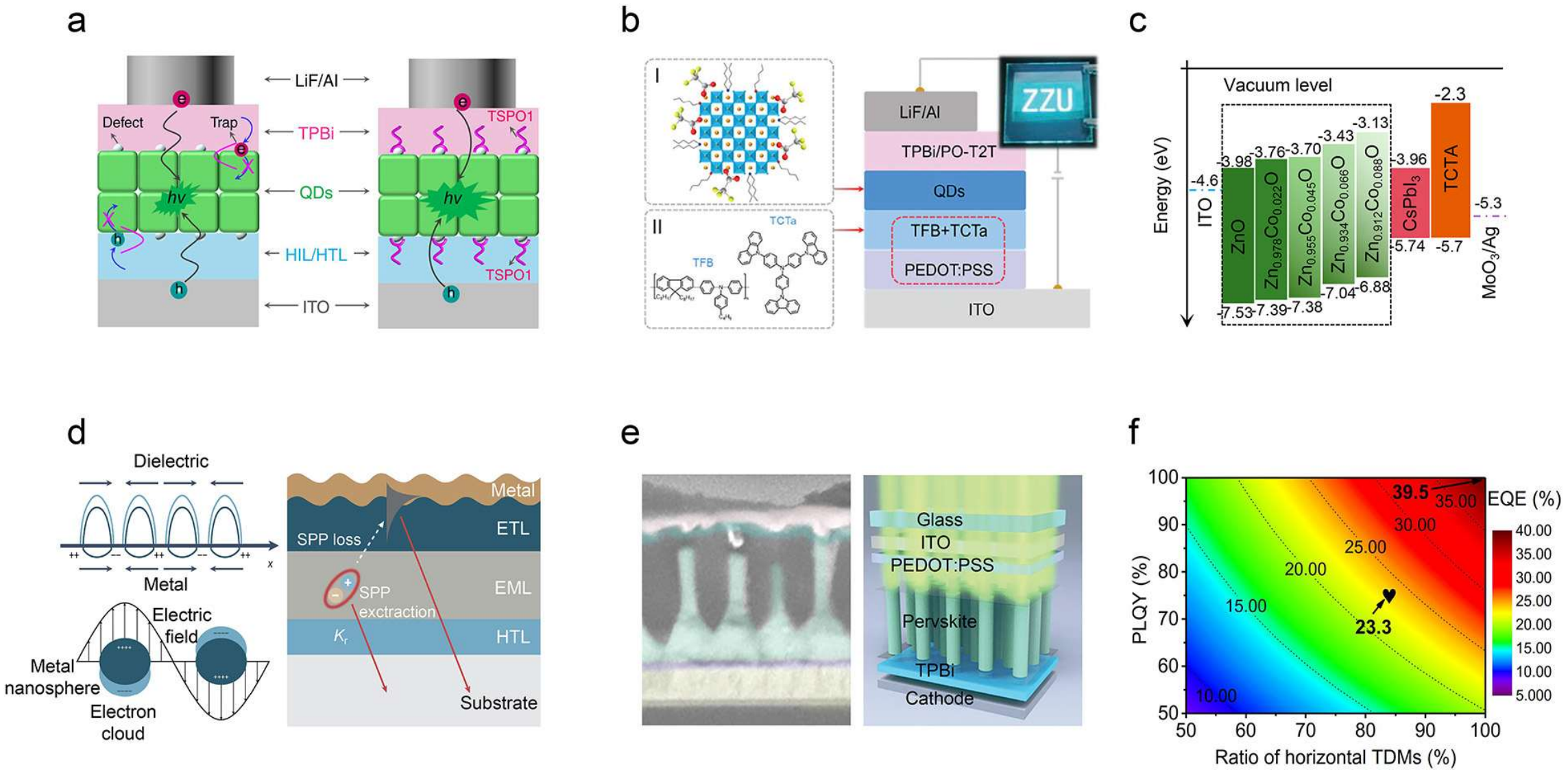

**Figure 12. Performance optimization of perovskite NCs LEDs.** (a) Device structure of NC LEDs based on NCs films without and with bilateral interfacial passivation. Reproduced and adapted from ref.[202]. Copyright 2020, Springer Nature. (b) Schematic concept for constructing perovskite NC LEDs with a mixed hole-transport layer.[203] (c) Energy level band diagram of ZnO with varying $Co^{2+}$ doping. Reproduced and adapted from ref.[204]. Copyright 2021, American Chemical Society. (d) Improved emission with surface plasmon via enhancement of the light outcoupling efficiency. Reproduced and adapted from ref.[200]. Copyright 2023, Springer Nature. (e) Perovskite nanophotonic wire LEDs structure and microstructure. Reproduced and adapted from ref.[205]. Copyright 2020, American Chemical Society. (f) Contour plot of the simulation results of EQE as a function of PLQY and ration of horizontal TMDs of the perovskite NCs layer. Reproduced and adapted from ref.[206]. Copyright 2021, Wiley.

### 5.1.2 Ligand engineering and conductivity balance in NCs layers

As is well known, surfactants are essential for synthesizing high PLQY NCs, as they well control the shape and size of the NCs and passivate the surface defects induced by the dangling bonds to enhance radiative recombination.[207] However, these surfactants are typically bulky insulating organics, such as OLA and OA, which impede carrier transport in the NCs layers, resulting in non-functional or low-efficiency NCs LEDs. To overcome this obstacle, strategies must be adopted that preserve the high PLQY of the perovskite NCs layers while also achieving high electrical conductivity. A straightforward method to partially meet this goal is the accurate washing of colloidal NCs, which can remove residual organic surfactants and byproducts, thereby minimizing the concentration of insulating ligands within the perovskite NCs and improving the performance of LEDs.[208] Generally, the PLQY of perovskite NCs can be directly affected by the dielectric constant of the antisolvent used in purification, with a trend of decreasing PLQY observed with increasing dielectric constant.[209] Polar solvents with high dielectric constants have been shown to be detrimental for the stability of perovskite NCs, resulting in complete emission quenching of NCs. On the other hand, solvents with lower dielectric constants, such as butyl acetate, ethyl acetate, and butanol, are more benign for purifying NCs. In 2015, Zeng's group used a mixture of ethyl acetate and hexanes to lower ligand density of OA/OLA-capped $CsPbBr_3$ NCs while maintaining stability during the purification process.[208a] The final EQE of the purified $CsPbBr_3$ NCs-based device reached 6.27%, a 50-fold enhancement over the control device. Subsequently, Juni Kido et al. demonstrated the role of butyl acetate (AcOBu) in NCs purification, with the AcOBu-washed NCs-assembled LEDs achieving an EQE of 8.73%.[210] By carefully controlling the ratio between NCs solution and ethyl acetate and the washing cycles, Li et al. boosted the EQE of DDAB-treated NC-based LEDs to 13.4%.[208b]

From the perspective of fabricating electroluminescent LEDs, achieving high-quality emitting layer with high PLQY and favorable conductivity properties is usually the first step towards high-efficiency LEDs.[22a, 211] To this aim, short chain length ligands, zwitterionic ligands, conductive aromatic conjugated ligands, as well as inorganic ligands, which are promising in both enhanced surface passivation and improved electrical coupling, have been probed as the capping ligands at the NC surface to balance the PLQY and conductivity of the NC layer.[207, 212] In 2017, T-W Lee et al. first attempted to use short-chain amines in the $FAPbBr_3$ NCs synthesis. $FAPbBr_3$ NCs with n-butylamine exhibited high PLQY, and the corresponding LEDs showed higher current efficiency compared to those with OLA ligands, due to the good surface passivation and improved conductivity by the n-butylamine layer.[213] Octylphosphonic acid (OPA) was used as a substitute for OA to fabricate $CsPbBr_3$ NCs for LEDs applications by Sun et al.[214] This relatively short alkyl ligand not only improved the NC film conductivity but also provided a strong interaction between the phosphonate group and surface lead atoms, playing the role of defect passivation to improve radiative recombination. The OPA-capped NCs maintained high PLQY and colloidal stability after multiple rounds of purification, facilitating their incorporation into LEDs, which achieved an EQE of 6.5% and a $T_{50}$ of 30 min.[214] Zwitterionic ligands can anchor on the NC surface more robustly, allowing for more thorough washing of the NCs solution, resulting in cleaner NCs films with high PLQY. Using zwitterionic sulfobetaine ligands, Kovalenko's group achieved a 1% EQE for a pure blue NCs device.[215]

Short conjugated ligands containing aromatic groups such as phenyl, thienyl, or carbazolyl can improve charge transport properties through π-π stacking interactions between the functional aromatic groups, thereby better injecting charge carriers into the NCs layers. Dai et al. reported that using the conjugated ligand 3-phenyl-2-propen-1-amine (PPA) in $MAPbBr_3$ NCs synthesis increased the carrier mobility in the final NCs layer by 22 times compared to that of the control, thus substantially improving the EL performance.[216] Gao et al. demonstrated that strong cation-π interactions between the $PbI_6$-octahedra of perovskite units and the electron-rich indole ring of tryptophan (TRP) molecules not only helped to remove the undercoordinated $Pb^{2+}$ at the 'imperfect' surface sites, thus inducing a surface reconstruction, but also markedly increased the binding affinity of the ligand molecules.[217] This leads to high PLQY and greatly enhanced spectral stability of the $CsPb(Br/I)_3$ NCs. Moreover, the incorporation of small-sized aromatic TRP ligands ensures superior charge transport properties of the NCs layers. The TRP-treated $CsPb(Br/I)_3$ NCs LEDs demonstrated a champion EQE of 22.8% and outstanding spectral stability, ranking among the best-performing Rec. 2020 pure-red perovskite LEDs reported so far.

Shorter, strongly bound ligands can be introduced during the NCs synthesis or after that, in a post-synthesis treatment, to boost the PLQY and conductivity of the NCs.[207, 218] DDAB is a ligand that can replace the original, longer OLA/OA ligands on $CsPbX_3$ NCs to improve charge carrier transport and passivate surface defects.[208b] The first LED incorporated with DDAB-passivated $CsPbBr_3$ NCs (ITO/PEDOT/PVK/$CsPbBr_3$ NCs/TPBi/LiF/Al) showed a lower turn-on voltage and an enhanced EQE of 3.0% compared to the OA/OLA-capped NCs counterpart.[219] In Yao's work, the synthesized 5 nm $CsPbI_3$ NCs were reacted in sequential steps with 1-hydroxy-3-phenylpropan-2-aminium iodide (HPAI) and tributylsulfonium iodide (TBSI) ligands in a post-synthesis treatment.[60a] The $CsPbI_3$ NCs films exhibited improved optoelectronic properties, high PLQY, and inhibited ion migration, which enabled the fabrication of a pure-red perovskite NCs LED with a peak EQE of 6.4% and a stable EL emission centered at 630 nm.[60a] Chen et al. developed a facile ligand-exchange method using amino acids to replace the long-chain ligands on the surface of $CsPbI_3$ NCs, improving the efficiency and stability of LEDs made from these NCs.[60b] An EQE of 18% was achieved, accompanied by a long $T_{50}$ of 87 minutes from this device.[60b]

Rather than casting films from pre-prepared NCs suspensions, in-situ or on-substrate growth of NCs through a one-step spin-coating fabrication method is an alternative way to obtain high PLQY NCs films with superior conductivity, since traditional long-chain insulating ligands are absent in this procedure.[220] For example, Barry P. Rand et al. studied the influence of additives FPMAI or PEABr in $MAPbX_3$ precursors in the in-situ growth of NCs films.[221] This strategy resulted in highly efficient $MAPbI_3$ red/NIR LEDs and $MAPbBr_3$ green LEDs, achieving EQEs of 7.9% and 7.0%, respectively.[221] Phenylalanine Bromide (PPABr)-passivated $CsPbBr_3$ NCs were directly synthesized by in-situ spin-coating the precursor solution, prepared by dissolving PPABr, CsBr, and $PbBr_2$ in DMSO, onto a preheated substrate, as reported by Liao et al.[222] The resulting green perovskite NCs LED delivered a high EQE of 15% with ideal operational stability of 1.2 h at initial luminance at 1000 cd $m^{-2}$.[222]

### 5.1.3. Interfacial modification and carrier balance in LEDs

Although the device performance can be greatly improved by carefully balancing PLQY and conductivity of the NCs layers, effective carrier transport between the perovskite layer and the HTL/ETL layer interface and carrier balance in the device are also critical to device performance.[223] However, numerous interface defects are formed during device fabrication, which severely affect carrier injection, transport, recombination, and ultimately deteriorate the LED performance. Studies have shown that proper interfacial passivation or various other treatments can eliminate traps and enhance LED performance.[202, 224] For instance, Li et al. treated the NCs film top surface with trimethylaluminum (TMA) to reduce the uncoordinated ion defects formation during HTL layer deposition, resulting in near-complete NCs film coverage and a remarkable EQE of 5.7%.[225] Zeng et al. proposed a bilateral passivation strategy by passivating both the top and bottom interfaces of the NCs film with organic molecules (oxide-4-(triphenylsilyl)phenyl, TSPO1), which drastically enhanced the efficiency and stability of LEDs (**Figure 12**a).[202] The key role of such bilateral passivation is to reduce the interface defects between the NCs layer and the transport layers effectively and thus enhance the charge carrier injection. Ultimately, the device with bilateral passivation achieved a maximum EQE of 18.7%, and the operational lifetime was improved by 20-fold.[202]

Regarding carrier balance, this is closely related to the energy level alignment between different layers in perovskite -LEDs and the discrepancy in hole/electron mobility in the HTL and ETL layers.[226] Undoubtedly, choosing the proper HTL/ETL layer according to the carrier mobility and energy level of the NCs emitting layer is essential. However, due to the limited options for transport layers, buffer layers such as perfluorinated ionomer (PFI), polyethylenimine (PEI), and LiF are used to adjust the energy levels at contacts.[227] Rogach et al. incorporated a PFI interlayer between the HTL and the perovskite NCs layer in their device, which resulted in 0.34 eV increase of the VBM of HTL.[227a] Meanwhile, a PFI layer can suppress charging of perovskite NCs emitters, thus preserving their outstanding emission properties, which results in a three-fold increase in maximum luminance, reaching 1377 cd $m^{-2}$. Similar to common problems with traditional Cd-based quantum dot LEDs, the large band gap of blue emitting perovskite NCs impedes carrier injection, and hole injection of blue perovskite NCs LEDs is particularly challenging.[86b] Specifically, the deep VBM of blue perovskite NCs make the HTL used in green or red LEDs unsuitable for blue devices, as high barrier between blue emitting layer and HTL results in a inhibition of hole injection into the perovskite layer. This unbalanced charge injection can lead to excessive electron accumulation at the interfaces, thereby degrading the device performance by Auger quenching or redox chemistry.[228] In 2024, Song et al. designed a mixed HTL consisting of poly[(9,9-dioctylfluorenyl-2,7-diyl)-co-(4,4'-(N-(4-sec-butylphenyl) diphenylamine)] (TFB) and tris(4-carbazoyl-9-ylphenyl) amine (TCTA) to modulate the carrier injection transportation ability and improve carrier balance (Figure 12b).[203] Such mixed HTL also can decrease carrier leakage and enhance radiative recombination in the device, enabling a record-high EQE of 23.5% for perovskite NCs LEDs. Zhang and co-workers replaced the HTL of PEDOT:PSS/TFB with the PEDOT:PSS/TFB: black phosphorus quantum dots having deeper work function and high mobility to fabricate perovskite NCs LEDs, greatly increasing current density.[229] This confirms that if more holes are injected into the NCs layers, this leads to more balanced carriers inside the LEDs. As a result, the resulting devices exhibited a reduced turn-on voltage of 2.6 V, an enhanced luminance of 128,842 cd $m^{-2}$, and an increased EQE of 25.32%.

Introducing dopants in the charge transport layer is also a useful strategy to regulate the carrier transport characteristics for efficient perovskite NCs LEDs.[230] For example, a NaCl-doped PEDOT:PSS and Li-doped $TiO_2$ layers were demonstrated to effectively balance the charge injection in the devices to enhance the LED performance.[231] ZnO is often used as an ETL due to its excellent electron mobility. However, most organic HTL materials exhibit lower hole mobility compared to the electron mobility observed in ZnO; thus, excess electrons tend to accumulate at the ZnO/NCs interface, resulting in carrier imbalance in devices.[232] To tackle this problem, Bai et al. proposed a strategy of $Co^{2+}$ doping into ZnO, which reduces the electron mobility of ZnO by passivating oxygen vacancies and electron trapping by the introduction of $Co^{2+}$-induced deep impurity levels (Figure 12c).[204] Meanwhile, the $Co^{2+}$-doped ZnO prevents exciton quenching at the ZnO/NCs interface. All those factors were beneficial to reach a balanced charge injection during the device operation, which enabled a high EQE of 13.0% and low efficiency roll-off. Similarly, Rand et al. solved the problems of unbalanced electron and hole injection and the instability of ZnO/perovskite NCs interface by applying an ultra-thin layer of $Al_2O_3$ between ZnO and the perovskite NCs.[233] Also, by precisely controlling the thickness of the $Al_2O_3$ layer, the electron injection mediated by tunneling through the $Al_2O_3$ could be carefully adjusted. This enabled the fabrication of efficient perovskite NCs LEDs with a maximum luminance of 21815 cd $m^{-2}$, with similar enhancements in efficiency and operational lifetime.

### 5.1.4. Light outcoupling management in LEDs

Improving light outcoupling is another important aspect of enhancing the efficiency of perovskite NCs LEDs.[234] The internal quantum efficiency of some high-performance perovskite NCs LEDs has already approach 100% values.[235] Nevertheless, the non-ideal outcoupling of light results in approximately 80% of the generated photons remaining trapped within the device stack and ultimately losing energy through a series of lossy channels. The typical reasons of finite light outcoupling efficiency ($\eta_{out}$) in perovskite LEDs are: i) waveguide mode, because perovskites (~2.5) have a large refractive index relative to the commonly used charge transport materials (~1.7−1.8); ii) the total internal reflection of photons produced within the device over a wide range of angles due to the diverse refractive index of device substrate and air.[200] Therefore, most of generated photons are trapped in the waveguide and substrate mode. Strategies to enhance the light outcoupling from perovskite NCs LEDs can be divided into two groups: modification of i) intrinsic and ii) extrinsic optical properties. Previous works have shown that effectively managing the refractive index of the constituent layers in LEDs is one of the key determinants to improve the $\eta_{out}$.[200] Shih et al. implemented a strategy of controlling ligand hydrophobicity to regulate the refractive index of $FA_{0.5}MA_{0.5}PbBr_3$ NCs in the range of 1.64–1.75.[236] The $FA_{0.5}MA_{0.5}PbBr_3$ NCs-based LEDs exhibited a maximum $\eta_{out}$ of 26%, realizing a peak luminance of 3322 cd m$^{-2}$ and EQE of 9.01%. Snaith et al. synthesized $MAPb(I_xBr_{1-x})_3$ NCs with relatively low refractive index using a modified ligand-assisted re-precipitation approach.[237] A refractive index of 1.82 (at 620 nm), $\eta_{out}$ of 32.2%, and EQE of 20.3% were achieved from the $MAPb(I_xBr_{1-x})_3$ NCs LEDs. In addition, Kang et al. specified that the effectiveness of the ligand exchange is an effective strategy to reduce the large gap for refractive index between the transporting layer and the NCs layers.[238] The original $CsPbBr_3$ NCs films had a refractive index of 1.91 at 520 nm, while the value of the PEA-treated films decreased to 1.82. This reduction of the refractive index in the PEA-treated films improved the light coupling output efficiency and was the likely cause of the EQE improvement from 1.0% to 6.85%.

To overcome the refractive index mismatch problem, several nanopatterns/nanophotonics strategies have been employed to improve the outcoupling efficiency.[200] Tang et al. explored a facile route to effectively reduce the optical losses associated with the waveguide mode by embedding bioinspired moth-eye nanostructures at the front electrode/perovskite interface.[239] By utilizing moth-eye nanostructures technology, the trapping of photons within the waveguide mode was significantly eliminated. The moth-eye nanostructure-based perovskite NCs LEDs exhibited 1.5-times higher performance than the flat device, delivering a high EQE of 20%, due to the enhancement associated with the efficient outcoupling of the waveguide light.

A surface plasma (SP) mode refers to the collective oscillation of electrons in the electromagnetic field at the metal/dielectric interface, which can be divided into propagating SP polarization (SPPs) and local SPs (LSPs) (Figure 12d).[240] SPPs are non-radiative modes typically forming at the flat metal/dielectric interface and results in a loss of efficiency of at least 20% in the LEDs due to exciton−SP coupling.[241] LSPs typically act as scattering and absorption centers for the incident light, leading to both radiative and nonradiative processes.[200] To mitigate energy losses in SPPs propagating along metal surfaces, the integration of metal nanostructure or tailored plasmonic structure within the opaque metal electrode represents an effective strategy. The incorporation of these structures is designed to elicit LSPs that may improve light coupling efficiency. Another strategy is to employ the near field enhancement effects of SPs, which can also improve the efficiency of perovskite LED. As an example, by adding Au-Ag nanoparticles with optimal size and concentration to ETLs, the EQE was increased by 25% due to the near-field enhancement effects of LSPs.[242] A promising approach to further improve the light outcoupling efficiency is to combine SPs with other nanophotonic structures. Shi et al. constructed a coaxial core/shell heterostructure perovskite NCs LED based on ZnO nanowires as carrier injector. By incorporating plasmonic gold (Au) nanoparticles into the device architecture, they achieved a 1.55-fold enhancement in electroluminescence (EL).[243] These results demonstrate that the fabrication of hybrid plasmonic–photonic structures is beneficial for light extraction and EL improvement in perovskite NCs LEDs, thanks to the simultaneous light scattering and SP enhancement effects.

The directional growth of perovskite nanostructures can enhance light outcoupling in LEDs by activating radiative and directional emission by the nanostructures patterned templates with the high-index grading of perovskite.[200] Fan et al. explored a $CsPbBr_3$ nanophotonic wire array-based LED in a porous alumina membrane with a capillary-effect-assisted template approach (Figure 12e).[205] The template-assisted nanophotonic wire-based devices demonstrated 45% enhancement of EQE from 11% to 16% compared with the planar control LEDs, due to the substantial improvement on light extraction efficiency verified by optical modeling. Additionally, the optimized LEDs could reach a luminance of 12,016 cd $m^{-2}$ at 5.5 V, while the control devices only yielded 10,953 cd $m^{-2}$. It was also found that the nanostructured devices had an operational lifetime 3.89 times longer than the planar control device, due to the template protection.

The orientation of transition dipole moments (TDMs) has been demonstrated to significantly influence the light outcoupling of organic LEDs. Similar effects have been probed for perovskite NC LEDs. Since the radiation direction is perpendicular to the dipole moment, the horizontal direction of TDMs with respect to the substrate plane enhances the light outcoupling, while the vertical direction of TDMs tends to produce the captured light mode.[105] Thus, a horizontally oriented crystal with TMDs parallel to the substrate surface induces more light to be extracted from the substrate by reducing the guiding and substrate modes, which can compensate for the refraction mismatch problem. Some reports have investigated the effects of TDM orientation in perovskite nanostructures, including nanosheets and nanorods. Jin et al. demonstrated that 84% horizontally oriented TDMs are obtained from the $CsPbBr_3$ NPLs, which results in a light-outcoupling efficiency of ~31%, higher than that from the isotropic emitter (~23%) (Figure 12f).[206] The large portion of horizontal TDMs improved the EQE of devices to 23.6%. In addition, by carefully optimizing horizontal-directed TMDs, it is expected that the efficiency of LEDs can be increased to the highest level of ~40%. Shih et al. pointed out that regulating the aspect ratio of the anisotropic superlattices perovskite NCs is crucial for enhancing light outcoupling in LEDs.[244] They claimed that, for a given dipole, the emission radiative rate is the product of the local electric field vector and the transition dipole moment. The synthesized anisotropic perovskite NCs exhibit an emission radiative rate of approximately 90%. This enables a high outcoupling efficiency in anisotropic NCs LEDs, in which the maximum luminance of 5009 cd $m^{-2}$, EQE of 24.96%, and operation lifetime of 150 h were achieved.

## 5.2. Phosphor-converted WLEDs

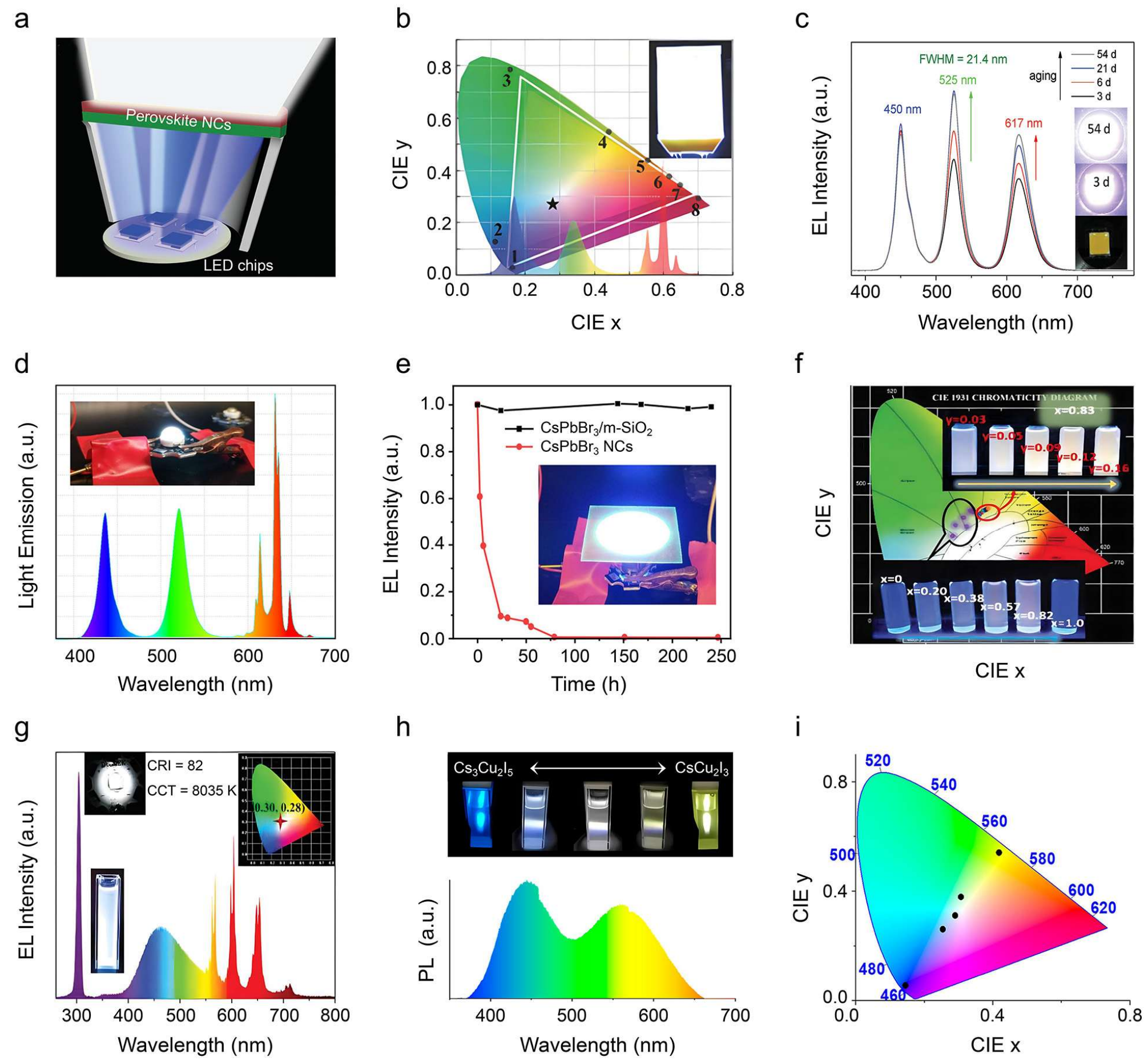

**Figure 13.** Phosphor-converted WLEDs based on perovskite NCs. (a) Scheme of phosphor-converted WLEDs. Reproduced and adapted from ref.[245]. Copyright 2016, Wiley. (b) CIE coordinates and EL spectra of WLEDs based on $MAPbBr_3$/PVDF NCs composites. Reproduced and adapted from ref.[246]. Copyright 2016, Wiley. (c) EL spectra and photographs (insets) at different aging times. (d) Emission spectrum of the WLEDs based on $CsPbBr_3$/m-$SiO_2$ polymer composites. Reproduced and adapted from ref.[247]. Copyright 2022, Wiley. (e) Time-dependent PL intensity of $CsPbBr_3$/m-$SiO_2$ and control $CsPbBr_3$ NCs under high flux remote application test. Reproduced and adapted from ref.[99d]. Copyright 2021, American Chemical Society. (f) CIE coordinates of white-emitting $Cs_2Ag_{1-x}Na_xIn_{1-y}Bi_yCl_6$ NCs with different extents of ion

doping. Reproduced and adapted from ref.[248]. Copyright 2019, Wiley. (g) EL spectra of WLEDs based on $Sm^{3+}$-doped $Cs_2NaInCl_6:Sb^{3+}$ NCs. Reproduced and adapted from ref.[150]. Copyright 2023, Wiley. (h) Photographs and PL spectra of Cs−Cu−I NCs under 310 nm excitation. (i) CIE coordinates of white-emitting Cs−Cu−I NCs showing the cold/warm white light tuning. Reproduced and adapted from ref.[155b]. Copyright 2019, American Chemical Society.

Phosphor-converted WLEDs are nowadays in a leading position in the field of solid-state lighting owing to their advantages of small and durable form factor, high luminous efficiency, long operational lifetime and simple manufacturing.[249] The schematic representation of a typical perovskite NCs WLEDs is depicted in **Figure 13**a. The device primarily consists of a gallium nitride (GaN) LED chip that emits ultraviolet (UV) or blue light,[99e, 131c] along with R/G/B-emitting perovskites or perovskites together with a mix of commercial phosphor materials, used for down-conversion. When a UV-emitting LED chip is employed, achieving white emission necessitates the use of phosphors that emit red, green, and blue light, combined to form the light conversion layer. Conversely, when a blue-emitting LED chip is used, only red and green phosphors are needed to produce white light emission.[100c] The Commission International de L'Eclairage (CIE) coordinates ($x$, $y$) are used to show the color of a light source perceived by the human eye. The standard (0.33, 0.33) is the ideal value for white light. In assessing the performance of WLED devices, three critical parameters are involved: luminous efficacy (LE), color rendering index (CRI), and correlated color temperature (CCT).[28] LE quantifies the energy efficiency of the light source, while the CRI measures the fidelity with which the light source renders colors, with values ranging from 0 to 100, where higher values denote superior color rendering capabilities. The CCT characterizes the spectral quality of the light source, typically indicating whether the light appears warm or cool. Specifically, WLEDs with a CCT between 2700 K and 3000 K produce warm yellow light, whereas those with a CCT between 5000 K and 6500 K emit cool white light. A comprehensive evaluation of these parameters facilitates a deeper understanding of device performance of phosphor-converted WLEDs.

Unlike electroluminescent LEDs, which require a balance between electrical, optical, and stability properties, the primary focus for phosphor-converted WLEDs based on perovskite NCs materials is on enhancing stability (chemistry/oxygen/moisture/thermal) and regulating spectral continuity.[250] Strategies based on encapsulation of perovskite NCs are important to improve the durability and reliability of phosphor-converted WLEDs. These mainly includes polymer encapsulation, inorganic oxide core-shell encapsulation, and composite encapsulation and have been discussed extensively in previous sections of this review.[250] In terms of spectral continuity, lead-free NCs characterized by broadband emission with a large Stokes shift are acquiring an increasingly important role the field of phosphor-converted WLEDs.[251] In this section, we will cove phosphor-converted WLEDs using lead-based and lead-free perovskite NCs.

### 5.2.1. Phosphor-converted WLEDs based on lead-based perovskite NCs

The merits of lead halide perovskite NCs, including high PLQY, high color purity, tunable emission covering the visible light region and facile synthesis, make them promising materials in phosphor-converted LEDs. The color gamut of phosphor-converted LEDs with lead halide perovskite NCs as emitting layers can reach over 110% of the NTSC standard. However, the poor stability limits their application. Encapsulating the NCs in protective shells represents an effective strategy to enhance the stability of the NCs, and has been discussed detailed in previous sections of this review. The reported phosphor-converted WLEDs based on lead halide perovskite NCs with improved stability are summarized in **Table 5**. Through encapsulation, in some cases WLEDs with relatively high luminous efficacy and good stability have been realized. For example, Zhong et al. reported an in-situ polymer encapsulation strategy to prepare $MAPbX_3$ NCs embedded in a PVDF matrix.[246] The resulting composite films exhibited improved stability against water and UV radiation even after a long aging time of 400 h. By integrating the green emitting composite films and $K_2SiF_6$ phosphor with the blue chip, they fabricated efficient a WLED with a LE of 109 lm $W^{-1}$ and a color gamut of 121% NTSC (Figure 13b).[246] Yoon et al. reported a highly stable $CsPbBr_3$ NCs−silica composite functionalized with surface phenyl molecules, which was synthesized by controlling the hydrolysis and condensation reactions of perhydropolysilazane followed by phenyl-functionalization.[247] The WLEDs underwent improvements in both green and red emission over time to reach an efficient performance of 38.8 lm $W^{-1}$ (Figure 13c).[247] The device could maintain 94% of the initial luminescence after operating at 20 mA for 101 h.[247] Mai et al. reported the use of $CsPbBr_3$/m-$SiO_2$ composites with superior stability against heat/water/oxygen as phosphors for WLEDs, which possess the CIE coordinate of (0.2985, 0.3076) and CCT of 7692 K.[99d] The device based on $CsPbBr_3$/m-$SiO_2$ composites polymer film could maintain 100% of the initial luminescence after operating at 200 mW/$cm^2$ for 240 h (Figure 13d, e).[99d] Zhu et al. synthesized double shell

$CsPbBr_3@SiO_2@PS$ NCs which combined the merits of the dense network of PS and robust inorganic shell of $SiO_2$. The WLEDs made of $CsPbBr_3@SiO_2@PS$ green NCs, KSF red phosphors, and blue chips had a high LE of 96 lm $W^{-1}$ and a long operational lifetime of 1200 h.[252]

**Table 5.** Summary of the device performances in phosphor-converted WLEDs based on lead halide perovskite NCs.

| Composite | Device structure | CIE coordinate | LE [lm $W^{-1}$] | Other indexes: CRI/CCT/NTSC | Device lifetime [maintain%/time (driving current)] | Refs. |
|---|---|---|---|---|---|---|
| $CsPbBr_3@$ m-$SiO_2$ | Blue chip/ composite/ $CsPb(Br_{0.4}I_{0.6})_3$ | (0.24, 0.28) | 30 | 113% NTSC | / | [109d] |
| | Blue chip/ composite/ $Sr_2Si_5N_8:Eu^{2+}$ | (0.3365, 0.3390) | 47.6 | CCT: 5318 K CRI: 72.3 | With increasing driving current (20 → 120 mA): LE: 47.6 →12.9 CCT: 5318 → 6253 CRI: 72.3 → 70.2 CIE coordinate: slightly blue-shifted | [109c] |
| | Blue chip/ transparent barrier polymer film/mixture (composite+$K_2SiF_6:Mn^{4+}$+$TiO_2$) enclosed in UV curable acrylate polymer (isobornyl acrylate based)/transparent barrier polymer films | (0.2985, 0.3076) | / | CCT: 7692 K 87% Rec.2020 | ~100%/240 h (operation under 200 mW/$cm^2$) | [99d] |
| $CsPbBr_3@$ PMAO | Blue chip/ composite/ commercial red phosphor (ER6436) | (0.390, 0.332) | 56.6 | CCT: 3320 K | / | [118] |
| $CsPbBr_3$@PECA-OVS | Blue chip/ composite/ commercial red phosphor | (0.33, 0.34) | 93 | CCT: 5300 K | little variation/ 800 min (15 mA) | [121] |
| $FAPbBr_3@$ PMMA | Blue chip/ composite/ N620 red phosphor | (0.37, 0.35) | 80.4 | CCT: 4000 K CRI: 90 | maintain good spectral stability | [253] |
| | | | | | under 20–80 mA driving current | |
| $CsPbBr_3:Al^{3+}$ and $CsPbBr_3$ | UV chip/ $CsPbBr_3:Al^{3+}$/ $CsPbBr_3$/ CdSe@ZnS NCs | (0.32, 0.34) | / | 116% NTSC | / | [16a] |
| $CsPbBr_3$@ $SiO_2$ | Blue chip/ composite/ CdZnSeS@ZnS−$SiO_2$ | (0.329, 0.336) | 38.8 | 120.1% NTSC | 94%/101 h (20 mA) | [247] |
| $CsPbBr_3$@ PE | Blue chip/ composite/ CdSe/CdS/ZnS | (0.324, 0.342) | 62.4 | CCT: 6028 K CRI: 37.01 95% NTSC | / | [117] |
| $CsPbBr_3$@ Sr/PbBr(OH)/m-$SiO_2$ | | (0.318, 0.341) | 86 | 124% NTSC | / | [109m] |
| $CsPbBr_3$@ Glass | Blue chip/ PiG plate (composed of $CsPbBr_3$@glass and $CaAlSiN_3:Eu^{2+}$) | (0.325 0.345) | 28 | CCT: 5849 K 115% NTSC | little variation with increasing driving current (60 → 160 mA) | [111c] |
| $CsPbMnX_3$@ $SiO_2$ | UV chip/ composite/ $CsPbBr_3$ NCs | / | 68.4 | CRI: 91 CCT: 3857 K | With increasing driving current (10 → 200 mA): LE: 68.4 →41.9 CCT: 3857 → 5934 CRI: 91 → 87 | [104c] |
| $CsPbX_3$@ $SiO_2$ | Blue chip/ $CsPbBr_3$@$SiO_2$/ $K_2SiF_6:Mn^{4+}$ | (0.336, 0.332) | 22.5 | CCT: 5340 K | / | [109i] |
| | Blue chip/ $CsPbBr_3$@$SiO_2$/ $CsPbBrI_2$@$SiO_2$ | (0.343, 0.328) | 7.6 | CCT: 4995 K | / | [109i] |
| $CsPbBr_3$@ $SiO_2$ | Blue chip/ composite /CIZS/ZnS/PVP | (0.352, 0.343) | 41.5 | CRI: 90.5 CCT: 4715 K | little variation with increasing driving current (20 → 120 mA) | [99e] |
| | Blue chip/ composite/ $K_2SiF_6:Mn^{4+}$ | (0.32, 0.30) | 63.5 | CRI: 83.3 CCT: 7425 K | no emission decay/ 13 h (6 mA) | [104e] |
| $CsPbBr_3$@ $Cs_4PbBr_6$ | | (0.39, 0.37) | 88 | 131% NTSC | / | [91c] |
| $MAPbBr_3$@ PVDF | | (0.272, 0.278) | 109 | 121% NTSC | / | [246] |
| $CsPbBr_3$@ PSZ | | (0.256, 0.304) | 71 | CCT: 9334 K | 92%/100 h | [100b] |
| $CsPbBr_3$@ zeolite Beta | | (0.288, 0.323) | 95 | 123.5% NTSC | / | [110] |
| $CsPbBr_3$@ $SiO_2$@PS | | (0.33, 0.32) | 100 | 128% NTSC | 91%/1200 h CCT: 5900 K CRI: 69 | [252] |
| $CsPbBr_3$/m-$SiO_2$@$SiO_2$ | | (0.33, 0.32) | 85 | 128% NTSC | / | [109f] |
| $CsPbBr_3$/ $CsPb_2Br_5$@$SiO_2$ | | (0.3421, 0.3373) | 31.5 | CCT: 5113 K 155% NTSC | Unchanged EL shape with increasing driving current (20 → 100 mA) | [109b] |
| $CsPbBr_3$@ Glass | | (0.338, 0.367) | 100.2 | 128.2% NTSC | / | [111a] |
| $CsPbBr_3$@ ZnO/$SiO_2$ | | (0.2966, 0.3323) | 52 | CCT: 7422 K 132% NTSC | 78% / 500 h (10 mA) | [104a] |
| $CsPbBr_3$@ ZIF-62 | | / | 50.4 | / | slight decrease/ 24 h (3 V) | [100a] |
| $CsPbBr_3$@ KIT-6 | | (0.32, 0.34) | 51.7 | 126.5% NTSC | 88%/30 days (5 mA) | [109k] |
| | | (0.327, 0.337) | / | 128% NTSC | / | [109l] |
| $CsPbBr_3$@ PMA | | (0.323, 0.345) | 58.4 | CRI: 83.2 CCT: 5916 K 125.3% NTSC | little variation/24 h (20 mA) | [123] |
| $CsPbX_3$@ PMA | Blue chip/ $CsPbBr_3$@PMA/ $CsPbBr_{1.6}I_{1.4}$@PMA | / | 4.5 | CRI: 72.4 CCT: 3665 K | CCT shifted from 3665 to 9300 K/ 5 min (58.8 mA) | [254] |

(KIT-6: mesoporous silica; MS: mesoporous molecular sieve; PiG: single phosphor-in-glass; PDMS: poly(dimethylsiloxane))

### 5.2.2. Lead-free metal halide NCs with broadband emissions for WLEDs

Down-converting materials exhibiting intrinsic broadband emissions are ideal candidates for applications in WLEDs since a continuous spectrum is required for artificial light sources to more closely mimic the natural light.[255] As previously discussed, various lead-free perovskite NCs can feature broadband emission and large Stokes shift, hence they are better suited, in principle, for white light emission. Artificial natural white light can be achieved based on mixtures of metal halide NCs with multicolor emissions, or single-phase white-light-emitting metal halide NCs. For example, a two-component strategy was proposed by several groups for white light emission by integrating blue-emitting $Cs_3Cu_2I_5$ NCs with yellow-emitting $CsCu_2I_3$ NCs.[141, 155b, 256] Both $Cs_3Cu_2I_5$ and $CsCu_2I_3$ have broadband emission that is sufficient to cover wide spectral ranges and so achieve high color rendering. Controlling the mixing ratio of $Cs_3Cu_2I_5$/$CsCu_2I_3$ led to a composite with a tunable cold/warm white light, with CIE color coordinates ranging from (0.418, 0.541) to (0.145, 0.055) (Figure 13h, i).[155b] The resulting WLEDs reached a record CRI of 91.6 for lead-free metal halide systems and demonstrated a long-term working stability.[155b]

Materials with single-component white light emission are more desired if one wants to decrease efficiency losses caused by self-absorption, prevent undesired changes in the composition over time in WLED, and simplify the device structure. Such materials have therefore the potential to overcome issues encountered in more conventional multicolor phosphor mixtures, such as efficiency losses caused by photon self-absorption, changes in the white light quality over time owing to the different degradation rates of the mixed phosphors, and increases in costs due to elaborate mixing processes.[251] White light emission from single phase metal halide NCs can be generally obtained from intrinsic STEs related broadband emission, or by the introduction of various dopants. For example, Locardi et al., reported the synthesis of broadband-emitting $Cs_2AgInCl_6$ NCs with a large FWHM of ~250 nm and Stokes shift of ~300 nm.[257] Subsequently, Yella et al. employed $Bi^{3+}$ doped $Cs_2AgInCl_6$ NCs with PMMA encapsulation to construct a single-component WLEDs that delivered an ideal warm light emission with CIE coordinates of (0.36, 0.35), CCT 0f 4443 K, and CRI of ~91.[152] Tang et al. further implemented the strategy of $Bi^{3+}$ doping and $Na^{+}$ alloying in $Cs_2AgInCl_6$ NCs, achieving a regulation of tunable white light emission from warm white to cool white with the CIE coordinates ranging from (0.246, 0.362) to (0.321, 0.445) and CCT from 9759.7 to 4429.2 K (Figure 13f).[248] Due to the lack of effective green and red light contributions, such single-component WLEDs suffered from the problem of low CRI. To overcome the shortcomings, Bai et al. introduced rare earth ions ($Sm^{3+}$) into $Cs_2NaInCl_6{:}Sb^{3+}$ NCs (see also earlier sections of this review). The red emission of $Sm^{3+}$ ions compensated for the lack of red light component in white light.[150] The WLEDs fabricated by combining the 310 nm chip and $Sm^{3+}$-doped $Cs_2NaInCl_6{:}Sb^{3+}$ NCs emitted ideal white light with CIE coordinates of (0.30, 0.28), LE of 37.5 lm $W^{-1}$, CCT of 8035 K, and CRI of 82 (Figure 13g).

An undeniable fact is that the reported literature does not always provide the LE parameter of the lead-free perovskite NCs-based WLEDs. Generally, the LE of WLEDs largely depends on the synergistic effects of the luminescence properties of phosphors and the selection of excitation source.[258] Most lead-free perovskite NCs have an optimal excitation wavelength shorter than 400 nm, which cannot match the main output wavelength ranges of 400‒450 nm for commercial GaN-related excitation sources with high output efficiency.[259] The LED chip with short-wavelength (< 400 nm) as the light source for WLED has a small light output efficiency, which leads to a large efficiency loss. Therefore, future efforts can be directed toward lead-free perovskite NCs with low excitation energy that can be excited by near-UV or blue light, while maintaining high PLQYs. These recommendations are amply discussed in the Conclusion and Outlook section.

### 5.3 Micro-LED displays

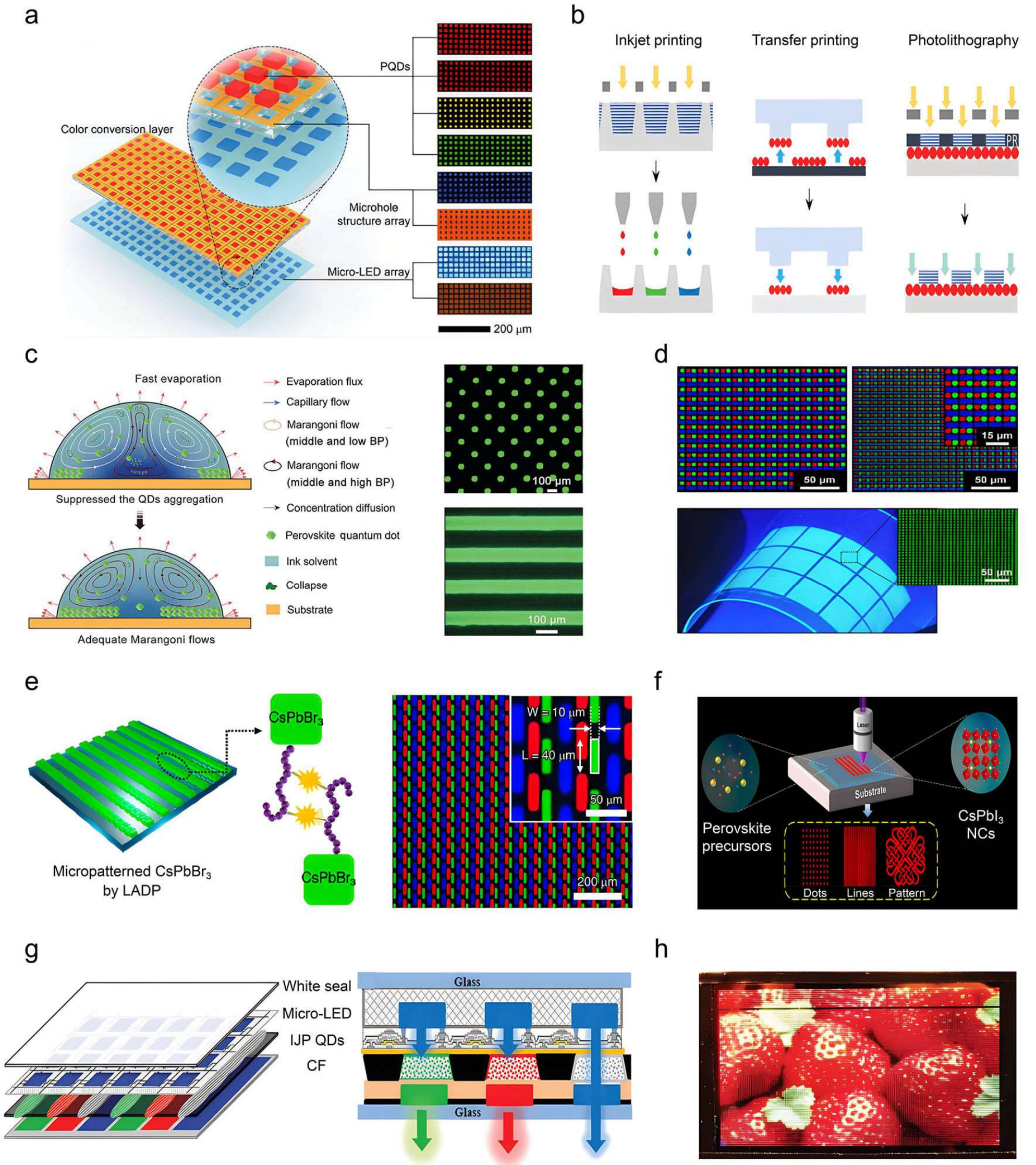

**Figure 14.** Micro-LED displays based on perovskite NCs. (a) Application of perovskite NCs in a micro-LED array. Reproduced and adapted from ref.[260]. Copyright 2023, Wiley. (b) Schematic diagram of inkjet printing, transfer printing, and photolithography for fabricating patterning perovskite NCs color conversion layers. Reproduced and adapted from ref.[261]. Copyright 2023, Wiley. (c) Ternary solvent engineering of inkjet printing. Reproduced and adapted from ref.[262]. Copyright 2022, Wiley. The right panels are the fluorescence optical microscopy images of nanodot arrays (upper) and stripe arrays (bottom). (d) PL image of the transfer-printed perovskite NCs layer under UV irradiation. Reproduced and adapted from ref.[21b]. Copyright 2022, AAAS. (e) Schematic diagram of ligand-assisted direct photolithography in perovskite NCs systems. Reproduced and adapted from ref.[263]. Copyright 2021, American Chemical Society. Right panel: PL images of the R/G/B perovskite NCs array. (f) Laser direct writing technology for preparing patterning perovskite NCs arrays. Reproduced and adapted from ref.[264]. Copyright 2021, American Chemical Society. (g) Schematic diagram and cross-section structure of the perovskite NCs-based micro-LED prototype devices. (h) A photograph of micro-LED operation. Reproduced and adapted from ref.[265]. Copyright 2020, Wiley.

Halide perovskite NCs have also been proposed for high-definition and high-brightness display technologies. The current main application route is that of enhanced color conversion layer (CCL) in LCD TVs.[246] Additionally, novel display technologies, such as micro-LED array displays, are exploring the integration of micro-LEDs with a pixelated perovskite NCs color conversion structure, which is poised to become a next-generation technological solution.[7b] In the traditional LCD display, a blue LED is combined with a traditional yellow phosphor such as Ce: YAG to generate white backlight.[266] However, owing to the inferior color purity of the YAG phosphor, the color gamut of traditional LCDs is barely satisfactory (70-80% NTSC). Although the adoption of perovskite NCs in LCD has expanded the color gamut, the issue of up to two-thirds of light waste in LCDs remains largely unresolved, since a polarizer and color filters are used to convert this mixture of white light into pixeled red, green, and blue color for display. Micro-LED displays are composed of millions of tiny LED pixels, each of which can be individually controlled.[7b] These self-luminous characteristics of micro-LEDs endow the display with excellent contrast and brightness, which does not require the polarizers, resulting in more vivid and clear images.[267] By adopting a technical approach that combines blue micro-LED with a perovskite NCs color conversion array, full-color display can be achieved. This display technology requires the manufacturing of a blue micro-LED display with an extremely high pixel density, and by patterning the perovskite NCs color conversion array some of the blue pixels are converted into red and green (F**igure 14**a).[260] To integrate perovskite NCs into full-color display panels, the fabrication of perovskite NCs pixelated patterns is crucial, while alleviating problems such as color cross-contamination and reduction in luminous efficiency. Currently available pixelation methods for perovskite NCs CCL used for micro-LED arrays include techniques such as inkjet printing, transfer printing, photolithography, and in-situ laser direct writing (Figure 14b). **Table 6** summarizes the patterning techniques, R/G/B multi-color pixelization, advantages, challenges, and resolutions of perovskite NCs patterning, illustrating the distinct characteristics of diverse patterning methods. Detailed strategies for the patterned perovskite NCs CCLs and LEDs implementation will be discussed in the following sections.

**Table 6.** Summary of patterning strategies based on perovskite NCs.

| Patterning strategy | R/G/B multi-color pixelization | Advantages | Challenges | Min. pattern size | Max. printing area | Ref. |
|---|---|---|---|---|---|---|
| Inkjet printing | Yes | i) Accurate alignment<br>ii) Orthogonality | i) Pattern size limitation<br>ii) Coffee ring effect | 1 μm | 1288 $m^2$ | [268] |
| Transfer printing | Yes | i) Small pattern size<br>ii) No organic solvents | i) Low throughput<br>ii) Sophisticated process | 400 nm | 630 $mm^2$ | [21b] |
| Photolithography | Yes | i) Accurate alignment | i) UV damage<br>ii) Photoresist damage | 5 μm | 25 $mm^2$ | [269] |
| Laser direct writing | Yes | i) Small pattern size<br>ii) Accurate patterning | i) Low throughput<br>ii) Damage to bottom layer | 0.5 μm | 34 $mm^2$ | [270] |

The basic principle of perovskite NCs inkjet printing technology is to spray inks containing red, green, and blue perovskite NCs onto a substrate separately, and to control the final pixel size by adjusting factors such as the size of the ink droplets, the characteristics of the printing substrate, and the size of the dam.[271] The quality of printed NCs film greatly determines the display effect. As a single droplet of ink evaporates and dries on the substrate, uneven accumulation of perovskite NCs often occurs owing to the capillary flow from the interior of the droplet to the edge, the so-called coffee ring effect.[272] Therefore, in the inkjet printing process, the suppression of the coffee ring effect is indispensable for uniform converter film luminescence and final display effect. Adjusting the formulation of the perovskite NC ink including solvent ratio, additives etc., is particularly critical to the quality of the final patterning CCL.[271] Sun et al. employed a dual-solvent adjustment strategy to regulate the evaporation rate and surface tension of $CsPbBr_3$ NCs inks, resulting in a uniform NCs film with a surface roughness of 2.19 nm.[273] Combing the inkjet printing and UV-curing method by adding UV-curing additives into the perovskite NCs ink could inhibit the formation of coffee rings and result in a 6 μm patterned perovskite NCs film with uniform morphology.[274] Zeng et al. proposed a ternary solvent-ink strategy to prepare a highly dispersible and stable $CsPbX_3$ NCs ink, which had better printability and film-forming ability than binary solvent systems (Figure 14c).[262] They pointed out that the addition of a low-boiling-point solvent (*n*-nonane) to the ink can greatly suppress the aggregation of perovskite NCs and accelerate the evaporation of the solvent as well as inhibit the coffee ring effect compared with the binary-solvent-system. Based on such inkjet printing technology, a maximum luminance of 43,883.39 cd $m^{-2}$ and peak EQE of 8.54% were achieved. This strategy showed general applicability in red and blue perovskite NCs LEDs as well as traditional Cd-based LEDs, and this represented a promising avenue for the inkjet-printed LEDs towards commercialization.[262]

Transfer printing is another promising technology compatible with micro-LED arrays, which uses embossing or imprinting of elastomers to duplicate NCs patterns on target substrates.[271] This method has several advantages, including the absence of additional organic additives and the avoidance of solvent exposure during pattern formation. Transfer printing typically involves two steps: a rapid pick-up process from a donor substrate with low surface energy and a release process using a viscoelastic stamp to the target substrates. Among the several steps associated with the process, the importance of uniform film transfer with high transfer yield has been often emphasized. Li et al. proposed a transfer printing method by directly spin-coating perovskite NCs onto the surface of the poly(dimethylsiloxane) (PDMS) stamps, which prevented internal cracking of the perovskite NCs films during the transfer printing process.[275] In addition, the introduction of an ultra-thin branched-polyethyleneimine (B-PEI) layer with amine functional groups as the interfacial chemical bonding layer could enhance the adhesion and electrical contact between perovskite NCs and HTLs. Such transfer printing of perovskite NCs enabled a monochromic pattern resolution of 1270 PPI and reaches micro-LED pixels with an EQE of 10.5% for the red (680 nm) emission.[275] Recently, Kwon et al. solved the problem of internal cracking of perovskite NCs films during dry transfer printing using double-layer transfer printing of perovskite NCs/ETLs.[21b] The weak adhesion between the PDMS stamps and ETLs allowed the easy removal of the perovskite NCs/ETLs bilayer from the stamp without damaging the NCs films. This double-layer transfer printing could reach a high-definition R/G/B pixelated perovskite NCs pattern with 2550 PPI and monochrome line pattern with a width of 400 nm and a transfer yield close to 100% (Figure 14d).

Photolithography is a mature patterning technology, which is widely applied in the fabrication of patterns at the micrometer scale.[276] In the photolithography process, the steps of photoresist spin-coating, alignment and exposure, residue lift-off are involved. The pattern formed by the photoresist is then used to determine the size and position of the pixels. After the lithography process, it is necessary to lift off the remaining photoresist. This process, with its complex solvent system, always damages the perovskite NCs layer, leading to a decrease in its optical performance.[277] To solve it, it is possible to mix perovskite NCs into photosensitive materials for direct patterning.[278] This method avoids the complex photolithography lift-off solvent treatment process, thereby preserving the performance of the perovskite NCs to some extent. However, due to the substantial incorporation of photoresist, this approach can lead to a lower light conversion rate in the conversion film, resulting in blue light leakage. Lin and co-workers employed an orthogonal lithography strategy utilizing fluorinated polymers and solvents to fabricate multi-color composite perovskite NCs patterns.[279] A fluorinated polymer resist was coated on the substrate as sacrificial layer and patterned by conventional photolithography processes. The utilization of fluorinated polymers solved the problems associated with polar-nonpolar solvent limitations, resulting in the successful formation of highly stable and high-definition perovskite NCs patterns with a resolution of 1000 PPI.[279] By employing post-treatment tactics involving anion and/or ligand exchange, the optical properties of patterned perovskite NCs can be improved. Bang et al. demonstrated photopatternable $CsPbBr_3$ NCs using multifunctional polymer ligands of poly(2-cinnamoyloxyethyl methacrylate).[263] Under UV irradiation, the cinnamoyl groups of polymer crosslink with each other, forming stable perovskite NCs patterns. These patterns displayed good stability under environmental conditions and chemical exposure. The terminal groups of the polymer can be easily exchanged with ammonium halides, allowing the versatile multicolored patterning of perovskite NCs through ion exchange (Figure 14e).

Laser direct writing technology can be considered a maskless photolithography technique.[264] In this process, the preparation of pixel points in the film is accomplished by controlling the patterned movement of the laser beam, and the pixel size should be more controllable, which can meet the requirements of miniaturization and scalability of micro-LEDs. Additionally, due to the low formation energy of perovskite NCs, laser direct writing technology can facilitate the in-situ formation of NCs with desired patterns.[264] For example, by using the in-situ direct laser writing fabrication of patterned $CsPbI_3$ NCs, Zhong et al. demonstrated various patterned display arrays with bright red emission with high PLQYs up to 92% (Figure 14f).[264] By changing the power and scanning speed of laser, the optical grating line width of the patterned $CsPbI_3$ NCs can be adjusted to 0.9 μm. The method provides a pathway for fabricating patterned NCs with rational designed structures for micro-LED displays.[264]

The combination of blue micro-LED and perovskite NCs to achieve high-resolution full-color micro-LED display is considered to be a promising display solution since it integrates the advantages of high-resolution of micro-LEDs with the wide color gamut of perovskite NCs.[280] Meng et al. reported full-color micro-LED display prototypes by using patterned perovskite NCs as the CCLs (Figure 14g), which was fabricated by applying inkjet printing technology.[265] The assembled micro-LED displays exhibit the colorful image of the strawberry very well (Figure 14h), which basically meets the requirements of commercialization. Compared to conventional purely R/G/B LEDs-based micro-LED displays, the resulting display prototype exhibited a significantly higher color gamut, which could reach up to 129% NTSC. Notably, the color gamut of 126% NTSC or 94% BT. 2020 can be reached by only using perovskite NCs as the CCLs.

### 5.4 Active-matrix LED displays

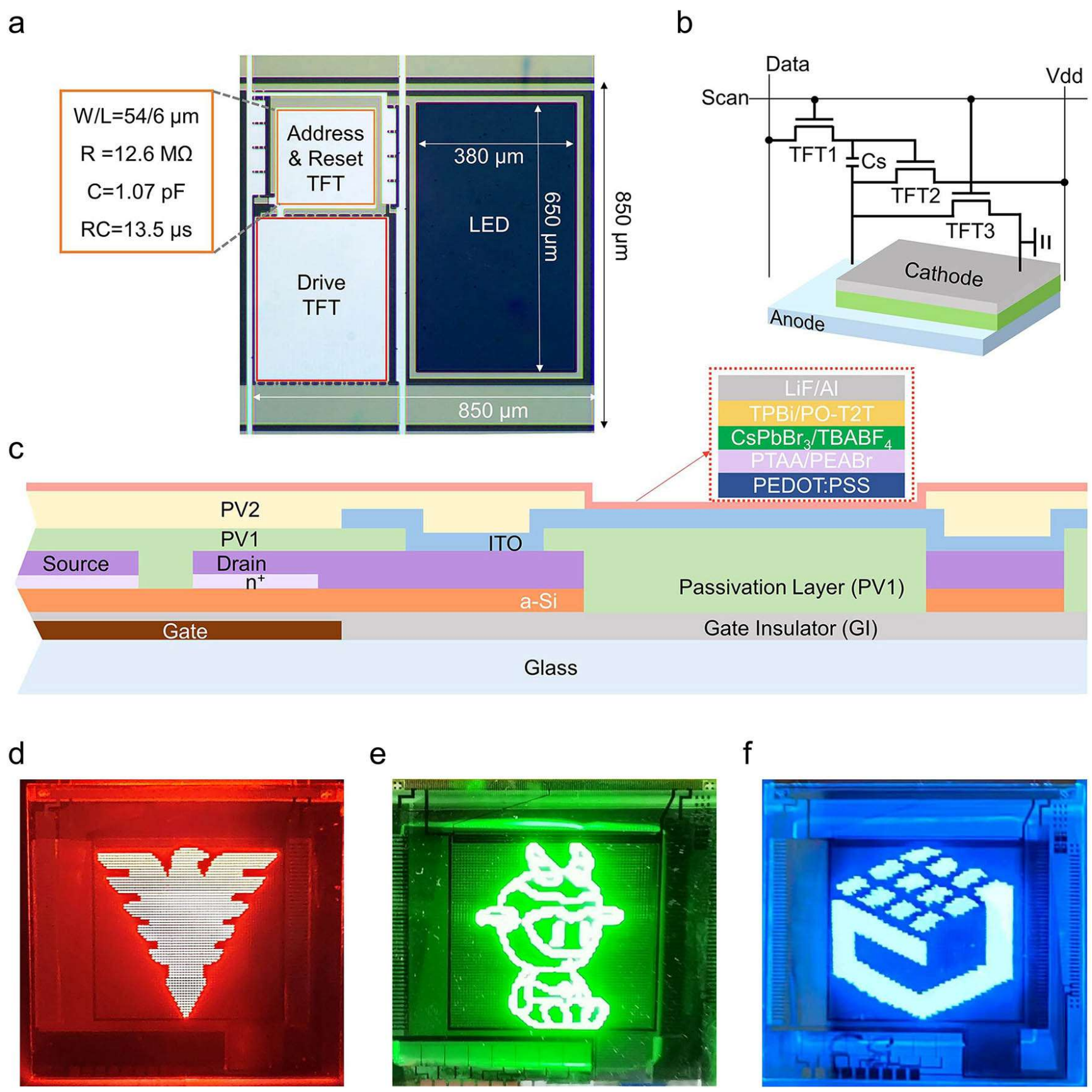

**Figure 15.** Active-matrix LED displays based on perovskite NCs. (a) Micrograph of one single pixel in the TFT backplane, exhibiting the architectural design of TFT and LED. (b) Schematic diagram of TFT driven display based on perovskite NCs systems. (c) Structure diagram of the drive TFT circuits and LEDs. (d−f) Digital photographs of the red, green, and blue emissive

active-matrix perovskite NCs LEDs. Reproduced and adapted from ref.[7a]. Copyright 2024, Springer Nature.

An active-matrix display is a self-emissive device in which each pixel of a LED is independently controlled using thin-film transistor (TFT) circuits. Compared with conventional LCDs, active-matrix displays possess the advantages of wider viewing angle, lower power consumption, and higher contrast.[281] The transition from laboratory-scale individual LED device to an industry-level active-matrix LED display is a necessary step for the commercialization of perovskite NCs.[23, 282] Achieving efficient perovskite NCs LEDs in a single device requires the development of high-quality NCs and advanced engineering of device architectures, as described in the previous sections. Combining the state-of-the-art perovskite NCs and device technologies with precise patterning technology is a key step towards realizing active-matrix perovskite NCs LED displays. Dai et al. specifically designed a TFT circuit with a pixel density of 30 pixels per inch to integrate perovskite NCs LEDs into each pixel and deposited a thick passivation layer to protect the TFT.[7d] The upper passivation layer was sputtered to expose ITO for integration with the perovskite NCs LEDs to achieve an emissive unit with an area of 380 × 650 μm. Such configurations allowed the realization of high-definition images and videos with accurate and continuous grayscale representation. To improve the refresh rate and response speed of perovskite NCs displays, they further integrated an individual-particle-passivated perovskite NCs film into an active-matrix LEDs (**Figure 15**a−c).[7a] Using this configuration, the response time of the resulting active-matrix NCs LED was reduced from milliseconds to microseconds, with a corresponding EQE of more than 20.0%. The concept is universal and can be extended to halide-mixed perovskite NCs systems, leading to bright R/G/B displays (Figure 15d−f).

Active-matrix LED screens based on colloidal perovskite NCs have been rarely reported, and lessons learned from commercialized organic LEDs are needed. The development of active-matrix perovskite NCs displays also requires additional considerations, such as i) backplane selection, ii) pixel-defining layer (PDL) design, iii) prevention of color crosstalk, iv) R/G/B full-color cavity structure design, and v) sub-pixel arrangement and shape. These problems do not arise in individual LED studies but become important in the implementation of active-matrix displays. Through unremitting exploration, active-matrix perovskite NCs screens will be successfully commercialized in the foreseeable future.

## 6. Conclusion and Outlook

In this review we have covered several aspects of halide perovskite NCs and related metal halide NCs in lighting applications, with focus on LEDs. What follows is a summary of the main outstanding issues in the field, along with recommendations for further strategies aimed at addressing such issues and an outlook on future developments.

1) *Optimization of perovskite NCs to improve the optoelectronic performance*. Although significant progress has been made in understanding the surface physicochemical properties of perovskite NCs in the past few years, many issues related to ligand-assisted colloidal synthesis, surface states formation, and surface structure evolution remain largely unexplored.[283] For instance, how ligands affect the nucleation, growth, and shape evolution of perovskite NCs is not fully understood, although ligands have been shown to play a crucial part in controlling their synthesis. Moreover, due to the weak and dynamic binding of the ligands to the NCs surface, the commonly used OLA/OA ligand pairs are easily removed from the NCs during the washing process, leading to the emergence of surface defects which reduce their PLQY.[284] Much of the research has focused on the use of various organic ligands and inorganic metal halide ligands for surface passivation to improve PLQY and colloidal stability. Unfortunately, the current knowledge of how different ligands influence the surface termination and passivation effects of NCs at the atomic level is still limited. Equally important, there are differences in the surface chemistry and ligand interactions of perovskites with different halide compositions, making it difficult to generalize the surface passivation mechanism and strategies to all the halide perovskite NCs and even less to other metal halide NCs.

Hence, further research on modeling and experimental studies of ligand binding is needed. Although many studies have claimed that reducing the length of the ligands favors charge transport in the NCs films, some reports have shown that the device works well even if the ligands have poor conductivity.[285] However, this is only possible if there is an effective electronic coupling between the LUMO of the ligand and the CBM of the perovskites.[283] Therefore, to improve the electrical conductivity of the perovskites NCs films, effective electronic coupling should be the focus of future research. Furthermore, it is of great interest to investigate the influence of ligands on the electronic band structure of perovskite NCs. Many studies have aimed to improve the optoelectronic properties of perovskite NCs by doping or alloying them with various ions in the host lattice. However, in many cases it is still unclear whether these impurity ions are fully doped into the lattice or are only/primarily on the surface. Moreover, gaining a more accurate structural information on the dopants, including the exact lattice site occupation and the distribution of the dopants in the structure, is still elusive in most situations. To achieve a proper understanding of the performance changes caused by doping, additional efforts are needed to characterize the material structure, especially in the local environment of the dopant. In addition, most A-site doping is limited to a few organic molecules, such as MA and FA.[286] In principle, additional organic groups should be explored to stabilize the phase structure of perovskite. At present, the reports on B-site doping are still controversial, and there is an urgent need to clarify the relationship between the electronic structure and optoelectronic properties of the doped perovskite NCs under the guidance of theory. Computational modeling might help to identify effective doping/alloying compositions to improve the performance of perovskite NCs. Another unresolved question concerns the effect of dopants on the surface termination and passivation of perovskite NCs and their interaction with surface ligands. Understanding these effects is crucial for optimizing the performance of doped perovskite NCs in LED applications.

2) *Tackling problems with the stability of metal halide NCs LEDs*.

i) Emission stability of metal halide NCs in harsh environments (e.g., high temperatures, photoirradiation, humidity, oxygen, and strong acids/bases). Encapsulation of metal halide NCs in various organic polymers or inorganic matrixes has been actively researched, however it is unclear whether this protective "shell" completely prevents the penetration of oxygen and water into the NCs over long periods of time. Although various encapsulation strategies can effectively improve the NCs thermal stability,[287] their thermal quenching is still higher than that of commercial phosphors.[288] In-depth theoretical calculations of degradation processes and mechanisms are of great significance for gaining a deeper understanding on the instability and designing more stable perovskite NCs. Targeted research on each degradation pathway would help to effectively improve the stability of metal halide NCs. In addition, strategies to improve the stability of metal halide NCs might degrade other properties. For example, although the use of an insulated matrix package enhances the stability of metal halide NCs, its charge transfer is suppressed as the NCs are basically insulated from each other. The use of conductive encapsulation materials or ligand engineering may represent promising strategies to improve the stability and maintain the conductivity of metal halide NCs. For example, Li et al. replaced oleate with diisooctylphosphinic acid as the coordinating ligand and "repaired" the NC surface by etching with hydriodic acid to stabilize small sized $CsPbI_3$ NCs. With this, they could fabricate a pure red LED with a EQE of 28.5% and an operational half-lifetime surpassing 30 h.[7d]

ii) Operational lifetime and spectral stability of metal halide NCs LEDs. At present, the longest operational lifetime ($T_{50}$) of perovskite NCs LEDs is 30,000 h under the initial luminance of 100 cd $m^{-2}$, while those of organic LEDs and cadmium-based quantum dot LEDs have exceeded 1,000,000 h.[289] Strategies should target the following aspects in order to improve the operational lifetime. Firstly, a uniform and compact surface morphology of NCs layer can reduce localized leakage current that generates Joule heating. Secondly, charge injection balance can lead to uniform radiative recombination throughout the NCs layers, avoiding ion migration behavior induced by local charge accumulation. Thirdly, defect passivation can effectively prevent ion migration in the NCs layers and suppress the increase of non-radiative recombination centers, thereby improving the device lifetime. In addition, as for mixed halide NCs LEDs, the EL spectra are usually shifted under forward drive voltages or long periods of operation due to ion migration and phase separation, resulting in poor spectral stability and seriously affecting the chromaticity of the display.[290] The primary strategy to inhibit ion migration in mixed halide NCs is to increase their defect formation energies by passivation and ion-blocking layers, and by addition of small molecules with functional groups. Also, the introduction of new device architectures, such as top-emitting structures or tandem structures, which can offer high luminance under a low bias voltage, is a promising strategy to improve the stability of mixed halide NCs LEDs.

3) *Deeper understanding of optoelectronic properties for designing lead-free perovskite NCs used in LEDs*. According to the RoHS (Restriction of Hazardous Substances) directive, the maximum allowable concentrations of cadmium (Cd) and lead (Pb) in electronic consumer products are 100 ppm and 1000 ppm, respectively. Assuming the use of $CsPbBr_3$ NCs as the emissive layer (~20 nm in thickness) in a 75-inch display screen, the total lead content in the device is approximately 123 mg. To adhere to RoHS regulations concerning lead content, the total weight of the device must exceed 123 grams. In comparison, devices using CdSe NCs as the emissive layer require a substantially higher total weight of approximately 2.46 kilograms to meet the regulations, which is significantly more challenging than achieving compliance with lead-based materials. In the current stage, the weight of a typically 75-inch display screen in the market is significantly greater than the threshold value for lead, so using Pb based nanocrystal as emitting material in screen is allowable at present. However, as the demand for lightweight electronic devices continues to increase, coupled with the likelihood of more stringent regulations under the RoHS directive, there will arise a compelling necessity to engineer non-lead based materials for application in electronic displays and other consumer electronics. Besides, we still need to clarify better the mechanism of emission in these materials. Research in this direction will provide important insights for electronic structural engineering to improve the PLQY. Moreover, as for the luminescence mechanism of STEs with broadband emission, it is difficult to directly observe the occurrence and evolution of the STEs, and this can only be inferred from the measured optical properties. The distribution of singlet and triplet STEs, as well as the factors affecting the reverse intersystem crossing (RISC) process between them, are still unclear. To further investigate these photophysical mechanisms, advanced techniques such as ultrafast spectroscopy to study the dynamics of excited states and the observation of transient lattice distortions by ultrafast transmission electron microscopy need to be utilized.[29, 291]

For most lead-free perovskite NCs developed so far, their emission bandwidth is wider than that of lead-halide perovskites, making them unsuitable for display applications. However, the emissions originated from STEs have great competitiveness in WLED applications due to their non-self-absorbing and broadband emission properties, as already mentioned. Nevertheless, it is still important to consider the development of lead-free perovskite NCs-based WLEDs with suitable excitation wavelengths and appropriate device encapsulation technologies for practical applications. For electroluminescent LEDs, developing NCs with high PLQY and excellent carrier mobility is the ultimate goal. Furthermore, achieving a high quality of the NCs layers is an aspect that cannot be disregarded. Additive engineering and post-treatment are two strategies that can be used to improve the quality of NCs layers. The low-dimensional electronic structure of many lead-free perovskites leads to a large carrier effective mass and thus to low charge carrier mobility and poor conductivity. In addition, the poor charge injection caused by the energy level mismatch is one key reason for insufficient EL properties. To promote charge injection into the emitters, the selection of suitable HTLs and ETLs or the introduction of additional charge transport functional layers as bridges are possible approaches to reduce the injection barriers and increase the EL efficiency.

4) *Developing key strategies to improve the efficiency limit of perovskite NCs LEDs*. Although the EQE of perovskite NCs-based LEDs is close to 30%, there is still much room for improvement in device performance, which is still far behind that of commercial organic LEDs and cadmium-based quantum dot LEDs (EQE over 40%).[55] The main factors affecting the luminescence efficiency of LEDs include high density of defects in the NCs emitting layers, unbalanced carrier injection, and poor light extraction efficiency. Most of the intrinsic point defects in perovskites tend to form shallow carrier traps, which is considered tolerable.[292]

However, some defects can form traps that are deep in the band gap, which is detrimental for radiative recombination of perovskite NCs LEDs. The defects in perovskite NCs layers can be minimized by adjusting the composition of ligands on the crystal surface, introducing passivators, and doping/alloying ions in crystal. From the device structure point of view, the energy mismatch of NCs layer/charge injection layer can create an injection barrier and affect the injection of charge carriers. Modifying the HTLs to effectively increase the work function can reduce the injection barrier between the NCs layers and HTLs, thereby balancing the carrier injection. We believe that exploring new HTLs with high charge carrier mobility and deep valence band energy levels may be necessary to push perovskite NCs LEDs to their theoretical maximum. Another issue is that heterojunction interfaces are susceptible to defects, leading to the quenching of charge carriers at the interface. To passivate these defects, small molecule passivators can be applied to the top and bottom of the NC layers. Inspired by the success of cadmium-based quantum dot LEDs, the development of inorganic core-shell perovskite NCs may become a solution to improve the efficiency of perovskite NCs LEDs. In principle, the formation of core-shell structure can passivate defects, suppress ion migration, enhance charge injection and transport, and confine excitons within the core by forming a wide band gap shell of type-I heterostructures, which promotes efficient radiative recombination. Due to the ionic properties of perovskite, the formation of covalently bonded inorganic semiconductor shells on perovskite NCs core may pose synthesis challenges, requiring further efforts from the chemical community to identify new ingenious synthesis avenues and more rigorous structural characterization of the NCs that are synthesized.

The introduction of photonic structures into devices can improve light extraction and thus significantly increase the device efficiency of LEDs.[200] Due to the refractive mismatch between the NCs layers and the glass substrates, part of the light generated inside the device is restricted or lost in waveguide modes. Moreover, total internal reflection at the interface between the glass substrate and air results in outcoupling losses. Therefore, developing effective outcoupling technologies to harvest internal light is a promising way to enhance the EQE of perovskite NCs LEDs. This strategy will become increasingly effective in improving the luminescence efficiency of the devices.

5) *Development of scalable and high-resolution fabrication towards commercialization*. Commercial demand for high-resolution displays, including micro-LEDs and active-matrix LEDs, requires scalable and high-resolution fabrication of perovskite NCs.[271] Therefore, it is vital to develop patterning technologies that can accurately deposit high-resolution and mass-produced perovskite NCs films. During the patterning process, the native electrical and optical properties of perovskite NCs and charge transport layer should not be compromised. For example, inkjet printing and photolithography have problems with organic residues that hinder efficient charge transport in NC patterns. Moreover, photolithography technology requires the introduction of photoresist, which can cause damage to the NCs layer or introduce other impurities during the process, thereby affecting the optoelectronic performance of the device. Transfer printing technology provides a way to these problems because it avoids the harmful effects of solvents on the substrate. For example, Kwon et al. developed a double layer transfer printing of perovskite NCs to solve the problem of internal cracking of the NCs patterns in the transfer printing process.[21b] They thus created a high-definition RGB pixelated perovskite NCs patterns with 2550 PPI and monochromic patterns of 33,000 line pairs per inch with a transfer yield of close to 100%.[21b] The transfer printing technology is still at the laboratory research stage, and is currently difficult to apply to large-scale fabrication of the patterned perovskite NCs films. In addition, there is an urgent need to design and optimize programmable elastomeric stamps in transfer printing processes, which is crucial for the effective assembly of complex pattern structures.

With the development of perovskite NCs synthesis, device fabrication, and NCs pattern technology, the ultimate goal of next-generation perovskite NCs lighting and displays is getting closer and closer. We hope that this review will help readers getting a broader view of the current research progress on perovskite NCs LEDs and encourage them to invest more efforts in this field.

**Acknowledgements**

Y.L. and Z.M. contributed equally to this work. The work was supported by the National Natural Science Foundation of China (No. 12304458, 12004298, 12404466, 12304473 and 12074347), the Key Project for Science and Technology Development of Henan Province (No. 232102231039), China Postdoctoral Science Foundation (Grant No.2023M743222, BX20230330 and 2024M752915), the Key Project for Science and Technology Development of Xianyang City under Grant L2024-ZDYF-ZDYF-GY-002, and the Postdoctoral Fellowship Program of CPSF (No. GZB20240594). L. M. acknowledges funding from the programme MiSE-ENEA under the Grant "Italian Energy Materials Acceleration Platform – IEMAP" and from the European Research Council though the ERC Advanced Grant NEHA (contract n. 101095974).